\documentclass[aps,prd,twocolumn,groupedaddress,showpacs,eqsecnum,floatfix]{revtex4-1}
\usepackage{amsmath}
\usepackage{graphicx}
\usepackage{bm}
\usepackage{braket}
\usepackage{tensor}
\usepackage{slashed}
\usepackage{color}
\usepackage[colorlinks]{hyperref}

\definecolor{CiteColor}{rgb}{0,0.5,0}
\hypersetup{citecolor=CiteColor}
\definecolor{RefColor}{rgb}{0.55,0,0}
\hypersetup{linkcolor=RefColor}
\definecolor{darkgreen}{rgb}{0.2,0.7,0.2}

\newcommand{\bg}{\bar{\gamma}}
\newcommand{\dg}{\delta \gamma}
\newcommand{\bh}{\bar{h}}
\newcommand{\barf}{\bar{f}}
\newcommand{\df}{\delta f}
\newcommand{\bpsi}{\bar{\psi}}
\newcommand{\bA}{\bar{A}_0}

\newcommand{\beq}{\begin{equation}}
\newcommand{\eeq}{\end{equation}}

\begin{document}

 \title{Stability of black holes in Einstein-charged scalar field theory in a cavity}

\author{Sam R.~Dolan}
\email{s.dolan@sheffield.ac.uk}

\author{Supakchai Ponglertsakul}
\email{smp12sp@sheffield.ac.uk}

\author{Elizabeth Winstanley}
\email{e.winstanley@sheffield.ac.uk}

\affiliation{Consortium for Fundamental Physics, School of Mathematics and Statistics,
The University of Sheffield, Hicks Building, Hounsfield Road, Sheffield S3 7RH, United Kingdom.}

\date{\today}

\begin{abstract}
Can a black hole that suffers a superradiant instability evolve towards a `hairy' configuration which is stable? We address this question in the context of Einstein-charged scalar field theory. First, we describe a family of static black hole solutions which possess charged scalar-field hair confined within a mirror-like boundary. Next, we derive a set of equations which govern the linear, spherically symmetric perturbations of these hairy solutions. We present numerical evidence which suggests that, unlike the vacuum solutions, the (single-node) hairy solutions are stable under linear perturbations. Thus, it is plausible that stable hairy black holes represent the end-point of the superradiant instability of electrically-charged Reissner-Nordstr\"om black holes in a cavity; we outline ways to explore this hypothesis.
\end{abstract}


\maketitle

\section{Introduction}\label{sec:intro}

The belief that a `typical' galaxy hosts a supermassive black hole, of mass $M \sim 10^5$--$10^{10} M_{\odot}$, is supported by dynamical evidence from nearby galaxies and extrapolation of the black hole mass--velocity dispersion relation \cite{King:2003ix}. Recent surveys suggest that a supermassive black hole may possess significant angular momentum \cite{Risaliti:2013cga, Reynolds:2013rva}. Black hole spin is conjectured to power relativistic jets in quasars through the Blandford-Znajek process \cite{Blandford:1977ds}, featuring an accretion disk and a force-free magnetosphere \cite{Gralla:2014yja}.

The Blandford-Znajek process is just one example of a Penrose process \cite{Penrose:1971uk}, in which a black hole may liberate energy and angular momentum (and/or charge) whilst still increasing its horizon area. Penrose processes are consistent with -- indeed, encouraged by -- the second law of black hole mechanics \cite{Bardeen:1973gs} and thus, it would appear, the second law of thermodynamics \cite{Hawking:1974sw, Bekenstein:1974ax}. One intriguing example of a Penrose process is \emph{superradiance} \cite{Starobinskii:1973}, in which a low-frequency electromagnetic or gravitational wave packet is amplified by a black hole (see \cite{Brito:2015oca} for a recent review). In the `black hole bomb' scenario \cite{Press:1972zz}, an exponentially-growing instability is stimulated by reflecting a superradiant field back onto the black hole. In scenarios with light bosonic fields (e.g.~axions \cite{Arvanitaki:2010sy} or massive photons \cite{Pani:2012vp}), the instability may develop in the gravitationally-bound modes of the field \cite{Damour:1976kh, Zouros:1979iw, Detweiler:1980uk, Cardoso:2005vk, Furuhashi:2004jk, Dolan:2007mj, Rosa:2009ei, Kodama:2011zc, Witek:2012tr, Dolan:2012yt, Shlapentokh-Rothman:2013ysa}, and thus arise spontaneously \cite{Berti:2015itd}. Variations on this theme involving only electromagnetic fields and accretion disks have also been discussed \cite{VanPutten:1999vda}.

In this article we address a key question: can a superradiant instability, pursued into the non-linear regime, lead to a new \emph{stable} hairy black hole configuration?

Superradiant instabilities appear to pose a challenge to the `no-hair' (Israel-Carter) conjecture \cite{Israel:1967za, Carter:1971zc, Ruffini:1971bza}, which asserts that a perturbed black hole should settle back into a stationary state, changing only a small number of  parameters (mass $M$, angular momentum $J$, and charge $Q$). The conjecture has been codified in a number of theorems in asymptotically-flat scenarios involving massless scalar, electromagnetic and gravitational fields under certain minimal assumptions \cite{Bekenstein:1971hc, Bekenstein:1972ky, Bekenstein:1995un, Bekenstein:1996pn, Chrusciel:2012}; there also exist  stability theorems for Kerr spacetime \cite{Whiting:1988vc, Shlapentokh-Rothman:2013hza, Dafermos:2014cua}. Nevertheless, as was recently shown in Refs.~\cite{Herdeiro:2014goa, Herdeiro:2015gia}, there exists an asymptotically-flat family of Kerr black holes endowed with (complex, massive) scalar-field `hair', which reduce to well-known `solitonic' boson star solutions in a well-defined limit. Crucially, these new solutions lie beyond the scope of the no-hair theorems, as the scalar field is only helically-symmetric, rather than being stationary and axisymmetric. More precisely, the scalar field is only invariant under a single Killing field, which is tangent to the null generators of the horizon. Novel solutions with a single Killing field were described in Ref.~\cite{Dias:2011at}. For a succinct summary of the relationship between various asymptotically-flat scalar-hairy solutions, the no-hair theorems, and the violated assumptions, see Table I in Ref.~\cite{Herdeiro:2015waa}.

In the small-amplitude regime, superradiance arises for charged scalar perturbations of the Kerr-Newman spacetime which have frequency $\sigma $ satisfying $\sigma (\sigma - \sigma_c) < 0$. The critical frequency $\sigma _{c}$ is given by
\begin{equation}
\sigma_c = m \Omega_H + q \phi_H,
\end{equation}
with  $\Omega_H = J/2M^2r_+$ and $\phi_H = Q / r_+$ the angular frequency and electric potential of the black hole horizon at $r=r_+=M+\sqrt{M^2-(J/M)^2-Q^2}$, and $m$ and $q$ the azimuthal mode number and charge of the scalar field, respectively. At the critical frequency $\sigma = \sigma_c$, linear perturbations are stationary: they do not decay or grow (see e.g.~Ref.~\cite{Hod:2012px}). In the limit of small field amplitude, the Kerr-scalar solutions in \cite{Herdeiro:2014goa, Herdeiro:2015gia} reduce to a Kerr black hole ($Q=0$) with a co-rotating dipolar ($\ell = m = 1$, where $\ell $ is the total angular momentum mode number) perturbation in the massive scalar field at the critical superradiant frequency, $\sigma = \sigma_c$.

Analysing the stability of the non-linear Kerr-scalar solutions (with $J\neq0$, $Q=0$) is challenging, principally because such solutions are only helically-symmetric (as well as being numerically-determined, i.e. not known in closed form). Here, we consider a simpler \emph{spherically-symmetric} model, with $J=0$, $Q\neq0$, of scalar electrodynamics coupled to gravity \cite{Gundlach:1996vv}. In this scenario, superradiance is associated with charge, rather than angular momentum. It was shown by Bekenstein (see Ref.~\cite{Bekenstein:1971hc}, Sec.~IV) that asymptotically-flat finite-energy configurations with charged scalar-field hair are forbidden. Instead, we consider an analogue of the black hole bomb scenario of Press and Teukolsky \cite{Press:1972zz}, in which the black hole is enclosed by a reflecting mirror.

It was shown in Refs.~\cite{Herdeiro:2013pia, Degollado:2013bha, Hod:2013fvl} that, in the linear (small-amplitude) regime, a charged scalar field with a mirror on a Reissner-Nordstr\"om black hole background ($J=0$, $Q\neq0$) suffers exponential growth due to superradiance, provided the mirror is sufficiently far from the horizon. Here we consider the progression into the non-linear regime. We present charged-scalar black holes which are plausible endpoints for the above charge-superradiant instability, and examine their stability under perturbation.

The outline of this paper is as follows.
In Sec.~\ref{sec:setup} we describe our Einstein-charged scalar field model and briefly review the instability of Reissner-Nordstr\"om black holes under spherically symmetric charged scalar field perturbations \cite{Herdeiro:2013pia, Degollado:2013bha, Hod:2013fvl}.
We also present static, spherically symmetric black hole solutions with nontrivial charged scalar field hair.  The charged scalar field has zeros outside the event horizon; the reflecting mirror can be situated at any one of these zeros.
To see if these hairy black holes are plausible endpoints of the charge superradiant instability,  in Sec.~\ref{sec:stability}
we investigate their stability under linear, spherically symmetric perturbations.
If the mirror is located at the first zero of the charged scalar field, we present numerical evidence that at least some of the hairy black holes are stable.
Our conclusions are presented in Sec.~\ref{sec:conclusions}.

\section{Black hole solutions with hair}
\label{sec:setup}

\subsection{The model}
We consider a fully coupled system consisting of gravity, an electromagnetic field and a massless charged scalar field. The action is given by:
\begin{align}
S = \int \sqrt{-g}\left[\frac{R}{16\pi G}-\frac{1}{4}F_{ab}F^{ab} -\frac{1}{2} g^{ab} D^\ast_{(a} \Phi^\ast D^{}_{b)} \Phi\right]d^4 x, \label{action}
\end{align}
where $F_{ab} = \nabla_a A_{b} - \nabla_b A_{a}$ is the Faraday tensor and  $D_a = \nabla_a - i q A_a$, with $\nabla_a$ the covariant derivative, $A_a$ the electromagnetic vector potential and $q$ the charge of the scalar field $\Phi$. Tensor indices are lowered and raised with the metric $g_{ab}$ and its inverse $g^{ab}$, and $g$ denotes the metric determinant. Round and square brackets on indices denote symmetrized and anti-symmetrized combinations, $X_{(ab)} = \frac{1}{2} \left(X_{ab} + X_{ba}\right)$ and $X_{[ab]} = \frac{1}{2} \left(X_{ab} - X_{ba}\right)$.

By varying (\ref{action}), three equation of motions are obtained
\begin{subequations} \label{field}
\begin{align}
G_{ab} &= 8\pi G T_{ab}, \label{EFE}\\
\nabla_a F^{ab} &= J^b, \label{MW}\\
D_{a}D^{a}\Phi &= 0, \label{KG}
\end{align}
\end{subequations}
alongside the usual Bianchi identities for the Faraday and Riemann tensors, $\nabla_{[a} F_{bc]} = 0 = \nabla_{[a} \tensor{R}{^{b c} _{d e]}}$.
The stress-energy tensor is given by
$T_{ab} = T_{ab}^F + T_{ab}^{\Phi}$
where
\begin{subequations}
\begin{align}
T_{ab}^{F} &= F_{a c} \tensor{F}{_b ^c} - \frac{1}{4} g_{ab} F_{cd} F^{cd}, \\
T_{ab}^{\Phi} &= D^\ast_{(a} \Phi^\ast D^{}_{b)} \Phi  -\frac{1}{2} g_{ab}\left[ g^{cd} D^\ast_{(c} \Phi^\ast D^{}_{d)} \Phi \right],
\end{align}
\end{subequations}
and the field current $J^a$ is given by
\begin{equation}
J^a = \frac{iq}{2} \left[\Phi^\ast D^a \Phi - \Phi (D^a \Phi)^\ast \right] .
\end{equation}
The current and stress energy are covariantly conserved, $\nabla_a J^a = 0 = \nabla_a T^{ab}$.

The charged scalar field $\Phi$ and vector potential $A_a$ are defined up to the usual gauge freedom, as $F_{ab}$ and $D_a \Phi$ are invariant under the mapping
\begin{equation}
\Phi \rightarrow e^{i \chi} \Phi, \quad \quad A_a \rightarrow A_a + q^{-1} \chi_{,a} ,
\label{gauge-transform}
\end{equation}
where $\chi$ is any scalar field. We will make use of this freedom both when we consider static equilibrium solutions and time-dependent, spherically
symmetric perturbations.

\subsection{Linear perturbations in electrovacuum}
\label{sec:RNstab}

\begin{figure*}%
\begin{tabular}{c}
\includegraphics[width=0.95\columnwidth]{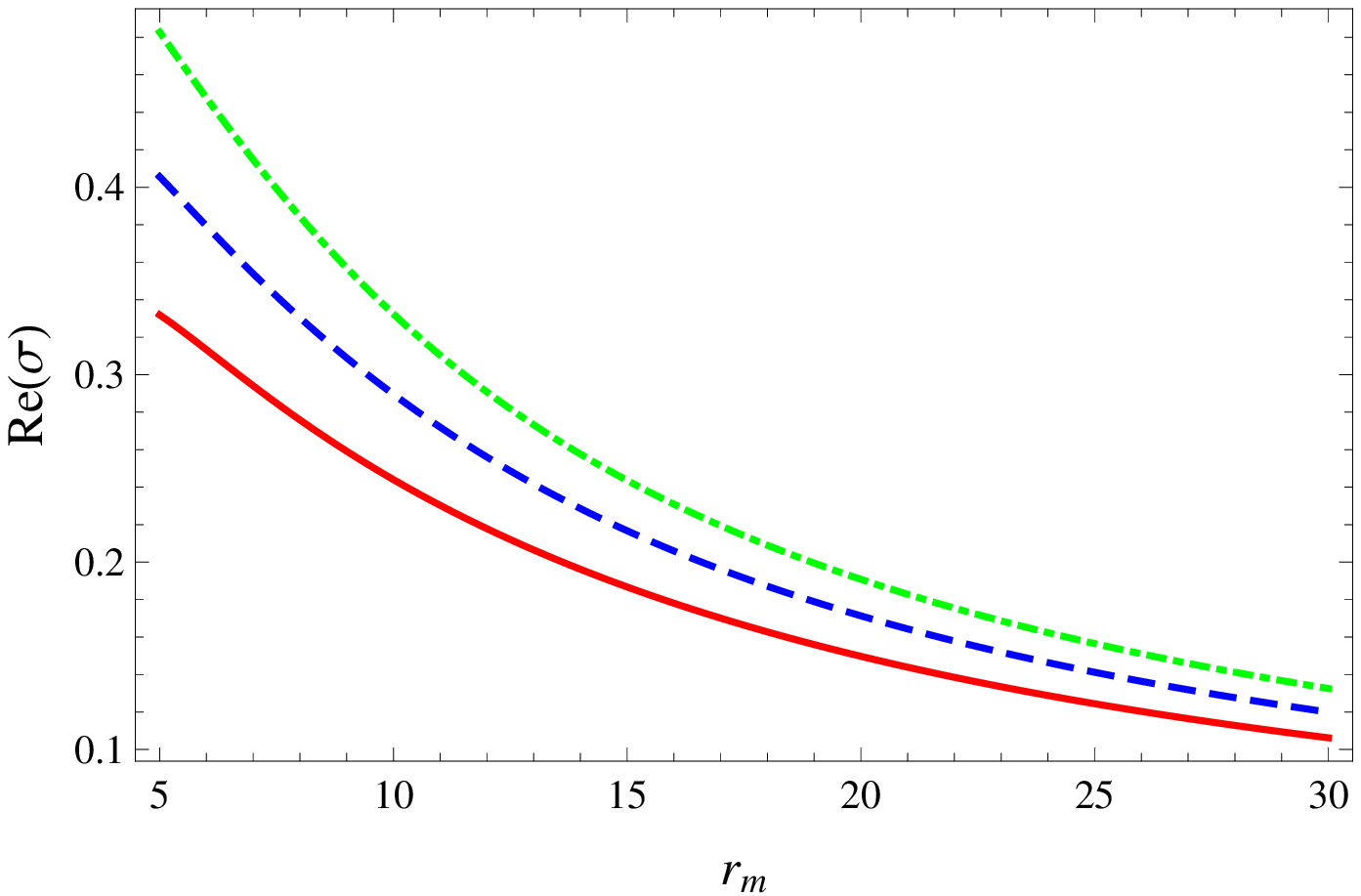}
\includegraphics[width=0.95\columnwidth]{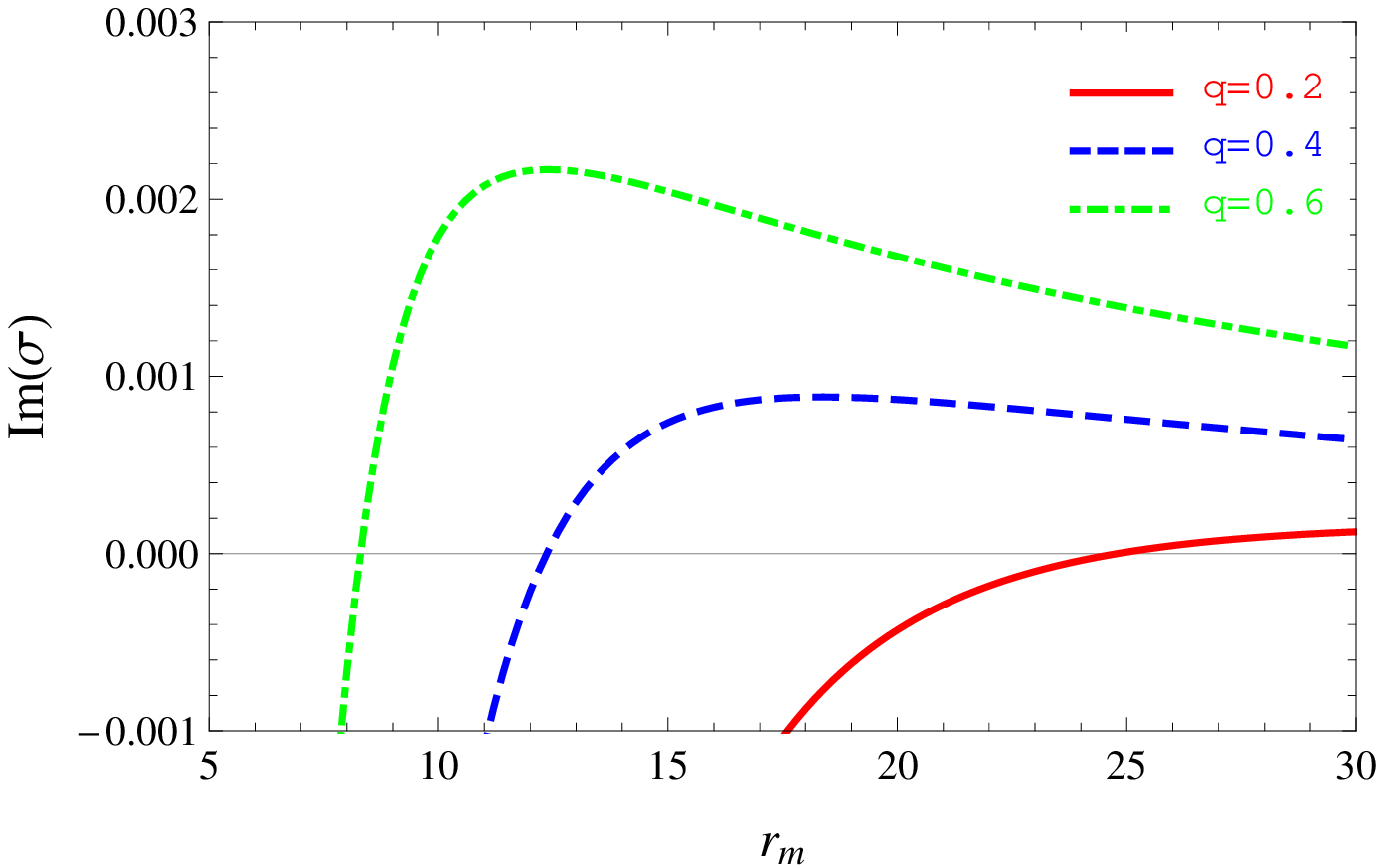} \\
\includegraphics[width=0.95\columnwidth]{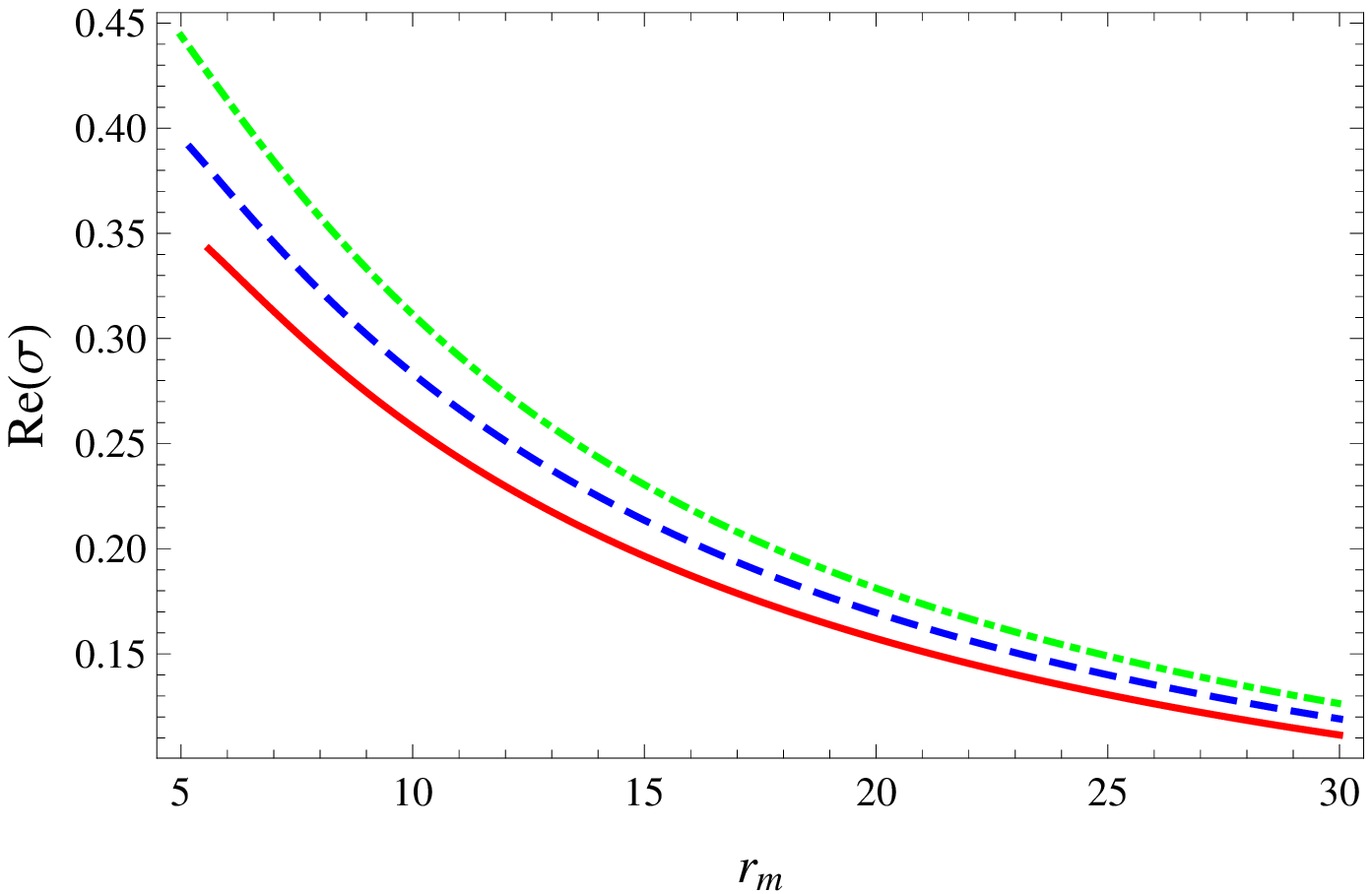}
\includegraphics[width=0.95\columnwidth]{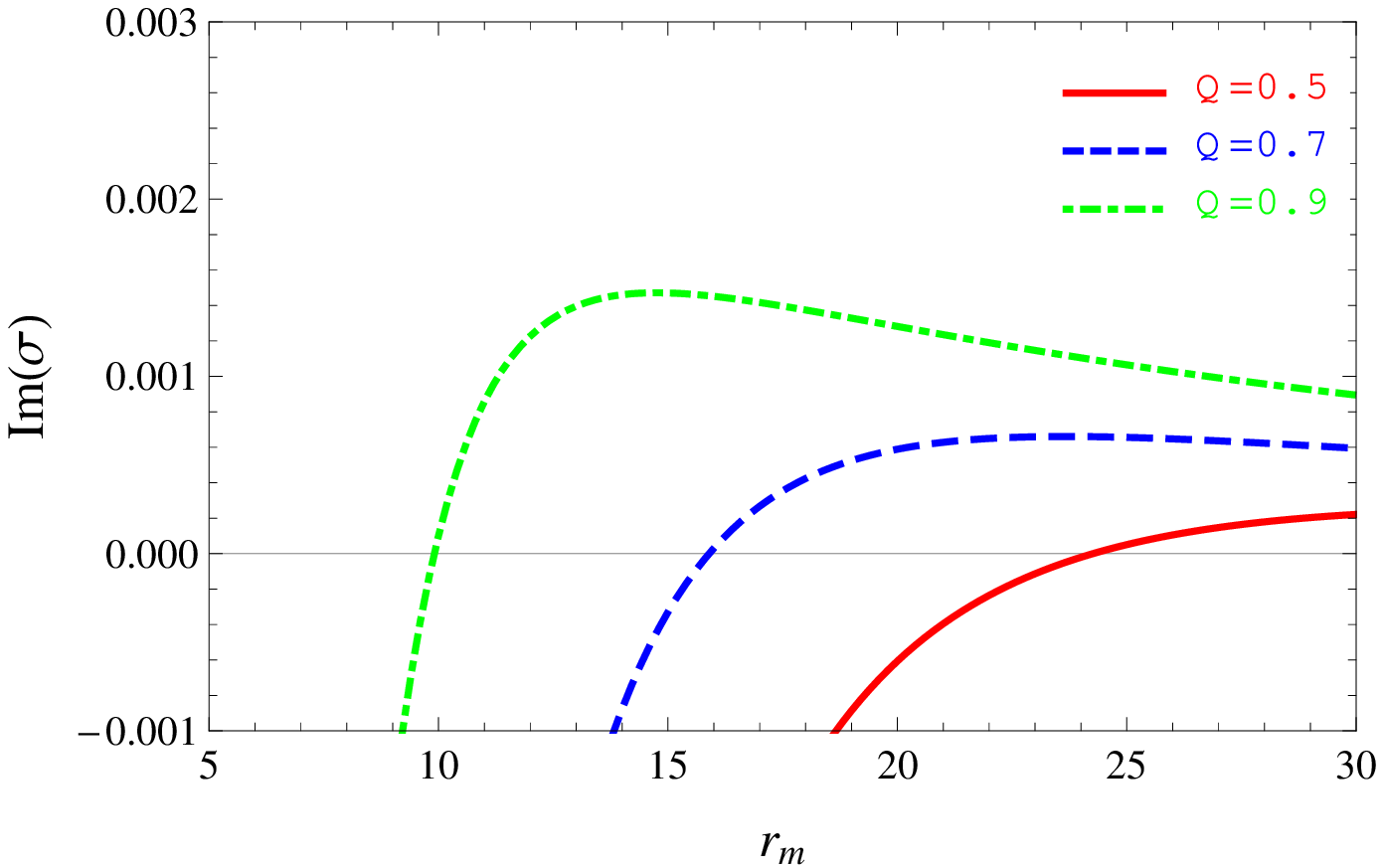} \\
\end{tabular}
\caption{The real (left) and imaginary (right) part of the frequency $\sigma_{0}$ of the fundamental massless scalar perturbation of Reissner-Nordstr\"{o}m spacetime with a reflecting mirror, plotted against the mirror radius $r_{m}$ with (top row) fixed black hole charge $Q=0.9$ and (bottom row) fixed scalar field charge $q=0.5$ (we use units in which the black hole mass $M=1$). The right-hand plots show the transition points where the imaginary part of $\sigma_{0}$ changes sign from negative (decaying) to positive (growing).}
\label{fig:RNfixedCharge}
\end{figure*}

In the case $\Phi = 0$, the spherically-symmetric solution of the field equations (\ref{field}) is the Reissner-Nordstr\"om black hole spacetime,
\begin{equation}
ds^2 = g_{ab} dx^a dx^b = -f_{RN} dt^2 + f_{RN}^{-1} dr^2 + r^2 d \Omega ^{2},
\end{equation}
where (in appropriate units),
\begin{equation}
f_{RN} = 1 - \frac {2M}{r} + \frac {Q^2}{r^2} = \frac{(r-r_+)(r-r_-)}{r^2}
\end{equation}
and the element of solid angle is
\begin{equation}
\label{dOmega}
d \Omega^2 = d\theta^2 + \sin^2 \theta \, d\varphi^2.
\end{equation}
The quantities
\begin{equation}
r_{\pm } = M \pm {\sqrt {M^{2}-Q^{2}}}
\end{equation}
are, respectively, the radii of the outer (event) and inner (Cauchy) horizons.

One may introduce a small-amplitude scalar field $\Phi$, and neglect the back-reaction on the electromagnetic and gravitational fields (as $T_{ab}$ and $J^a$ are quadratic in the scalar field amplitude). Let us consider a monochromatic, spherically-symmetric perturbation with frequency $\sigma $
\begin{equation}
\Phi = \frac{\phi(r)}{r} e^{-i \sigma t} ,  \label{Phi-ansatz}
\end{equation}
which, via Eq.~(\ref{KG}), satisfies
\begin{equation}
\frac{d^2 \phi}{d r_\ast^2} + \left[ \left(\sigma - \frac{qQ}{r} \right)^2 - \frac{f_{RN}}{r}\frac{df_{RN}}{dr} \right] \phi = 0 ,
\label{RNpert}
\end{equation}
where the tortoise coordinate $r_\ast$ is defined by $dr_\ast / dr = f_{RN}^{-1}$.  The scalar field perturbation should be ingoing at the horizon, that is,  regular in a (future) horizon-penetrating coordinate system, which implies that
\begin{equation}
\phi \sim e^{-i \sigma r_\ast} \quad \quad \text{as} \quad r_\ast \rightarrow -\infty .
\end{equation}
Imposing a `mirror' boundary condition $\phi(r_m) = 0$ at $r=r_m$ leads to a discrete spectrum of states with, in general, complex frequencies $\sigma_n$. A positive (negative) imaginary component of frequency corresponds to exponential growth (decay). The states are labelled with $n$, the number of nodes they possess in the region $r < r_m$.

An analytic approximation for the discrete frequencies $\sigma _{n}$ can be found in \cite{Cardoso:2004nk}. We used this approximation as an initial input value for the frequency $\sigma _{0}$ of the fundamental mode $n=0$, numerically integrating the radial perturbation equation (\ref{RNpert}) and searching for the value of $\sigma _{0}$ for which the scalar field perturbation vanishes on the mirror.

Figure \ref{fig:RNfixedCharge} shows the real and imaginary parts of the fundamental mode frequency $\sigma_0$ as a function of mirror radius $r_m$, for a selection of black hole and field charges, $Q$ and $q$. The plots illustrate the following point: with the mirror placed close the black hole, the $n=0$ mode decays exponentially; with the mirror placed far from the black hole, the $n=0$ mode grows exponentially, generating a superradiant instability; between these regimes is a `transition point', at exactly $\sigma = \sigma_c$, at which the scalar field is in equilibrium with the black hole.

In this paper we restrict our attention to a massless charged scalar field, but the superradiant instability shown in Fig.~\ref{fig:RNfixedCharge}
is also present when a massive charged scalar field is considered \cite{Herdeiro:2013pia, Degollado:2013bha, Hod:2013fvl}.
In Ref.~\cite{Herdeiro:2013pia}, a frequency-domain analysis was performed and a superradiant instability found for massive charged scalar field modes with
$\ell = m = 1$, where $\ell $ is the total angular momentum mode number and $m$ the azimuthal mode number.
A time-domain study was undertaken in \cite{Degollado:2013bha}, again for a massive charged scalar field.
In the $\ell = 1$ case, the results of \cite{Degollado:2013bha} show that at late times, the fundamental unstable mode found in \cite{Herdeiro:2013pia}
dominates the evolution.  They also find a superradiant instability for the $\ell =0$ (spherically symmetric) mode, which grows more quickly than the $\ell =m =1$ unstable mode.  The growth time of the $\ell =0$ unstable modes of the massless charged scalar field (whose frequencies are shown in Fig.~\ref{fig:RNfixedCharge}) is of a similar order of magnitude to those in \cite{Degollado:2013bha} for a charged scalar field with mass $0.1M$. The highly-explosive (yet still linear) regime was studied in Ref.~\cite{Hod:2013fvl}.

\subsection{Static hairy black holes}
\label{sec:solutions}
At first glance, perturbations at the critical frequency $\sigma = \sigma_c$ are time-dependent via ansatz (\ref{Phi-ansatz}), and thus not static. However, we may choose a gauge in which the scalar field \emph{is} static, by inserting $\chi = \sigma t$ into Eq.~(\ref{gauge-transform}). This gauge transformation removes the time-dependence from the field, and introduces a static constant term to $A_0$. This raises the possibility that static solutions may also exist for nontrivial scalar field $\Phi $.

\subsubsection{Field equations}
To investigate this possibility, we now consider the spherically symmetric spacetime defined as follows
\begin{align}
ds^{2} &= -f h \, dt^2 + f^{-1} dr^2 + r^2 d\Omega^2 ,
\label{metric}
\end{align}
where $f = f(r)$ and $h = h(r)$ and $d\Omega ^{2}$ is given by (\ref{dOmega}). We may write
\begin{equation}
f(r)\equiv 1 - \frac {2m(r)}{r}
\end{equation}
where $m = m(r)$ is interpreted as the total mass within the given radius $r$.
We assume that the static scalar field is real and depends only on the radial coordinate $r$, setting $\Phi = \phi (r)$.
Since we are considering a spherically symmetric spacetime, we can set the $A_{\theta }$ and $A_{\varphi }$ components of the electromagnetic gauge potential to zero, and then use a gauge transformation (\ref{gauge-transform}) to set $A_{r}$ to vanish. Thus the electromagnetic gauge potential takes the form $A_\mu=[A_{0}(r),0,0,0]$.

With the above ansatz, the equations of motion (\ref{field}) yield four non-trivial equations,
\begin{subequations}
\begin{align}
h' &= r\kappa \left[ \left(\frac{q A_0 \phi}{f} \right)^2 + h (\phi')^2 \right],  \label{eq-hprime} \\
\kappa E^2 &= -\frac{2}{r} \left[ f' h + \frac{1}{2} f h' + \frac{h}{r}(f - 1) \right], \label{eq-fprime}  \\
0 &= f A_0'' +  \left( \frac{2f}{r} - \frac{f h'}{2h} \right) A_0' - q^2 \phi^2 A_0, \label{eq-Aprimeprime} \\
0 &= f \phi'' + \left(\frac{2f}{r} + f' + \frac{f h'}{2 h}\right) \phi' + \frac{(q A_0)^2}{fh} \phi , \label{eq-phiprimeprime}
\end{align} \label{fieldss}
\end{subequations}
where $\kappa = 8\pi G$ and $E^2 = \left(A'_{0}\right)^2$. A prime $'$ denotes $d/dr$. For $\phi\neq0$, Eqs.~(\ref{fieldss}) cannot (apparently) be solved analytically, and some numerical analysis is required.

\subsubsection{Boundary conditions}
Let us now consider appropriate conditions to impose on the fields at the black hole horizon $(r=r_h)$ and at the mirror $(r=r_m)$.

We assume that there is a regular event horizon defined by $f(r_h) = 0$ and $f'(r_h) > 0$. Thus, $m_h \equiv m(r_h) = \tfrac{1}{2} r_h$.
We demand that all physical quantities are finite in a future-horizon-penetrating coordinate system. This implies that the vector potential is zero at the horizon, $A(r_h) = 0$; and $h(r_h)$ is finite. Without loss of generality, we set $h(r_h) = 1$, which corresponds to a gauge choice in the definition of the time coordinate $t$. The scalar field equation (\ref{eq-phiprimeprime}) implies that $\phi'(r_h) = 0$. Hence regular Taylor series expansions of the field variables about $r = r_h$ take the following form
\begin{align}\label{powerseriesnearthehorizon}
m &= m_h +  m^{\prime}_h(r-r_h) + O(r-r_h)^2, \nonumber \\
h &= 1 +  h^{\prime}_{h}(r-r_h) + O(r-r_h)^2, \nonumber \\
A_0 &= E_h(r-r_h) + \frac{A^{\prime\prime}_h}{2}(r-r_h)^2 + O(r-r_h)^3, \nonumber \\
\phi &= \phi_h + \frac{\phi^{\prime\prime}_h}{2}(r-r_h)^2 + O(r-r_h)^3,
\end{align}
where $E_h = A_0'(r_h)$ is the electric field on the horizon. Inserting these expansions back into the field equations (\ref{fieldss}) gives
\begin{align}\label{powerseriesnearthehorizon1}
m^{\prime}_h &= \frac{\kappa r^2_h E^2_h}{4}, \nonumber \\
h^{\prime}_h &= \frac{4\kappa q^2 r^3_h\phi^2_h E^2_h}{\left(\kappa r^2_h E^2_h-2\right)^2}, \nonumber \\
A^{\prime\prime}_h &= \frac{2 E_h}{r_h}\left[\frac{2q^2r^2_h\phi^2_h}{(\kappa r^2_h E^{2}_h-2)^2}-1\right], \nonumber \\
\phi^{\prime\prime}_h &= -\frac{2\phi_hq^2r^2_hE^{2}_h}{(\kappa r^2_h E^2_h-2)^2}.
\end{align}
At fixed $q, r_h$ and $E_{h}$, these expansions are determined by just one further constant $\phi_h$.

\begin{figure*}
\begin{tabular}{c}
\includegraphics[width=0.95\columnwidth]{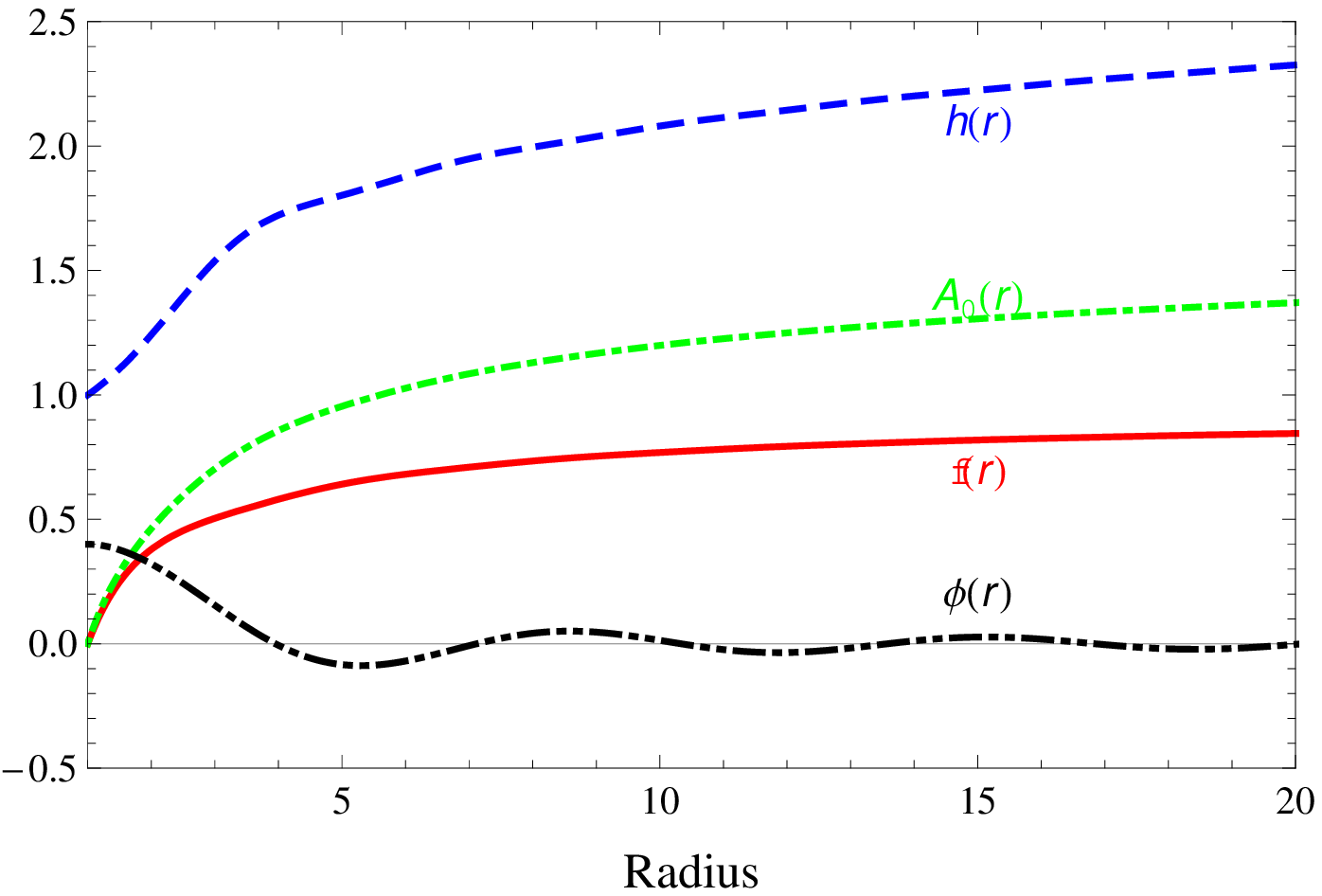}
\includegraphics[width=0.99\columnwidth]{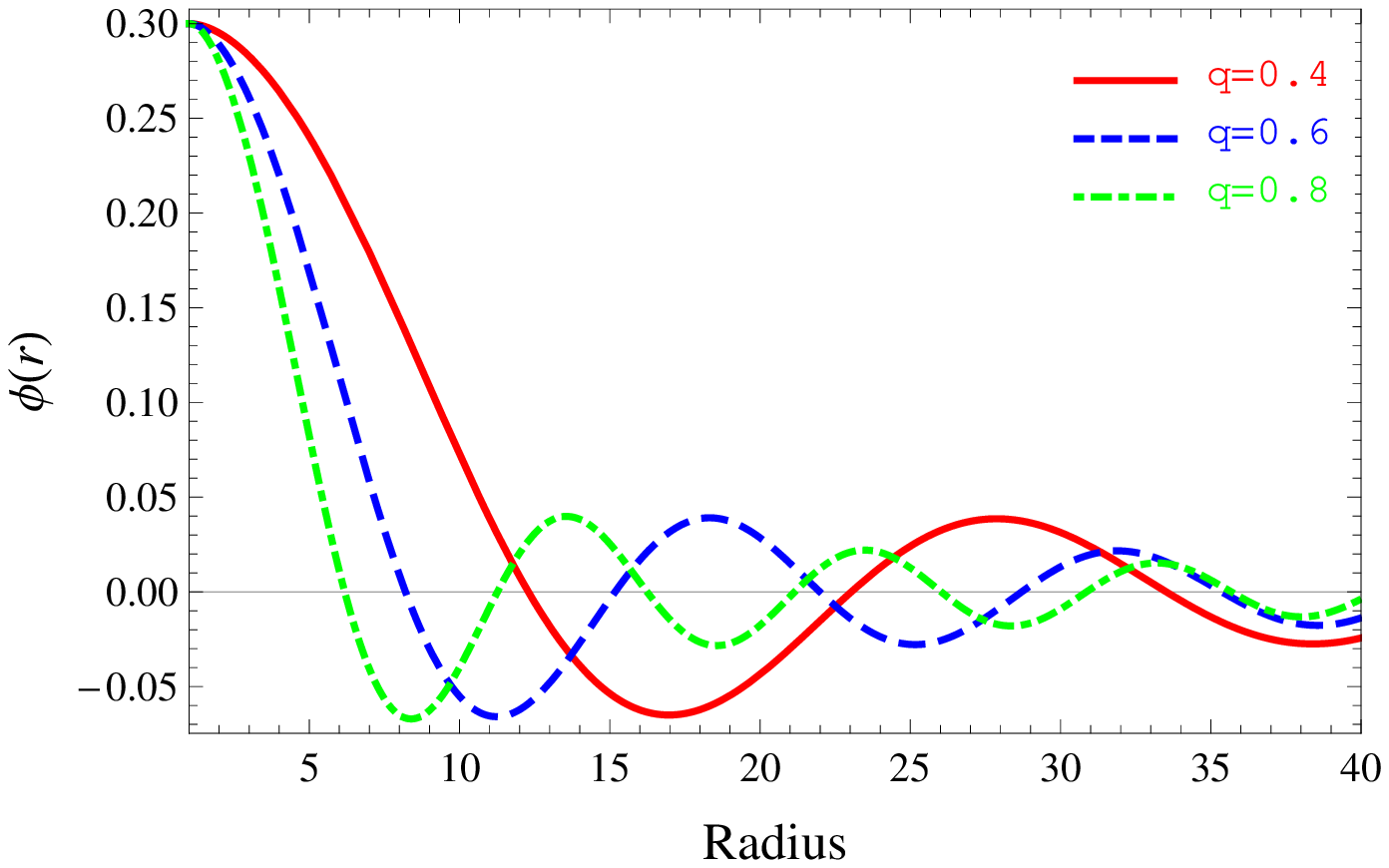}
\end{tabular}
\caption{Left: an example plot of the two metric functions $f,h$ and two matter functions $A_0$, $\phi$ for a particular static black hole solution with $q=0.9,\phi_h=0.4$ and $E_h=0.8$. Right: Scalar field profiles for three different black hole solutions for fixed $\phi_h=0.3$ and $E_h=0.6$ and three values of the scalar-field charge $q$.}
\label{fig:all2gether}
\end{figure*}

\begin{figure*}
\begin{tabular}{c}
\includegraphics[width=0.95\columnwidth]{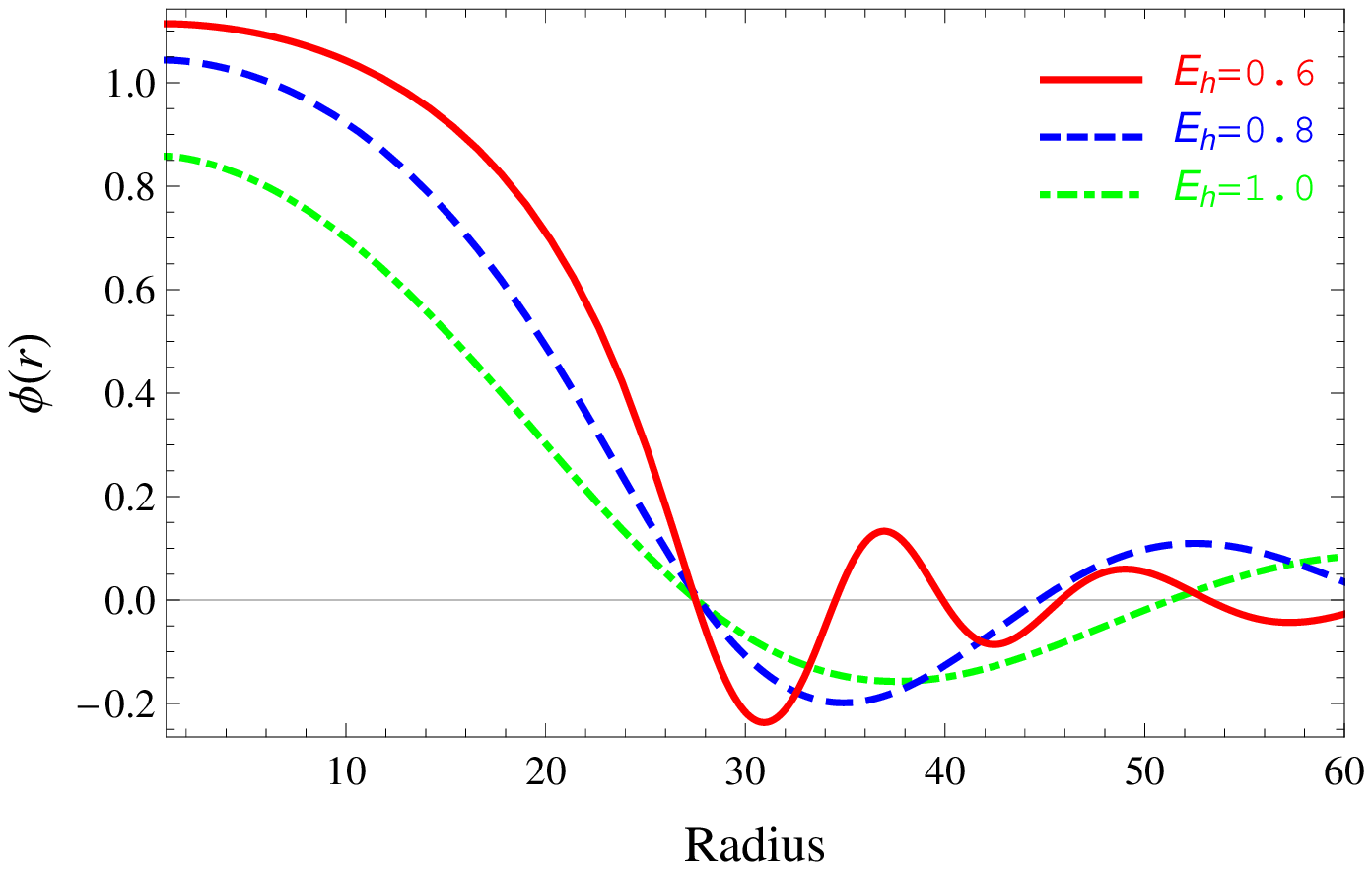}
\includegraphics[width=0.95\columnwidth]{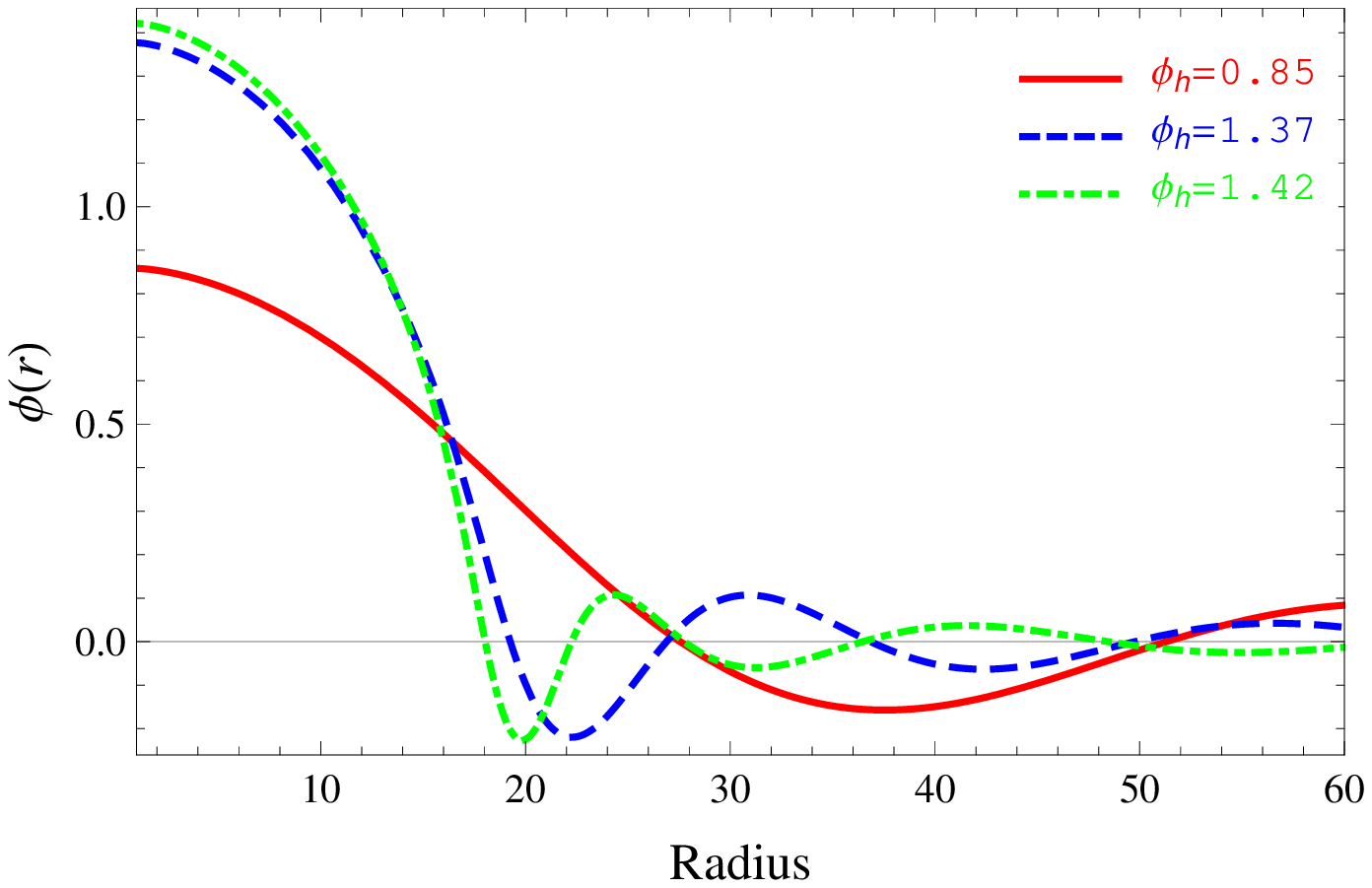} \\
\end{tabular}
\caption{An example plot of different black hole solutions with scalar charge $q=0.1$. Left: Three scalar field profiles which share the same location of their first node at $r_m\approx27$. Right: Three scalar field profiles with fixed $E_h=1$ with a common node; their first (red, solid), second (blue, dashed) and third (green, dot-dashed) nodes coincide at $r_m\approx27$.  }
\label{fig:StaticSameRm}
\end{figure*}

We insist that the scalar field vanishes at the location of the mirror
\begin{align}
\phi(r_m) &= 0.
\end{align}
No further conditions are applied at $r=r_m$, so the metric functions $f$, $h$ and the electric gauge potential $A_{0}$ are unconstrained at the mirror's location.

\begin{figure*}
\begin{tabular}{c}
\includegraphics[width=0.75\columnwidth,angle=270]{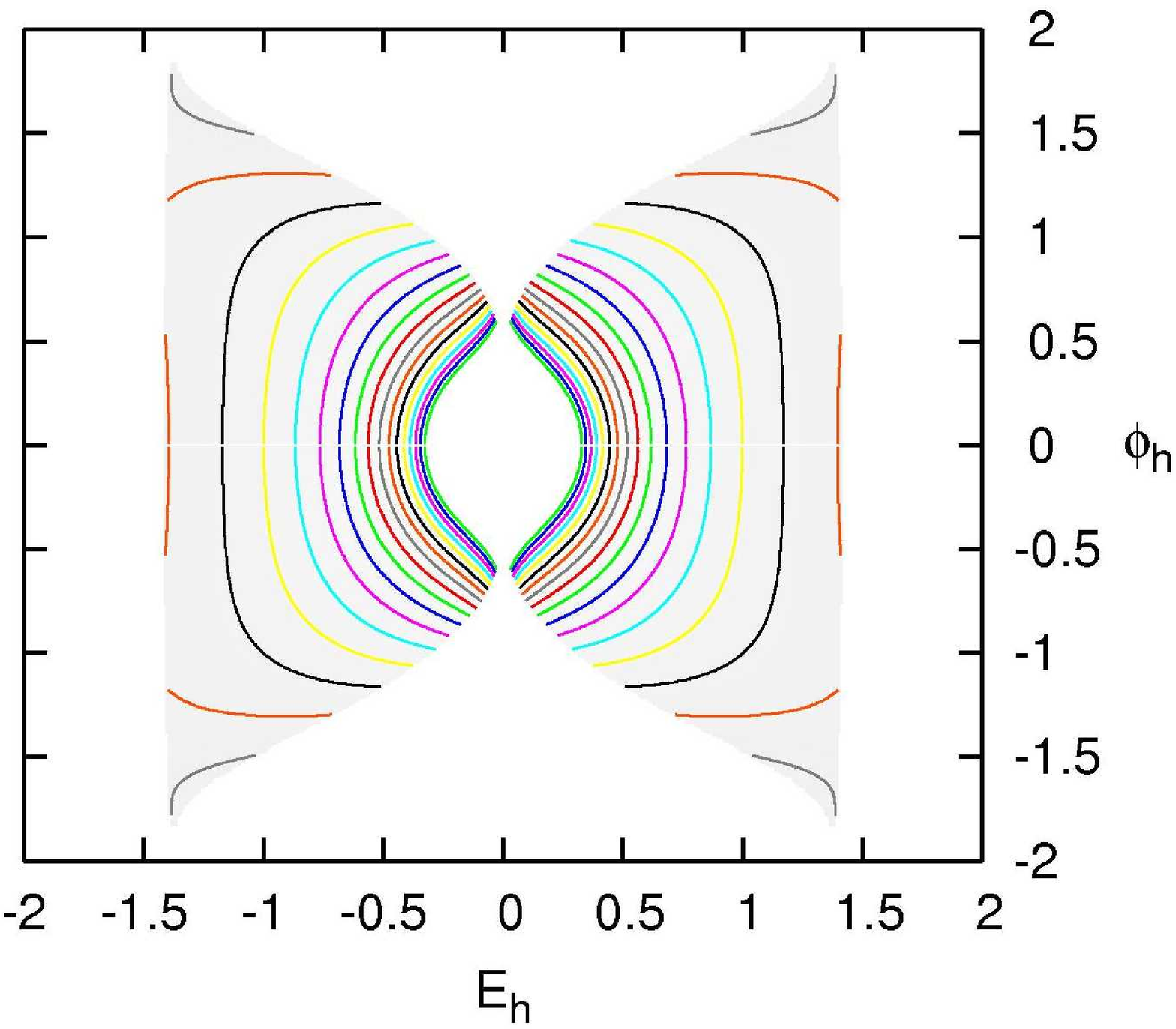}
\includegraphics[width=0.75\columnwidth,angle=270]{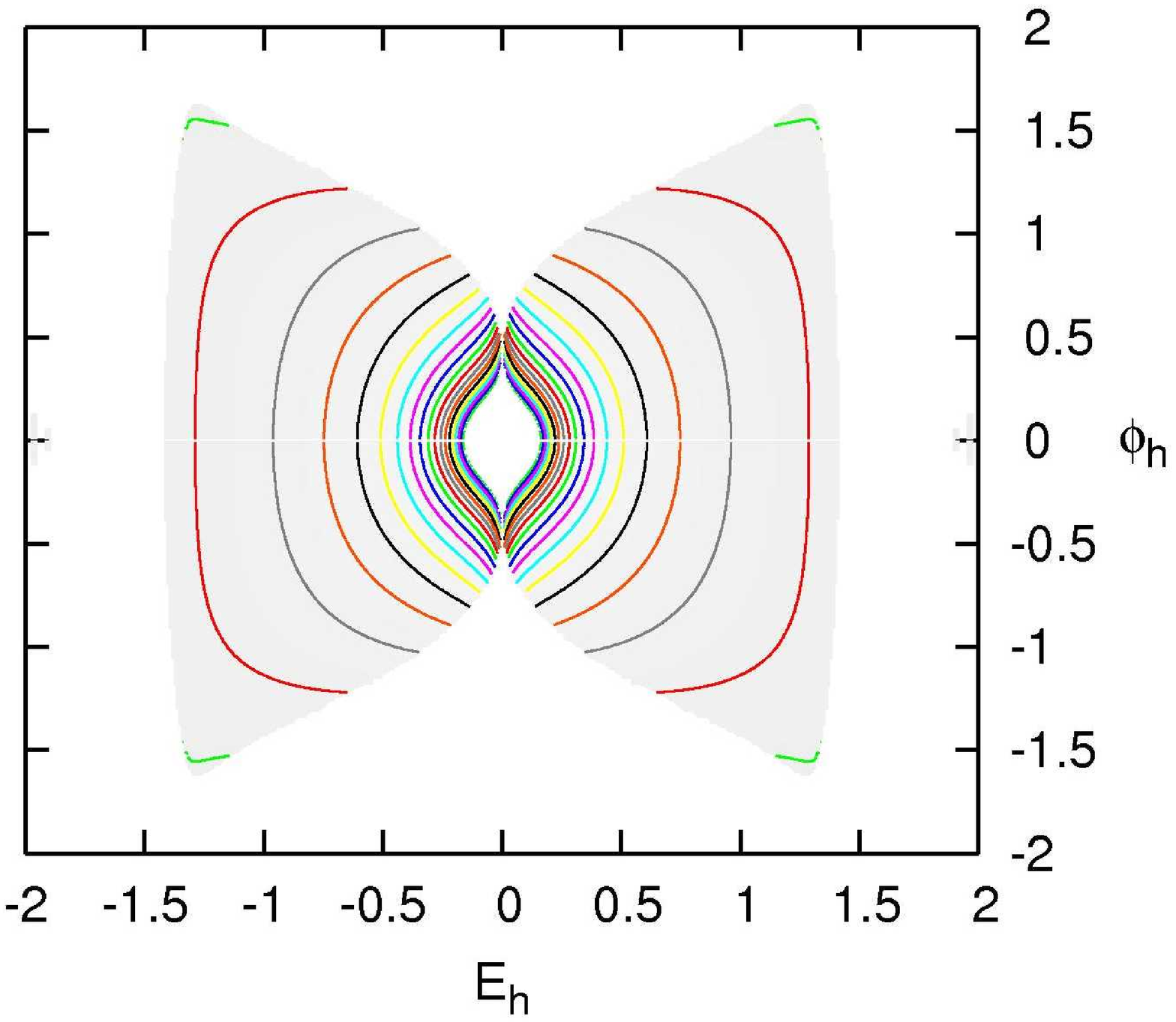} \\
\includegraphics[width=0.75\columnwidth,angle=270]{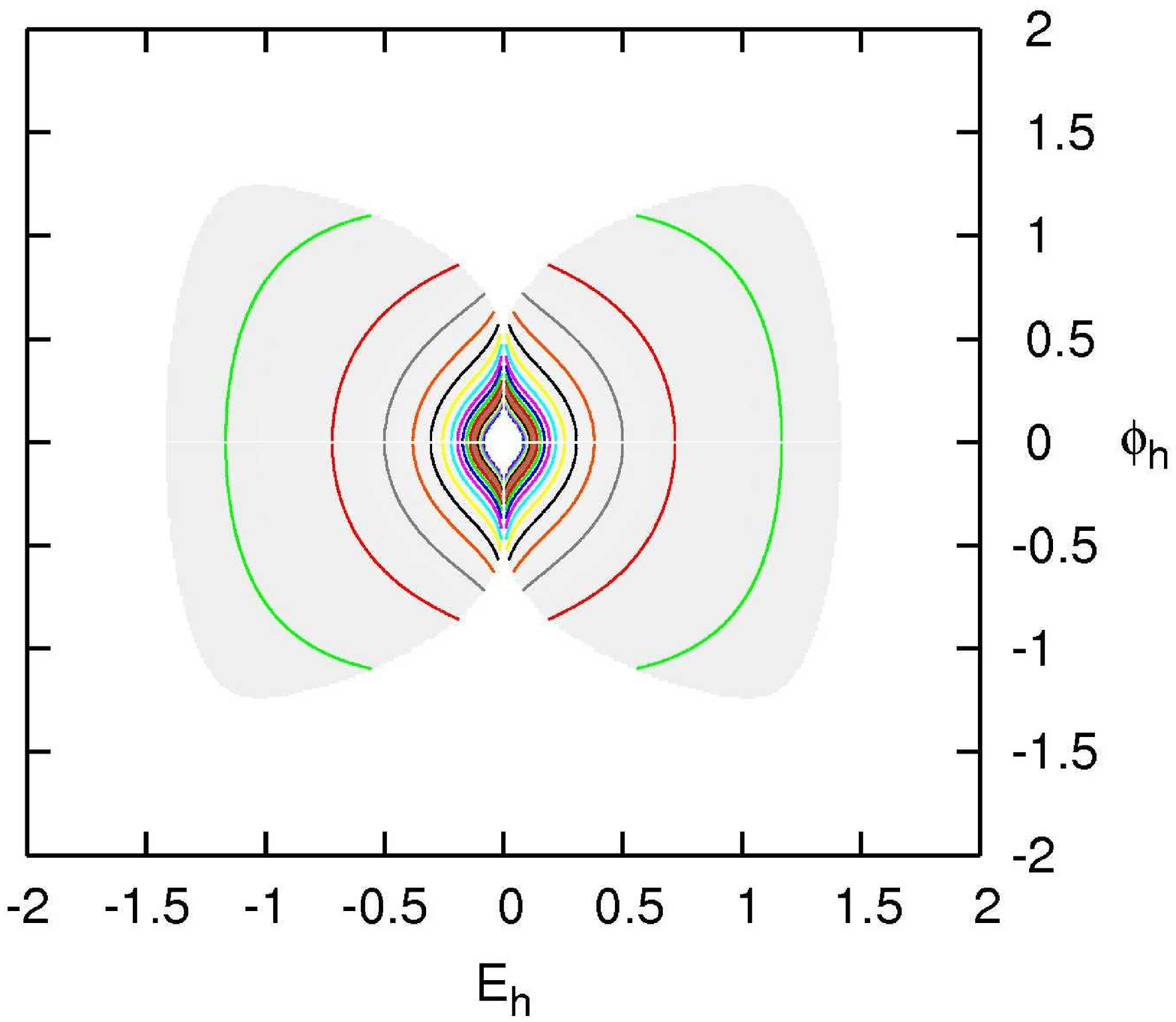}
\includegraphics[width=0.75\columnwidth,angle=270]{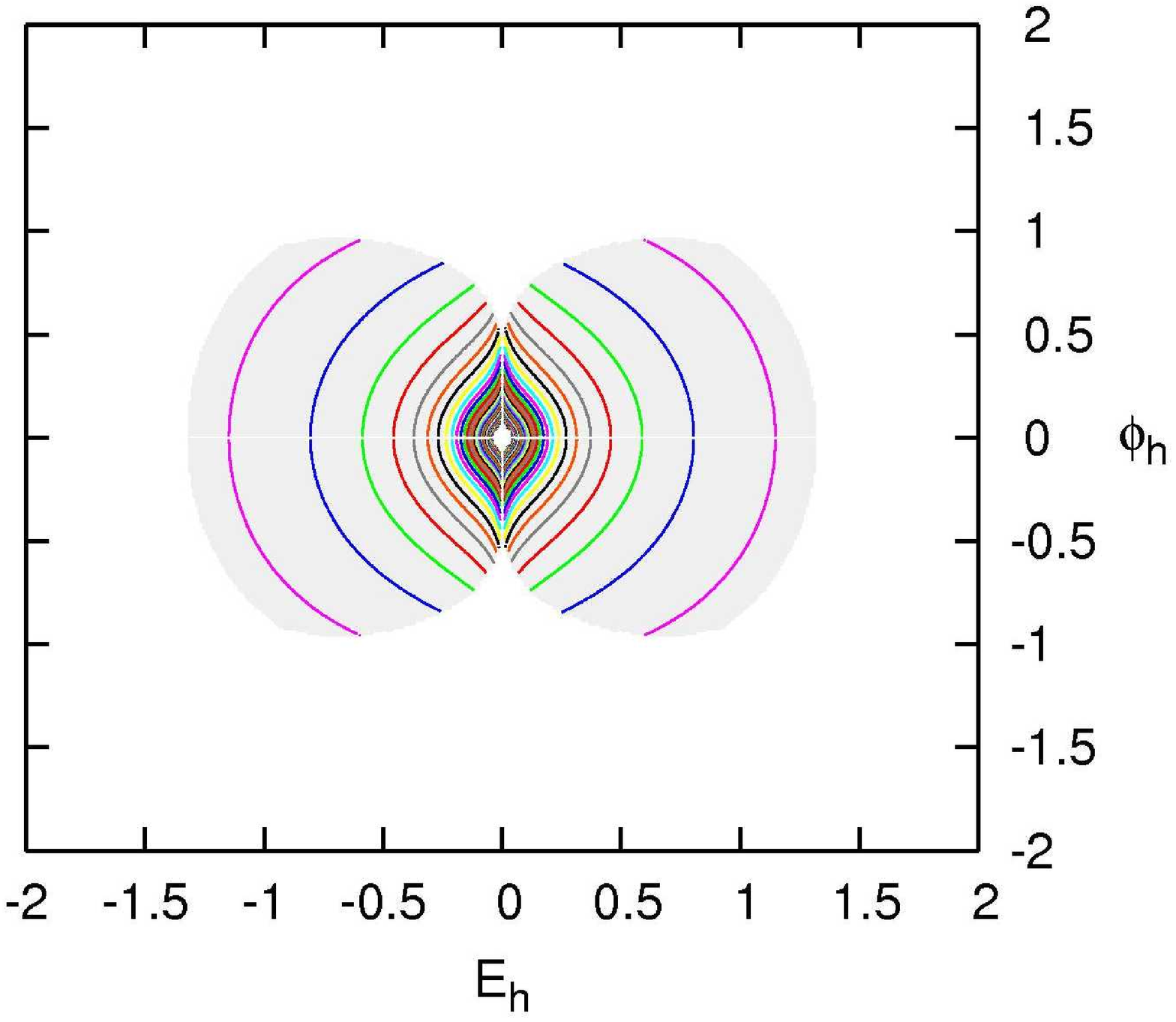}
\end{tabular}
\caption{Numerical exploration of the parameter space for static solutions with scalar field hair in a cavity. The  system has three free parameters: $q$ (the charge of the field), $\phi_h$ (the scalar field on the horizon), and $E_h$ (the electric field on the horizon). The plots show two-dimensional slices of the solution space with (top row, left to right) $q=0.1$, $q=0.2$, (bottom row, left to right) $q=0.4$ and $q = 0.8$. The shaded area indicates the region where solutions exist, with the scalar field having at least one node, and $f(r) > 0$ for $r > r_h$.  Note that solutions exist throughout the central region all the way towards $\phi_h \rightarrow 0$, $E_h \rightarrow 0$; except along the line $E_h = 0$. The coloured lines are contours of constant mirror radius $r_m$, where  $r_m$ lies at the first node of the scalar field $\phi$.
}
\label{fig:solution-space}
\end{figure*}

\subsubsection{Solutions}
We seek static solutions by numerically integrating the equation set (\ref{fieldss}).
To avoid the regular singular point at $r=r_h$, we use the series expansions (\ref{powerseriesnearthehorizon}--\ref{powerseriesnearthehorizon1}) as initial data, evaluated at $r=r_h + \epsilon$ (where typically $\epsilon \sim 10^{-12}$). We choose units such that $\kappa = 8\pi G = 1$.

Without loss of generality, one may rescale all dimensionful quantities by $r_h$ to obtain a dimensionless equation set. Equivalently, one may simply set $r_h=1$ (thus $m_h = 1/2$), leaving three free parameters: $q$, the scalar-field charge; $\phi_h$, the scalar-field magnitude on the horizon; and $E_h \equiv A'_{0}(r_h)$, the electric field on the horizon. Henceforth, these should be thought of as dimensionless quantities, in units of $r_h$. We now explore this three-parameter solution space.

The left-hand plot in Fig.~\ref{fig:all2gether} shows the four field variables $f(r)$, $h(r)$, $A_0(r)$ and $\phi(r)$ for the case of scalar field charge $q=0.9$, $\phi_h=0.4$ and $E_h=0.8$.
The scalar field $\phi (r)$ oscillates about zero; the other three field variables $f(r)$, $h(r)$ and $A_{0}(r)$ are monotonically increasing with $r$.
Note that we do not expect that $A_{0}(r)$, $f(r)$ and $h(r)$ will necessarily approach finite limits as $r\rightarrow \infty $, since there are no asymptotically flat black hole solutions in this model with nontrivial scalar field hair \cite{Bekenstein:1971hc}.
The right-hand plot in Fig.~\ref{fig:all2gether} shows an example of the scalar field profile outside the horizon for three values of $q$ with fixed $\phi_h=0.3$ and $E_h=0.6$. Here, the oscillating behaviour of the scalar field is more clearly seen. One could obtain a black-hole-in-a-cavity solution by placing the mirror at any of the nodes of the scalar field.  For the majority of this paper we will consider the case that the mirror is located at the first node.

It is possible to have different black hole configurations which share the same mirror radius, as illustrated by Fig.~\ref{fig:StaticSameRm}. In the left-hand plot in Fig.~\ref{fig:StaticSameRm}, we show three scalar field profiles $\phi (r)$ for different values of the electric field at the horizon $E_h$
and fixed scalar-field charge $q=0.1$. Each solution, despite having different values of $\phi_h$ and $E_h$, has a node of the scalar field at $r_m\approx 27$. A further three distinct scalar field profiles are displayed in the right-hand plot in Fig.~\ref{fig:StaticSameRm}. For a given $E_h=1$ (the scalar-field charge is still fixed to be $q=0.1$), the first, second and third nodes of the scalar field share the same location.

Figure \ref{fig:solution-space} illustrates the three-dimensional solution space of static hairy black holes in a cavity. The plots indicate that solutions exist in a contiguous region of $\{q, E_h, \phi_h\}$ parameter space, in which solutions with at least one node in the scalar field are permitted. Outside this region, we find that an excess of stress energy causes the metric function $f(r)$ to develop an additional zero \emph{before} the scalar field develops its first node (suggesting that an additional horizon forms). We note that (i) solutions with $\phi_h=0$ are well-known: these are the Reissner-Nordstr\"om black holes; and (ii) uncharged solutions, with $E_h = 0$ or $q=0$, are not possible, as the scalar field does not develop a node.

\section{Stability analysis}
\label{sec:stability}
In the previous section, we learned that a fully coupled system of gravity and a charged scalar field admits black hole solutions with scalar hair confined within a cavity. Are such hairy configurations stable or unstable? If such solutions are stable, it is plausible that they represent the end-point of the superradiant instability for a massless charged scalar perturbation on the Reissner-Nordstr\"om background with a mirror, as described in Sec.~\ref{sec:RNstab}.
In this section we examine the stability of the hairy black hole configurations under linear, spherically symmetric, perturbations.

\subsection{Dynamical equations}\label{subsec:dyn}
We begin by outlining the ansatz for the field variables.  We consider a spherically symmetric metric of the form (\ref{metric}), but with the functions $f(t,r)$ and $h(t,r)$ depending on time $t$ as well as the radial coordinate $r$. The scalar field $\Phi =\phi (t,r)$ similarly is time-dependent but is now complex.
 By virtue of spherical symmetry, we can set $A_{\theta} = 0 = A_{\varphi}$. Using (\ref{gauge-transform}), we may apply a gauge transformation to eliminate $A_r$, leaving $A_{\mu}$ with only a temporal component, $A_\mu = [A_{0}(t,r),0,0,0]$.

It is convenient to introduce a new metric variable,
\begin{equation}
\gamma =  fh^{1/2}.
\end{equation}
From the Einstein field equations (\ref{EFE}), and, in particular, from the combinations $G_{tt},~G_{tt}+\gamma^2 G_{rr}$ and $G_{tt}-\gamma^2 G_{rr}$, we obtain
\begin{subequations}
\begin{align}
\frac{f'}{f} &= -\frac{r}{2 \gamma^2} \left( \tau + f E^2 \right) + \frac{1}{fr} (1 - f), \label{dyn_eq1} \\
\frac{h'}{h} &= \frac{r\tau}{\gamma^2}, \label{dyn_eq2} \\
\frac{\gamma'}{\gamma} &= - \frac{r}{2 \gamma^2} f E^2 + \frac{1}{fr} \left(1 - f \right) , \label{dyn_eq3}
\end{align}
\label{dyneqns1}
\end{subequations}
where
\begin{align}
\tau \equiv |\dot{\phi}|^2 + |\gamma\phi'|^2 + q^2 A_0^2 |\phi|^2 + 2 q A_0 \text{Im}(\phi \dot{\phi}^\ast).
\end{align}
Here, the dot $\dot{}$ and prime $'$ denote partial derivatives with respect to $t$ and $r$, respectively, and the asterisk ${}^\ast$ denotes complex conjugation. We note the equation for $\gamma$ has no dependence on $\tau$, and thus it does not explicitly depend on the scalar field $\phi$.

There is one further independent, nontrivial component of the Einstein field equations (\ref{EFE}), namely the $G_{tr}$ component, which gives
\begin{align}\label{dyn_eq4}
-\frac{\dot{f}}{f} = r \text{Re} \left( \dot{\phi}^\ast \phi' \right) + r q A_0 \text{Im} \left( \phi'^{\ast} \phi \right) .
\end{align}
Note that in the static limit this component is identically zero.

From the $t$ and $r$ components of the Maxwell equations (\ref{MW}) we obtain two dynamical equations
\begin{subequations}
\begin{align}
\frac{\gamma}{r^2} \left(  \frac{r^2 A'_{0}}{h^{1/2}} \right)' &= J_t = q^2 |\phi|^2 A_0 - q \text{Im}\left(\dot{\phi} \phi^\ast\right), \label{dyn_eq5} \\
\frac{1}{r} \partial_t \left( \frac{r A'_{0}}{h^{1/2}} \right) &= \gamma J_r = - q \text{Im}( \gamma\phi' \phi^\ast). \label{dyn_eq6}
\end{align}
\label{dyneqns2}
\end{subequations}

The scalar field equation (\ref{KG}), first written in the form $r f h D^a D_a \phi = 0$, yields
\begin{align}
0 = & -\ddot{\psi} + \frac{\dot{\gamma}}{\gamma} \dot{\psi} + \gamma\left(\gamma\psi'\right)' - \frac{\gamma\gamma'}{r} \psi + 2 i q A_0 \dot{\psi} + i q \dot{A}_0 \psi \nonumber \\
& \qquad - i q \frac{\dot{\gamma}}{\gamma} A_0 \psi + q^2 A_0^2 \psi, \label{dyn_eq7}
\end{align}
where
\begin{equation}
\psi = r \phi.
\end{equation}
The equations (\ref{dyneqns1}--\ref{dyn_eq7}) govern how the spacetime metric, electromagnetic field and the massless scalar field evolve with time.

\subsection{Perturbation equations}
Our aim in this section is to study the stability of the hairy black holes found previously in Sec.~\ref{sec:solutions}.
We therefore now consider linear perturbations around a non-vacuum solution by introducing the following notation:
\begin{align}
f &= \bar{f}(r) + \delta f(t,r), \nonumber \\
h &= \bar{h}(r) + \delta h(t,r), \nonumber \\
\gamma &= \bar{\gamma}(r) + \delta \gamma(t,r), \nonumber \\
A_0 &= \bar{A}_0(r) + \delta A_0(t,r), \nonumber \\
\psi &= \bar{\psi}(r) + \delta \psi(t,r).
\end{align}
In this formalism, $\bar{f}$ is the equilibrium quantity and $\delta f$ is the perturbation. We assume that only $\delta\psi$ is a complex variable while all other quantities are real.
From (\ref{dyneqns1}, \ref{dyn_eq4}--\ref{dyn_eq7}) six independent dynamical equations can be obtained. For the remainder of this section, we work to
first order in the perturbations.
\begin{widetext}
\begin{subequations}
First, (\ref{dyn_eq1}) gives
\begin{align}\label{Pt-G00}
\frac{\delta f^{\prime}}{\bar{f}} + \left[\frac{1}{r \bar{f}^2}+ \frac{r (\bar{A}'_0)^{2}}{2\bar{\gamma}^2}-\frac{\bar{f}^{\prime}}{\bar{f}^2}\right]\delta f &= - \frac{r\bar{f}\bar{A}'_0}{\bar{\gamma}^2}\delta A'_0-\frac{\bar{A}_0q^2\bar{\psi}^2}{r\bar{\gamma}^2}\delta A_0 + \frac{1}{\bar{\gamma}^3}\left[\frac{q^2\bar{A}_0^2\bar{\psi}^2}{r} + r\bar{f}(\bar{A}'_0)^2\right]\delta\gamma  \nonumber \\
&\qquad -\frac{iq\bar{A}_0\bar{\psi}}{2r\bar{\gamma}^2}\left(\delta\dot{\psi}-\delta\dot{\psi}^*\right) + \left[\frac{\bar{\psi}}{2r^2}-\frac{\bar{\psi}'}{2r}\right]\left(\delta\psi'+\delta\psi^{\prime *}\right) \nonumber \\
&\qquad + \left[\frac{\bar{\psi}'}{2r^2} - \frac{\bar{\psi}q^2\bar{A}_0^2}{2r\bar{\gamma}^2} - \frac{\bar{\psi}}{2r^3}\right]\left(\delta\psi + \delta\psi^*\right).
\end{align}
A similar equation for $\delta h$ can be obtained from (\ref{dyn_eq2}), that is,
\begin{align}\label{Pt-G00+11}
\frac{\delta h^{\prime}}{\bar{h}} - \frac{\bar{h}'}{\bar{h}^2}\delta h - \frac{2\bar{A}_0q^2\bar{\psi}^2}{r\bar{\gamma}^2}\delta A_0 + \frac{2q^2\bar{A}_0^2\bar{\psi}^2}{r\bar{\gamma}^3}\delta\gamma &= \frac{iq\bar{A}_0\bar{\psi}}{r\bar{\gamma}^2}\left(\delta\dot{\psi} - \delta\dot{\psi}^*\right) + \left[\frac{\bar{\psi}'}{r} - \frac{\bar{\psi}}{r^2}\right]\left(\delta\psi^{\prime} + \delta\psi^{\prime *}\right) \nonumber \\
&\qquad + \left[\frac{\bar{\psi}}{r^3} + \frac{\bar{\psi}q^2\bar{A}_0^2}{r\bar{\gamma}^2} - \frac{\bar{\psi}^{\prime}}{r^2}\right]\left(\delta\psi + \delta\psi^{*}\right).
\end{align}
It will be useful for our later analysis to have an equation with the same structure for $\delta \gamma$. This can be derived from (\ref{dyn_eq3}), leading to
\begin{align}\label{Pt-G00-11}
\frac{\delta\gamma'}{\bar{\gamma}} - \frac{1}{\bar{\gamma}^2}\left[\frac{r\bar{f}(\bar{A}'_0)^2}{\bar{\gamma}}+ \bar{\gamma}'\right]\delta\gamma + \frac{r\bar{f}\bar{A}_0'}{\bar{\gamma}^2}\delta A'_0 + \left[\frac{1}{r\bar{f}^2} + \frac{r \bar{A}'_0}{2\bar{\gamma}^2}\right]\delta f &= 0.
\end{align}
Note that Eq.~(\ref{Pt-G00-11}) is not an independent equation because it can be derived directly from the definition of $\delta \gamma=\delta(fh^{1/2})$ and Eqs.~(\ref{Pt-G00}--\ref{Pt-G00+11}). The final component of the Einstein field equations (\ref{dyn_eq4}) takes the form
\begin{align}\label{Pt-G01}
-\frac{\delta \dot{f}}{\bar{f}} &= \left[\frac{\bar{\psi}'}{2r} - \frac{\bar{\psi}}{2r^2}\right]\left(\delta\dot{\psi} + \delta\dot{\psi}^*\right) + \frac{iq\bar{A}_0\bar{\psi}}{2r}\left(\delta\psi' - \delta\psi^{\prime *}\right) - \frac{iq\bar{A}_0\bar{\psi}'}{2r}\left(\delta\psi - \delta\psi^*\right).
\end{align}
The two components of the Maxwell equations (\ref{dyneqns2}) yield the following expressions,
\begin{align}
\frac{\bar{\gamma}}{\sqrt{\bar{h}}}\delta A^{\prime\prime}_0 + \frac{\bar{\gamma}}{\sqrt{\bar{h}}}\left[\frac{2}{r} - \frac{\bar{h}'}{2\bar{h}}\right]\delta A'_0 - \frac{q^2\bar{\psi}^2}{r^2}\delta A_0 &= -\frac{q^2\bar{A}_0\bar{\psi^2}}{r^2\bar{\gamma}}\delta\gamma + \frac{\bar{\gamma}\bar{A}'_0}{2\bar{h}\sqrt{\bar{h}}}\delta h'
+ \frac{1}{2\bar{h}}\left[\frac{\bar{A}_0\bar{\gamma}q^2\bar{\psi^2}}{\bar{\gamma}r^2} - \frac{\bar{\gamma}\bar{A}'_0\bar{h}'}{\bar{h}\sqrt{\bar{h}}}\right]\delta h \nonumber \\
&\qquad + \frac{iq\bar{\psi}}{2r^2}\left(\delta\dot{\psi}-\delta\dot{\psi}^*\right) + \frac{q^2\bar{A}_0\bar{\psi}}{r^2}\left(\delta\psi + \delta\psi^*\right), \label{Pt-EMt} \\
\frac{\delta\dot{A}'_0}{\sqrt{\bar{h}}} - \frac{\bar{A}'_0}{2\bar{h}\sqrt{\bar{h}}}\delta \dot{h} &=  \frac{iq\bar{\psi}\bar{\gamma}}{2r^2}\left(\delta\psi' - \delta\psi^{\prime *}\right) - \frac{iq\bar{\gamma}\bar{\psi}'}{2r^2}\left(\delta\psi - \delta\psi^*\right). \label{Pt-EMr}
\end{align}
Lastly, the Klein-Gordon equation (\ref{dyn_eq7}) yields
\begin{align}\label{Pt-KG}
0 &= -\delta\ddot{\psi} + \bar{\gamma}^2\delta\psi^{\prime\prime} + 2iq\bar{A_0}\delta\dot{\psi} + \bar{\gamma}\bar{\gamma}'\delta\psi' + \left[q^2\bar{A}_0^2 -
\frac{\bar{\gamma}\bar{\gamma}'}{r}\right]\delta\psi - \frac{iq\bar{A}_0\bar{\psi}}{\bar{\gamma}}\delta\dot{\gamma} + \left[\bar{\gamma}\bar{\psi}' - \frac{\bar{\gamma}\bar{\psi}}{r}\right]\delta\gamma' \nonumber \\
& \qquad + \left[2\bar{\gamma}\bar{\psi}^{\prime\prime} + \bar{\gamma}'\bar{\psi}' - \frac{\bar{\psi}\bar{\gamma}'}{r}\right]\delta\gamma + iq\bar{\psi}\delta\dot{A}_0 + 2q^2\bar{\psi}\bar{A}_0\delta A_0.
\end{align}
\label{perteqns}
\end{subequations}

It can be seen from (\ref{Pt-G01}, \ref{Pt-EMr}) that the imaginary part of the scalar field perturbation, $\text{Im}(\delta \psi)$, is out of phase with $\delta f,\delta h,\delta A_0$ and the real part of $\delta \psi$. For this reason, we decompose the perturbed scalar field in the following way
\begin{align}
\delta \psi (t,r) &= \delta u(t,r) + i\delta\dot{w}(t,r),
\label{deltapsi}
\end{align}
where $u(t,r)$ and $w(t,r)$ are real perturbations. This definition implies that $\delta w(t,r)$ is only determined up to an arbitrary function of $r$.
With the definition (\ref{deltapsi}), the Klein-Gordon equation (\ref{Pt-KG}) can be now separated into two independent equations, corresponding to the real and imaginary parts.

By integrating once with respect to time $t$, (\ref{Pt-G01}, \ref{Pt-EMr}) imply that
\begin{subequations}
\begin{align}
\frac{\df}{\barf} &= \frac{1}{r}\left[\frac{\bpsi}{r}-\bpsi'\right]\delta u - \frac{q\bA\bpsi'}{r}\delta w + \frac{q\bA\bpsi}{r}\delta w' + \delta {\mathcal {F}}(r), \label{Pt-df} \\
\frac{\delta h}{\bh\sqrt{\bh}} &= -\frac{2q\bg\bpsi'}{r^2\bA'}\delta w + \frac{2q\bg\bpsi}{r^2\bA'}\delta w' + \frac{2}{\sqrt{\bh}\bA'}\delta A'_0 + \delta {\mathcal  {H}}(r), \label{Pt-dh}
\end{align}
where $\delta {\mathcal {F}}(r)$ and $\delta {\mathcal {H}}(r)$ are arbitrary functions of the radial coordinate $r$. Thus it is straightforward to obtain
\begin{align}\label{Pt-dg}
\dg &= \frac{\bg}{r}\left[\frac{\bpsi}{r}-\bpsi'\right]\delta u - \frac{q\bg\bpsi'}{r^2\bA'}\left[\barf\bh+r\bA\bA'\right]\delta w + \frac{q\bg\bpsi}{r^2\bA'}\left[\barf\bh+r\bA\bA'\right]\delta w' + \frac{\bg}{\bA'}\delta A'_0 + \bg\delta {\mathcal {F}} + \frac{\barf\bh}{2}\delta {\mathcal {H}}.
\end{align}
\label{metricperts}
\end{subequations}
Equations (\ref{metricperts}) allow us to rewrite the metric perturbations $\delta f,\delta h$ and $\delta \gamma$ in terms of the matter perturbations $\delta u,\delta w$ and $\delta A_0$, and hence eliminate the metric perturbations from the Maxwell (\ref{Pt-EMt}--\ref{Pt-EMr}) and Klein-Gordon (\ref{Pt-KG}) equations. Moreover, it is possible to construct the following linear first-order differential equation, from (\ref{Pt-G00}, \ref{Pt-EMt})
\begin{align}
\delta {\mathcal {F}}' + \left[\frac{\barf'}{\barf}+\frac{\bh'}{2\bh}+\frac{1}{r}\right]\delta {\mathcal {F}} &= \frac{r\bA\bA'}{2\bg}\delta {\mathcal {H}}' + \frac{r\bA}{2\bg^2}\left[\frac{q^2\bA\sqrt{\bh}\bpsi^2}{r^2} + \frac{\bg\bA'^2}{\bA} + \frac{\barf\bA'\bh'}{2\sqrt{\bh}}\right]\delta {\mathcal {H}}.
\end{align}
The above equation is integrable, with solution
\begin{align}
\delta {\mathcal {F}} &= \frac{r\bA\bA'}{2\bg}\delta {\mathcal {H}}
\end{align}
up to an overall constant.
We can use this relation to eliminate $\delta {\mathcal {H}}$ from the perturbed field equations.

To find the equations for the matter perturbations, we begin with (\ref{Pt-G00}), obtaining the following equation once the metric perturbations have been eliminated:
\begin{align}\label{Pt-G00new}
0 &= \delta\ddot{w} - \bg^2\delta w'' + \left[-\bg\bg' + \frac{q^2\bpsi^2\bA}{r^2\bA'}\mathcal{A}\right]\delta w' + \left[-q^2\bA^2 + \frac{\barf\bh}{r^2}-\frac{\bg^2}{r^2} -\frac{\barf\bA'^2}{2}-\frac{q^{2}\bA\bpsi\bpsi'}{r^{2}\bA'}\mathcal{A}\right]\delta w \nonumber \\
&\qquad~~ + q\bA\left[-2 + \frac{\bpsi^2}{r^2} - \frac{\bpsi\bpsi'}{r}\right]\delta u + \frac{q\bA\bpsi}{\bA'}\delta A'_0 - q\bpsi\delta A_0
 \nonumber \\
 &\qquad~~
 -\frac{r\bg^2}{q\bA\bpsi}\delta {\mathcal {F}}' + \left[-\frac{\barf\bh}{q\bA\bpsi} -\frac{\barf r^2 \bA'^2}{2q\bA\bpsi}
+ \frac{r\barf\bA'}{q\bpsi\bA^2}\mathcal{A} + \frac{q\bpsi}{r\bA'}\mathcal{A}\right]\delta {\mathcal {F}},
\end{align}
where
\begin{equation}
\mathcal{A}\equiv\barf\bh + r\bA\bA'.
\end{equation}
The imaginary part of scalar field equation (\ref{Pt-KG}) can be integrated once with respect to time to give
\begin{align}\label{Pt-KGIm}
0 &= \delta\ddot{w} - \bg^2\delta w'' + \left[- \bg\bg' + \frac{q^2\bpsi^2\bA}{r^2\bA'}\mathcal{A} \right]\delta w' + \left[-q^2\bA^2 -\frac{q^2\bA\bpsi\bpsi'}{r^2\bA'}\mathcal{A} + \frac{\bg\bg'}{r}\right]\delta w  \nonumber \\
&\qquad~~ + q\bA\left[-2 + \frac{\bpsi^2}{r^2} - \frac{\bpsi\bpsi'}{r}\right]\delta u + \frac{q\bA\bpsi}{\bA'}\delta A'_0 - q\bpsi\delta A_0 + \delta {\mathcal {G}}(r),
\end{align}
where $\delta {\mathcal {G}}(r)$ is an arbitrary function of the radial coordinate $r$.
We next compare the two equations (\ref{Pt-G00new}, \ref{Pt-KGIm}). This gives another linear first-order equation, this time for $\delta {\mathcal {F}}$ and $\delta {\mathcal {G}}$:
\begin{align}\label{dF-dG}
0 &= \delta {\mathcal {F}}' + \left[r\left(\frac{\bar{\psi}}{r}\right)'^2 - \frac{\bA''}{\bA'} - \frac{\bA'}{\bA} - \frac{1}{r} + \frac{\barf'}{\barf}\right]\delta {\mathcal {F}} + \frac{q\bA\bpsi}{r\bg^2}\delta {\mathcal {G}}.
\end{align}
We will return to Eq.~(\ref{dF-dG}) in the next section where we eliminate the unknown functions $\delta {\mathcal {F}}(r)$ and $\delta {\mathcal {G}}(r)$.
For the last step in our derivation of the linearised perturbation equations, we use (\ref{Pt-G00+11}, \ref{Pt-G00-11}) to eliminate $\delta \ddot{w}$ and $\delta A^{''}_{0}$ from the real part of the Klein-Gordon equation (\ref{Pt-KG}).

Following these steps, we may obtain three equations governing three perturbations: $\delta u$, $\delta w$ and $\delta A_0$. The first equation is derived from the real part of the Klein-Gordon equation (\ref{Pt-KG}) and takes the form
\begin{subequations}
\begin{align}\label{Pt-Real}
0 &= \delta\ddot{u} - \bg^2\delta u'' -\bg\bg'\delta u' + \left[3q^2\bA^2+\frac{\bg\bg'}{r}-\barf\bh \left(\frac{\bar{\psi}}{r}\right)'^2+\frac{\barf\bA'^2}{2}\left(\left(\frac{\bar{\psi}}{r}\right)^2+\bpsi'^2\right)-\frac{\barf\bpsi\bpsi'\bA'^2}{r}\right]\delta u + 2q\bA\bg^2\delta w'' \nonumber \\
&\quad + q\barf\bA\left[2\sqrt{\bh}\bg'+\left(-\frac{\bA'}{\bA}\mathcal{A}+\frac{\bh}{r}+\frac{r\bA'^2}{2}\right)\left(\frac{\bar{\psi}}{r}\right)'\bpsi\right]\delta w' + q\bA\left[2q^2\bA^2 -\frac{2\bg\bg'}{r}+\bg\bpsi'\left(\frac{\bar{\psi}}{r}\right)'\left(\frac{\bg\bA'}{\bA}-\bg'-\frac{\bg}{r}\right)\right]\delta w \nonumber \\
&\quad + \frac{2r\bg^2}{\bpsi}\delta {\mathcal {F}}' + \bg\left(\left(\frac{\bar{\psi}}{r}\right)'+\frac{2}{\bpsi}\right)\left[\left(r\bg\right)'-\frac{r\bg\bA'}{\bA}\right]\delta {\mathcal {F}}.
\end{align}
The second equation is obtained from (\ref{Pt-KGIm}), after application of (\ref{dF-dG}),
\begin{align}\label{Pt-Im}
0 &= \delta\ddot{w} - \bg^2\delta w'' + \left[-\bg\bg' + \frac{q^2\bA\bpsi^2}{r^2\bA'}\mathcal{A}\right]\delta w' + \left[-q^2\bA^2 - \frac{q^2\bA\bpsi\bpsi'}{r^2\bA'}\mathcal{A}+\frac{\bg\bg'}{r}\right]\delta w -q\bA\left[2+r\left(\frac{\bar{\psi}}{r}\right)\left(\frac{\bar{\psi}}{r}\right)'\right]\delta u \nonumber \\
&\qquad + \frac{q\bA\bpsi}{\bA'}\delta A'_0 - q\bpsi\delta A_0 - \frac{r\bg^2}{q\bA\bpsi}\delta {\mathcal {F}}' + \frac{r\bg^2}{q\bA\bpsi}\left[\frac{1}{r}-\frac{\barf'}{\barf} + \frac{\bA'}{\bA} + \frac{\bA''}{\bA'} - r\left(\frac{\bar{\psi}}{r}\right)'^2\right]\delta  {\mathcal {F}}.
\end{align}
The third equation comes from the Einstein field equation (\ref{Pt-G00-11})
\begin{align}\label{Pt-Const}
0 &= \frac{q\bpsi}{\bA' r^2}\mathcal{A}\delta w'' + \frac{q\bpsi\bA}{r^2}\left[\frac{\bg'}{\bA\bA'\bg}\mathcal{A} - \frac{q^2\bpsi^2\bh}{r^2\bA'^2}\right]\delta w' + \frac{q\bpsi\bA}{r^2}\left[\frac{\mathcal{A}}{r\bA\bA'\bg}\left(-\bg'+\frac{rq^2\bA^2}{\bg}\right) + \frac{q^2\bh\bpsi\bpsi'}{r^2\bA'^2}\right]\delta w \nonumber \\
&\qquad -\left(\frac{\bar{\psi}}{r}\right)'\delta u' -\left[\left(\frac{\bar{\psi}}{r}\right)''+\left(\frac{1}{r}+\frac{\bg'}{\bg}\right)\left(\frac{\bar{\psi}}{r}\right)'\right]\delta u + \left[\frac{\delta A'_0}{\bA'}\right]' + \left[\frac{\mathcal{A}}{r\bA\bA'}\delta {\mathcal {F}}\right]' - \left[-\frac{1}{r\barf}+\frac{\bA'}{\bA}+\frac{r\bA'^2}{2\barf\bh}\right]\delta {\mathcal {F}}.
\end{align}
\label{perteqnsfinal}
\end{subequations}
\end{widetext}
To summarise, we have obtained two dynamical equations (\ref{Pt-Real}, \ref{Pt-Im}) which involve time derivatives, and one constraint equation (\ref{Pt-Const}) which contains only derivatives with respect to $r$.
Note that that all these equations (\ref{perteqnsfinal}) only contain radial derivatives of the electric potential perturbation $\delta A_{0}$, and not time derivatives. Essentially, this is due to residual gauge freedom (as discussed in Sec.~\ref{subsec:dyn}), which means that an arbitrary global function of time can be added to $\delta A_{0}$ without changing physical quantities such as the electromagnetic field.

\begin{figure*}
%\figuretitle{$q=0.1,E_h=0.8$ and $\phi_h=1.2$}
\begin{tabular}{c}
\includegraphics[width=0.67\columnwidth]{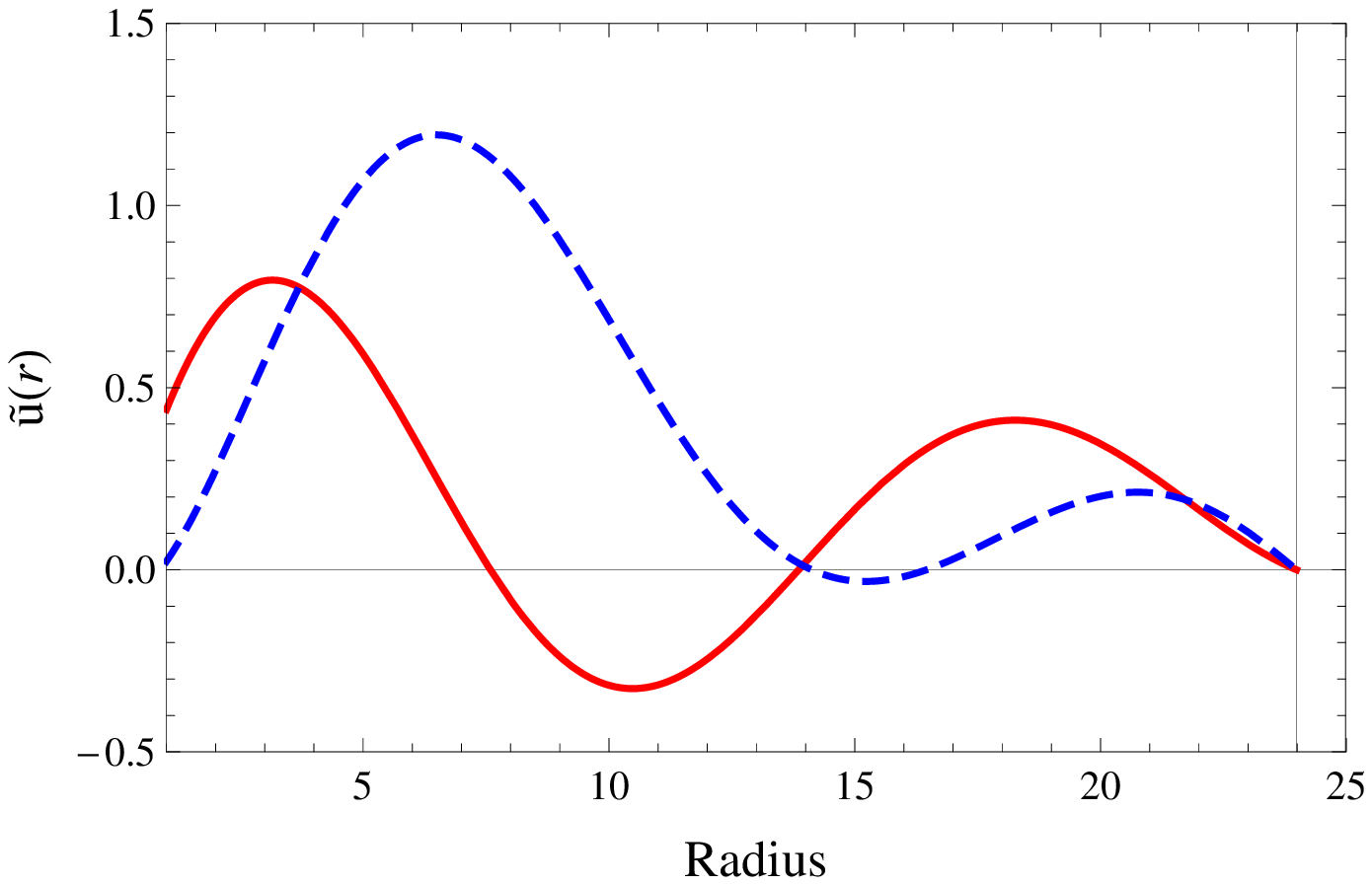}
\includegraphics[width=0.67\columnwidth]{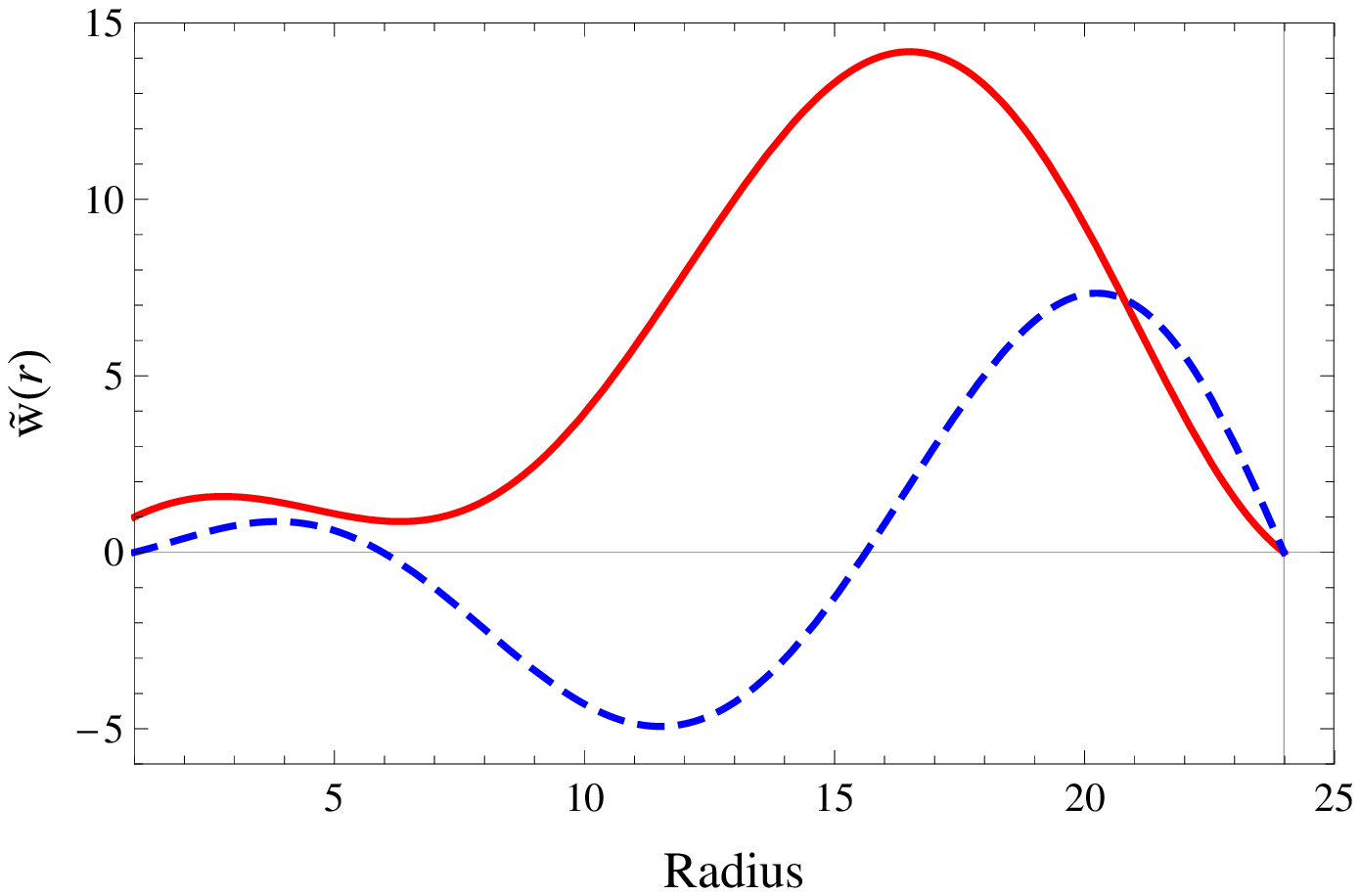}
\includegraphics[width=0.67\columnwidth]{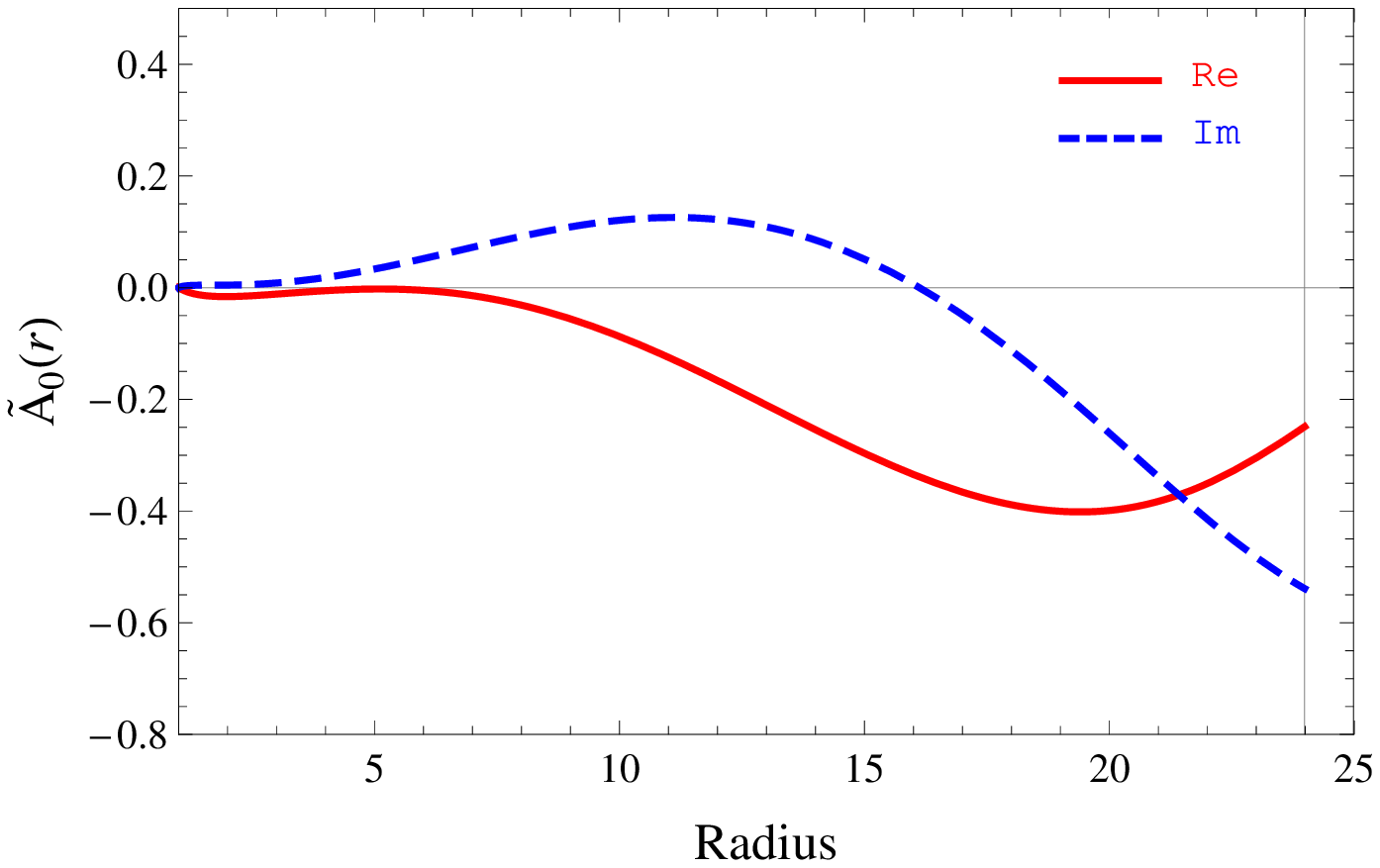}
\end{tabular}
\caption{An example plot of the three perturbation functions, $\tilde{u}$, $\tilde{w}$ and $\tilde{A_0}$ satisfying the perturbation equations (\ref{perteqnsfinal}) and boundary conditions (\ref{Pt-BCHorizon}, \ref{Pt-BCmirror}). The equilibrium solution parameters are: scalar-field charge $q=0.1$, electric field on the horizon $E_{h}=0.8$ and scalar field on the horizon $\phi _{h}=1.2$. The corresponding eigenvalue $\sigma$ is $0.1731-0.0038i$ and ${\tilde {u}}_0 = 0.4397+0.0231i$. The horizontal and vertical lines are included to help visualise where the perturbation functions vanish and the location of the mirror. For this example the mirror is situated at $r_m \approx 24 $, which is the first zero of the equilibrium scalar field. }
\label{fig:pert_functions}
\end{figure*}

\subsection{Boundary conditions}
We now consider the boundary conditions for the perturbed field variables at two boundaries, i.e. at the black hole horizon $(r=r_h)$ and at the mirror $(r=r_m)$.
Near the horizon, we impose ingoing boundary conditions
\begin{align}\label{Pt-BCHorizon}
\delta u(t,r) &= \text{Re}\left[ e^{-i\sigma (t+r_{*})}\tilde{u}(r) \right] , \nonumber \\
\delta w(t,r) &= \text{Re} \left[ e^{-i\sigma (t+r_{*})}\tilde{w}(r) \right] , \nonumber \\
\delta A_0(t,r) &= \text{Re} \left[ e^{-i\sigma (t+r_{*})}\tilde{A}_{0}(r) \right] ,
\end{align}
where $r_{*}$ is the tortoise coordinate defined by
\begin{equation}
\frac {dr_{*}}{dr}=\bg .
\end{equation}
Here, ${\tilde {u}}$, ${\tilde {w}}$ and ${\tilde {A}}_{0}$ are complex functions which depend only on the radial coordinate $r$ and have Taylor series expansions near the horizon of the form
\begin{align}\label{Pt-BCHorizon1}
\tilde{u} &= {\tilde {u}}_{0} + {\tilde {u}}_{1}(r-r_h) + O(r-r_h)^2, \nonumber \\
\tilde{w} &= {\tilde {w}}_{0} + {\tilde {w}}_{1}(r-r_h) + O(r-r_h)^2, \nonumber \\
\tilde{A}_0 &= \tilde{A}_{1}(r-r_h) + \tilde{A}_{2}(r-r_h)^2 + O(r-r_h)^3.
\end{align}

Before we proceed further, we noted earlier that adding an arbitrary function of $r$ to $\delta w$ makes no difference to the scalar field perturbation $\delta\psi$. This freedom allows us to set $\delta {\mathcal {G}}=0$ in (\ref{Pt-KGIm}).  Hence, (\ref{dF-dG}) is solvable using the conventional integrating factor method and the solution is given by
\begin{align}\label{int-factor}
\delta  {\mathcal {F}}&= {\mathcal {K}} \frac{r\bA\bA'}{\barf}\exp \left[ {-\int_{r_h}^{r} r \left(\frac{\bar{\psi}}{r}\right)'^2 dr } \right] ,
\end{align}
where ${\mathcal {K}}$ is a constant of integration. As ingoing boundary conditions are required for \emph{all} perturbations, including the metric variables $\delta f$, $\delta h$ (and thus $\delta \gamma$),
it must be the case from Eq.~(\ref{Pt-df}) that  $\delta {\mathcal {F}}=0$ at $r=r_{h}$.
Therefore we must set ${\mathcal {K}}=0$ in (\ref{int-factor}), and so $\delta {\mathcal {F}}$ vanishes identically. Thus $\delta {\mathcal {F}}'$ and $\delta {\mathcal {F}}$ are eliminated from the perturbation equations (\ref{perteqnsfinal}) by our choice of boundary conditions.

By inserting (\ref{Pt-BCHorizon}, \ref{Pt-BCHorizon1}) into the perturbation equations (\ref{perteqnsfinal}), we find that ${\tilde {u}}_1$, ${\tilde {w}}_1$, $\tilde{A}_1$ and $\tilde{A}_2$ can be expressed in terms of ${\tilde {u}}_0$, ${\tilde {w}}_0$ and $\sigma$. The simpler of these expressions are
\begin{align}
\tilde{A}_{1} &= -\frac{iq\sigma\phi_{h}\left(2+r^{2}_h E^{2}_h\right)}{r_h\left(-2+2ir_h\sigma+r^{2}_h E^{2}_h\right)}{\tilde {w}}_{0}, \nonumber \\
{\tilde {u}}_{1} &= \frac{-8qE_h\sigma^2r^{3}_h{\tilde {w}}_0 + \left(-2+r^{2}_h E^{2}_h\right)^2 {\tilde {u}}_0}{r_h\left(-2 + r^{2}_h E^{2}_h\right)\left(-2+4ir_h\sigma+r^{2}_h E^{2}_h\right)},
\end{align}
while the expressions for ${\tilde {w}}_1$ and $\tilde{A}_2$ are sufficiently complicated to be omitted here. Thus with given values for the background parameters $q,\phi_h$ and $E_h$, the boundary conditions (\ref{Pt-BCHorizon}) depend on three additional parameters, namely ${\tilde {u}}_0$, ${\tilde {w}}_0$ and $\sigma$.
We emphasize that the parameters ${\tilde {u}}_{0}$, ${\tilde {w}}_{0}$ and $\sigma $ are all complex.  The physical perturbations arise from taking the real
part in (\ref{Pt-BCHorizon}).

At the mirror $r=r_{m}$, the scalar field perturbation $\delta \psi$ (like the background scalar field $\bpsi$) must vanish. The perturbations of the metric functions and electric potential are unconstrained there. Since the real and imaginary parts of the scalar field perturbation (\ref{deltapsi}) take the form (\ref{Pt-BCHorizon}),
at the mirror the functions ${\tilde {u}}(r)$ and ${\tilde {w}}(r)$ must satisfy
\begin{align}\label{Pt-BCmirror}
\tilde{u}(r_m) &= 0=
\tilde{w}(r_m) .
\end{align}
We require both the real and imaginary parts of ${\tilde {u}}(r)$ and ${\tilde {w}}(r)$ to vanish at the mirror so that the real and imaginary parts of the
scalar field perturbation (given by (\ref{Pt-BCHorizon})) vanish for all time $t$, when the real part in (\ref{Pt-BCHorizon}) is taken.

In summary, using the form (\ref{Pt-BCHorizon}) for the matter perturbations, we now have three ordinary differential equations (\ref{perteqnsfinal}) for three unknown functions of $r$, namely: ${\tilde {u}}$, ${\tilde {w}}$ and ${\tilde {A}}_0$. Together with the boundary conditions (\ref{Pt-BCHorizon}, \ref{Pt-BCmirror}), we now have a system which can be solved numerically.

\subsection{Method and results}

We implement a shooting method to numerically solve the boundary value problem (\ref{perteqnsfinal}, \ref{Pt-BCHorizon}, \ref{Pt-BCmirror}). Since both the perturbation equations and boundary conditions are linear, we set the overall scale of the perturbations so that
${\tilde {w}}_0$ is fixed to be unity.
This leaves two free parameters, ${\tilde {u}}_0$ and $\sigma$, which we use as shooting parameters. The process of numerical integration is as follows. Firstly, we specify the background parameters $q,\phi_h$ and $E_h$, then integrate the static field equations. We obtain the numerical hairy black hole solution and find the location of the first zero of the equilibrium scalar field, setting this to be the mirror location $r_m$. Secondly, the three coupled perturbation equations (\ref{perteqnsfinal}) are solved by searching for values of ${\tilde {u}}_0$ and $\sigma$ such that the boundary conditions (\ref{Pt-BCHorizon}, \ref{Pt-BCmirror}) are satisfied.
We are particularly interested in the sign of the imaginary part of the frequency, ${\text{Im}}(\sigma )$. Perturbations for which $\text{Im}(\sigma )<0$ are stable and decay exponentially in time, whereas perturbations for which $\text{Im}(\sigma )>0$ are unstable, growing exponentially in time.

\begin{figure*}
\begin{tabular}{c}
\includegraphics[width=0.95\columnwidth]{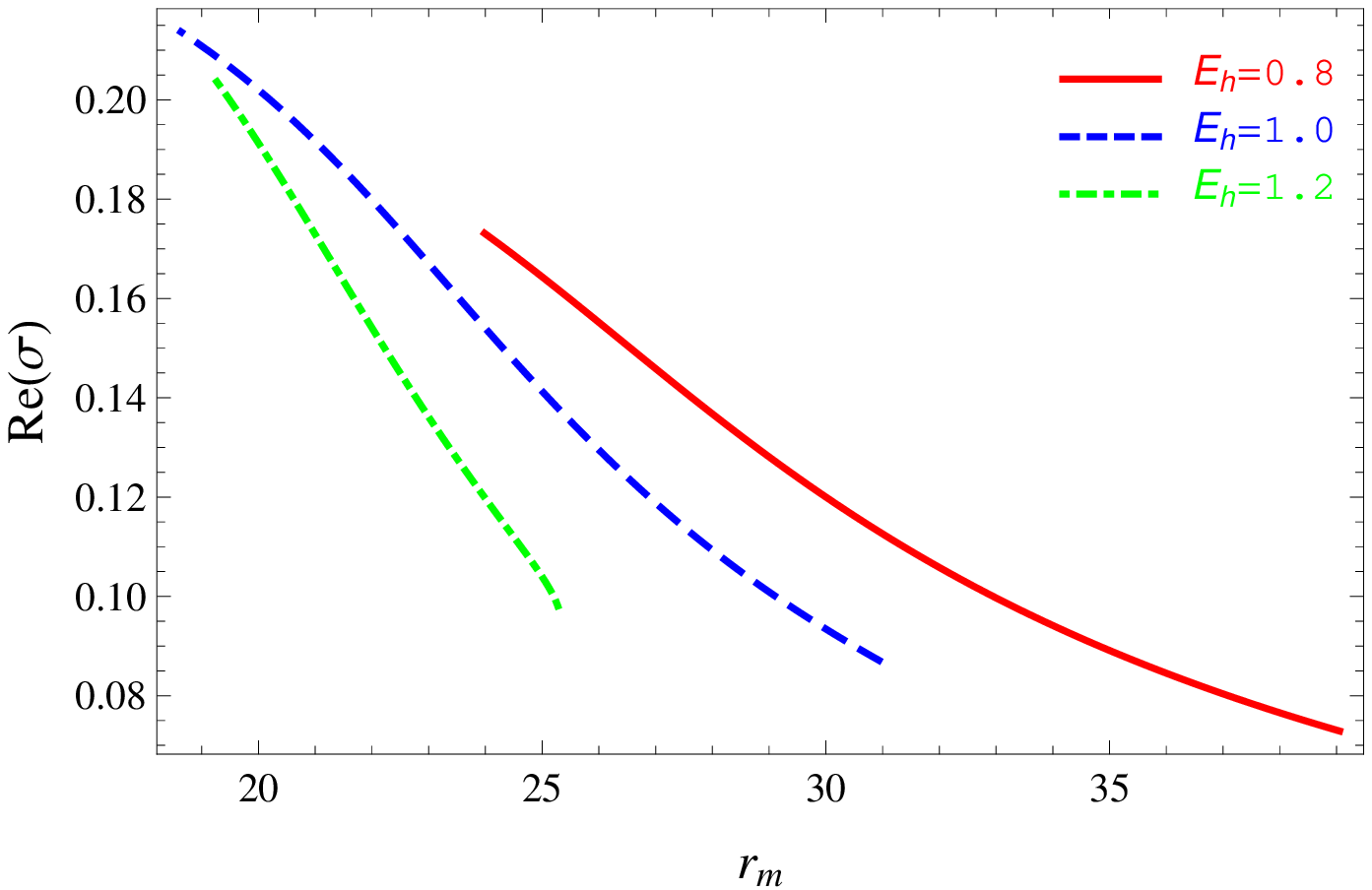}
\includegraphics[width=0.95\columnwidth]{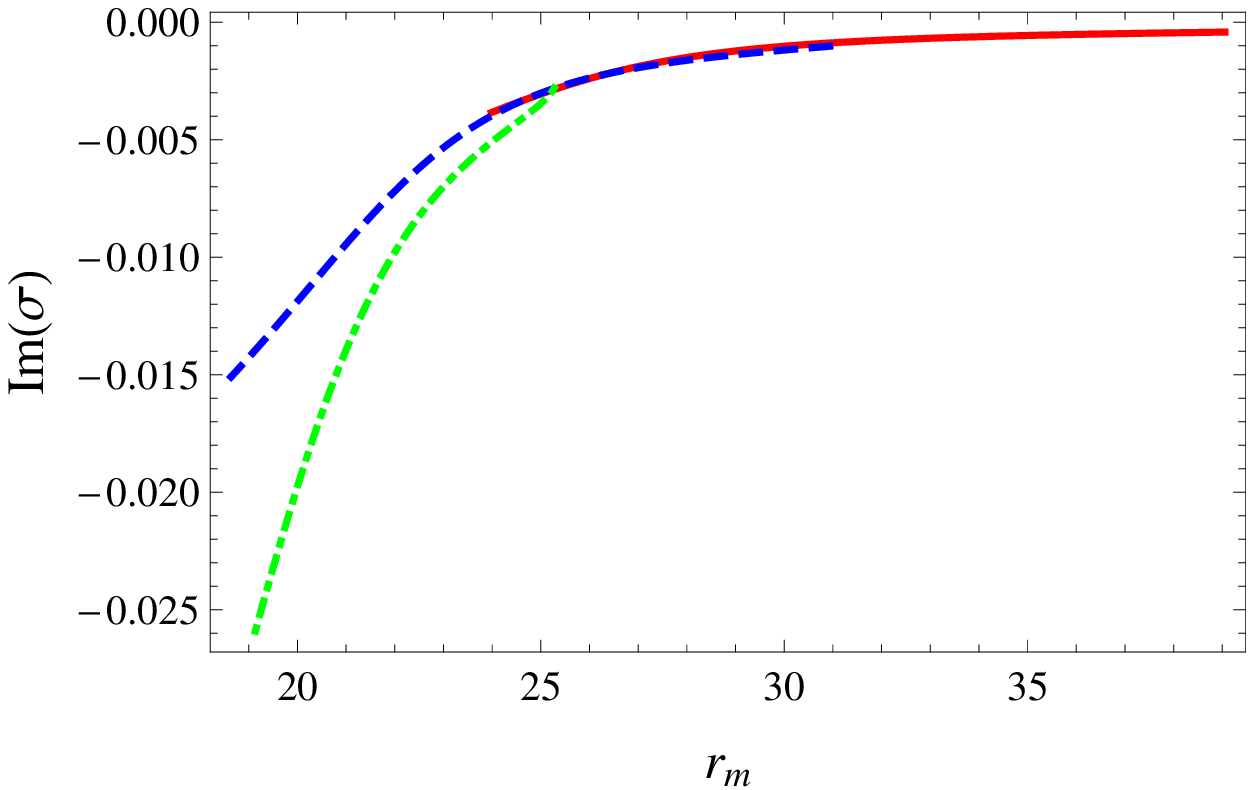} \\
\includegraphics[width=0.95\columnwidth]{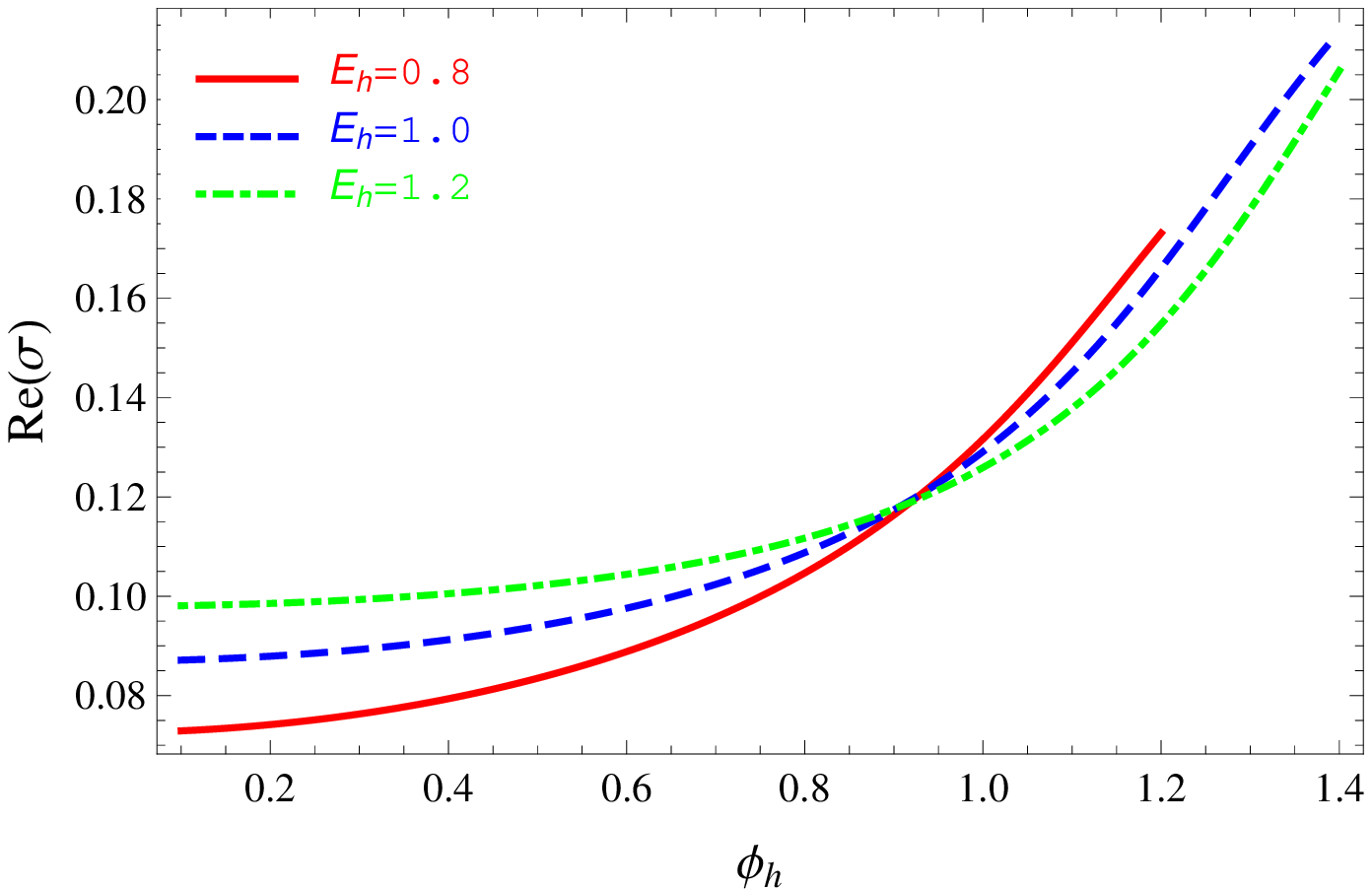}
\includegraphics[width=0.95\columnwidth]{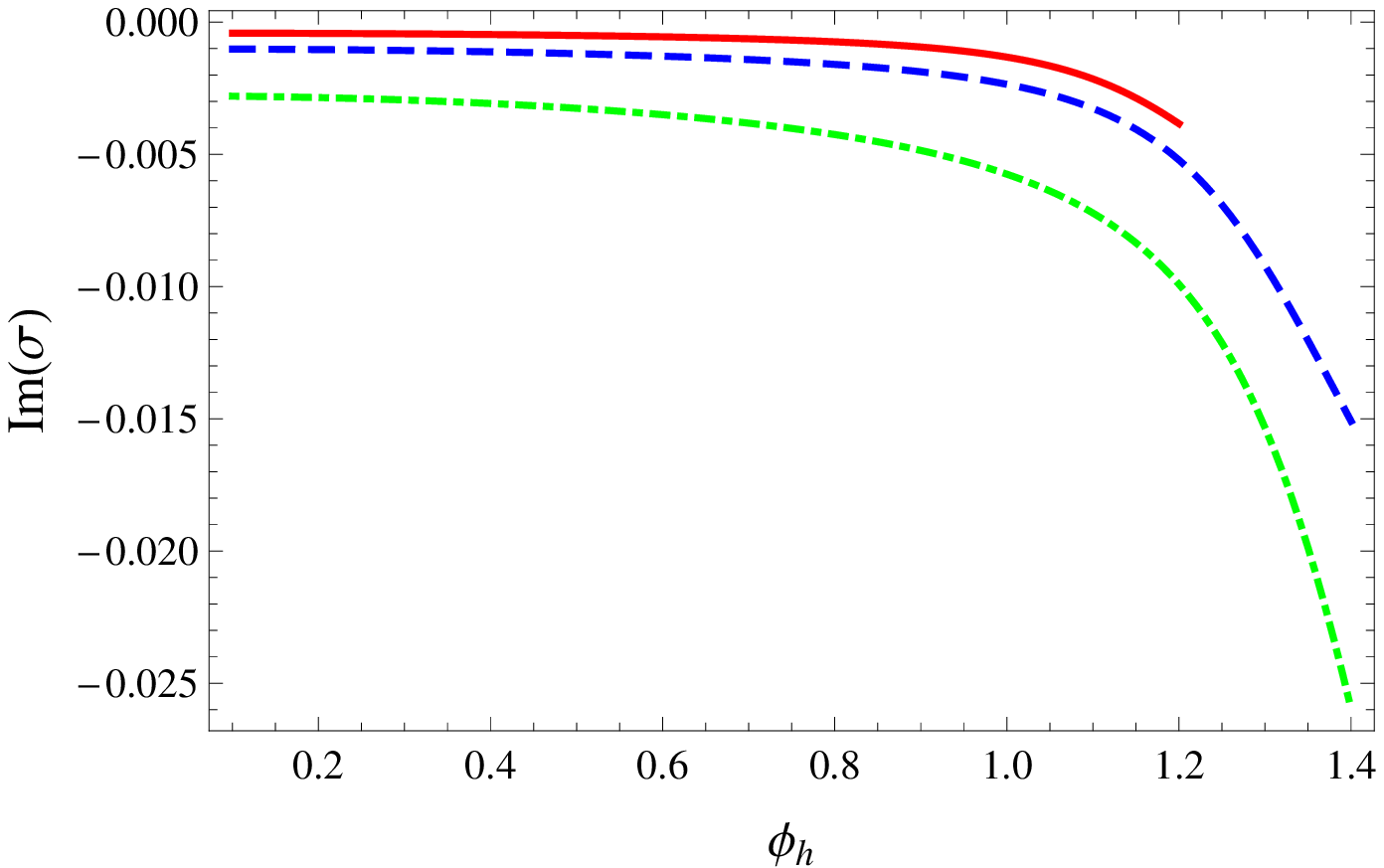} \\
\end{tabular}
\caption{The real (left) and imaginary (right) part of the mode frequency $\sigma$ is plotted as a function of (top row) the mirror radius $r_m$ and (bottom row) the equilibrium scalar field on the horizon $\phi_h$, for $q=0.1$, $\phi_h \in (0.1, 1.4)$ and various values of $E_h$. The mirror is located at the first zero of the equilibrium scalar field.  In all these plots, $\text{Im}(\sigma )<0$, and the perturbations decay exponentially in time.}
\label{fig:q1fixedEhvaryPhih}
\end{figure*}

\begin{figure*}
\begin{tabular}{c}
\includegraphics[width=0.95\columnwidth]{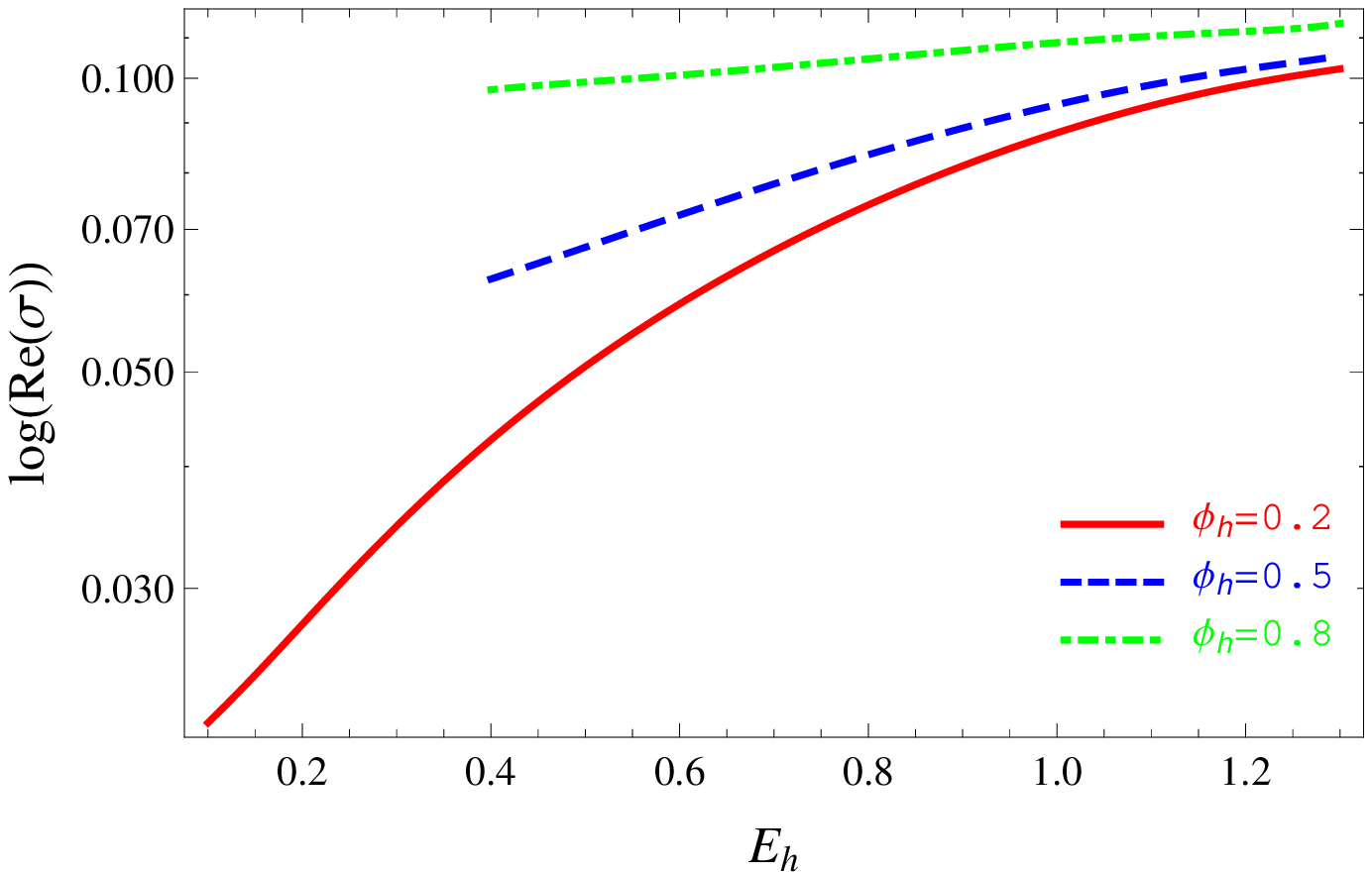}
\includegraphics[width=0.95\columnwidth]{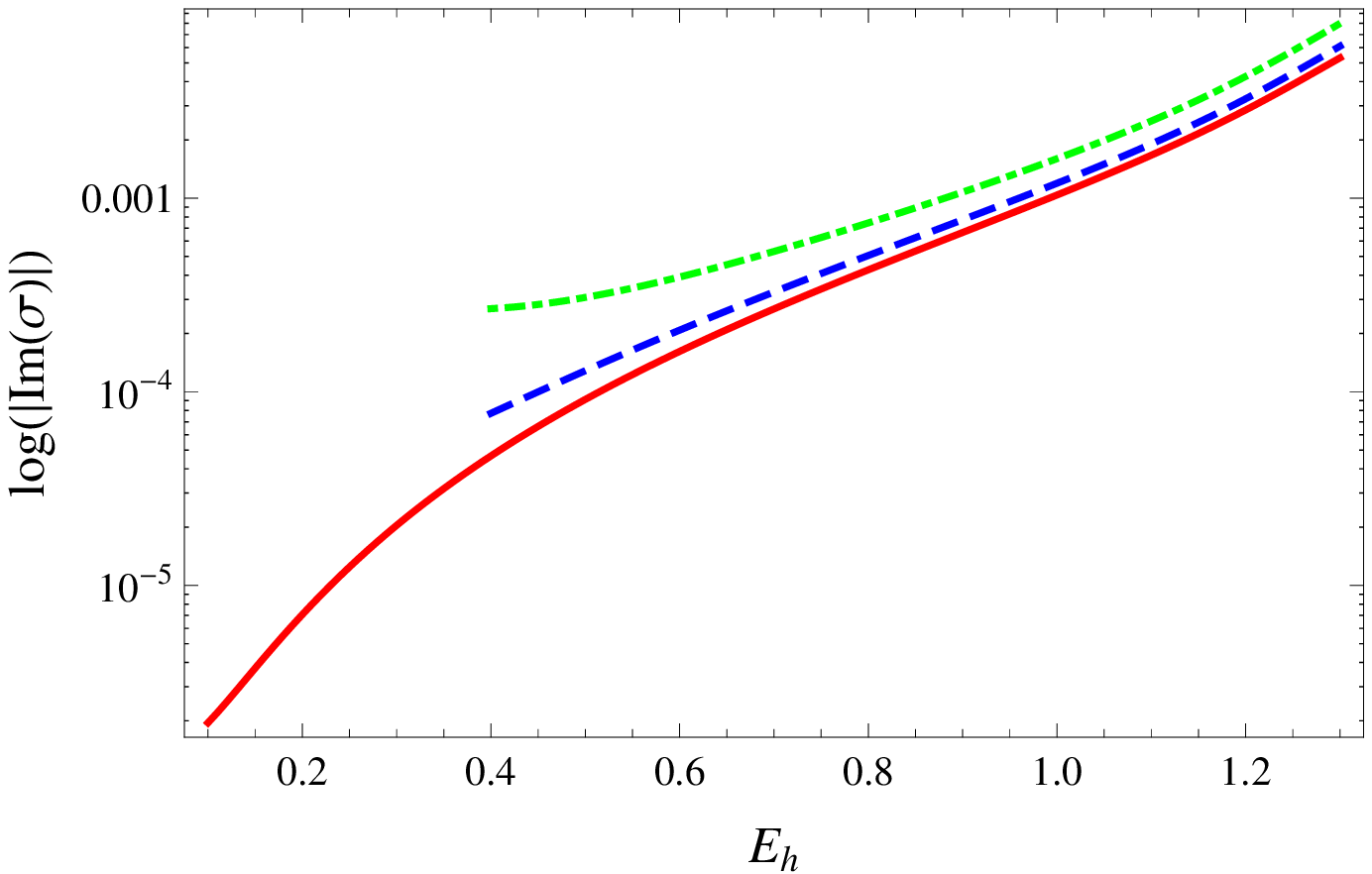}
\end{tabular}
\caption{The real (left) and imaginary (right) part of the mode frequency $\sigma$ is plotted as a function of the electric field at the horizon $E_h$, for $q=0.1$ and various values of $\phi_h$. The mirror is located at the first zero of the equilibrium scalar field.  In the right-hand plot, $\text{Im}(\sigma )<0$ and we have plotted the logarithm of the modulus of $\text{Im}(\sigma )$.}
\label{fig:q1fixedPhihvaryEh}
\end{figure*}

\begin{figure*}
\begin{tabular}{c}
\includegraphics[width=0.95\columnwidth]{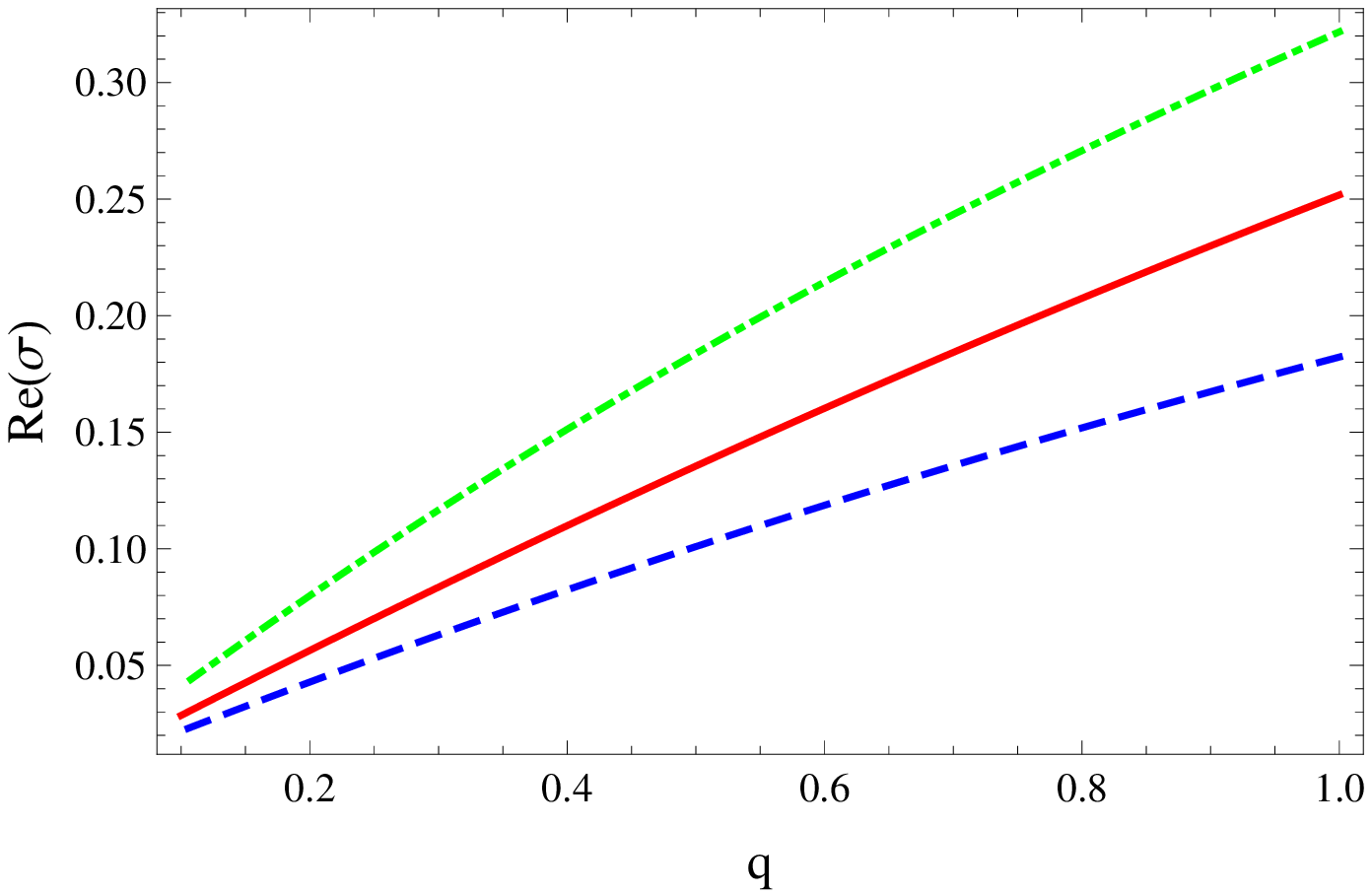}
\includegraphics[width=0.95\columnwidth]{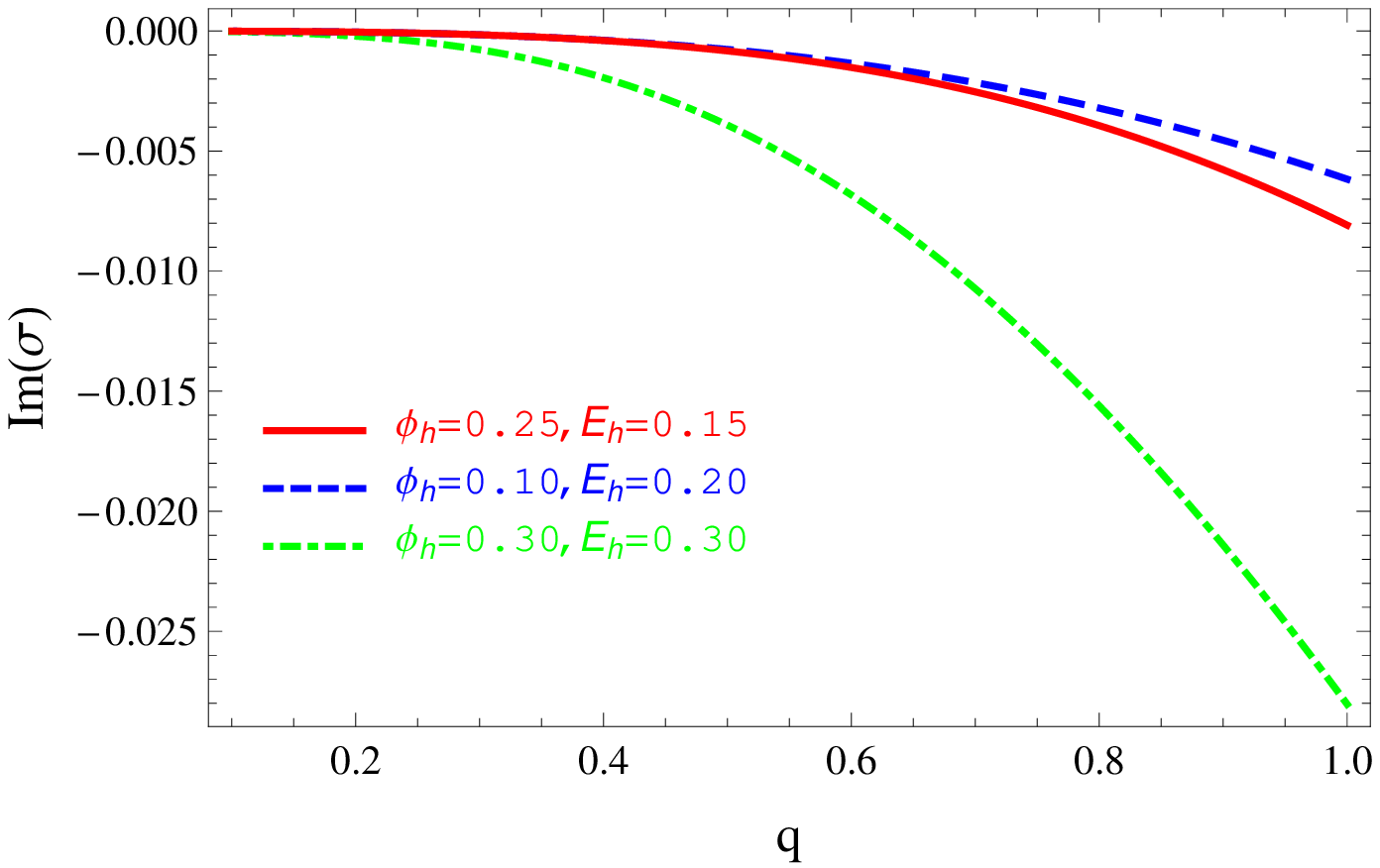}
\end{tabular}
\caption{The real (left) and imaginary (right) part of the mode frequency $\sigma$ is plotted as a function of the scalar charge $q$, for various values of $\phi_h$ and $E_h$. The mirror is located at the first zero of the equilibrium scalar field.  All values of $\text{Im}(\sigma )$ shown in the right-hand plot are negative.}
\label{fig:fixedPhihEhvaryq}
\end{figure*}

\begin{figure*}
\begin{tabular}{c}
\includegraphics[width=0.95\columnwidth]{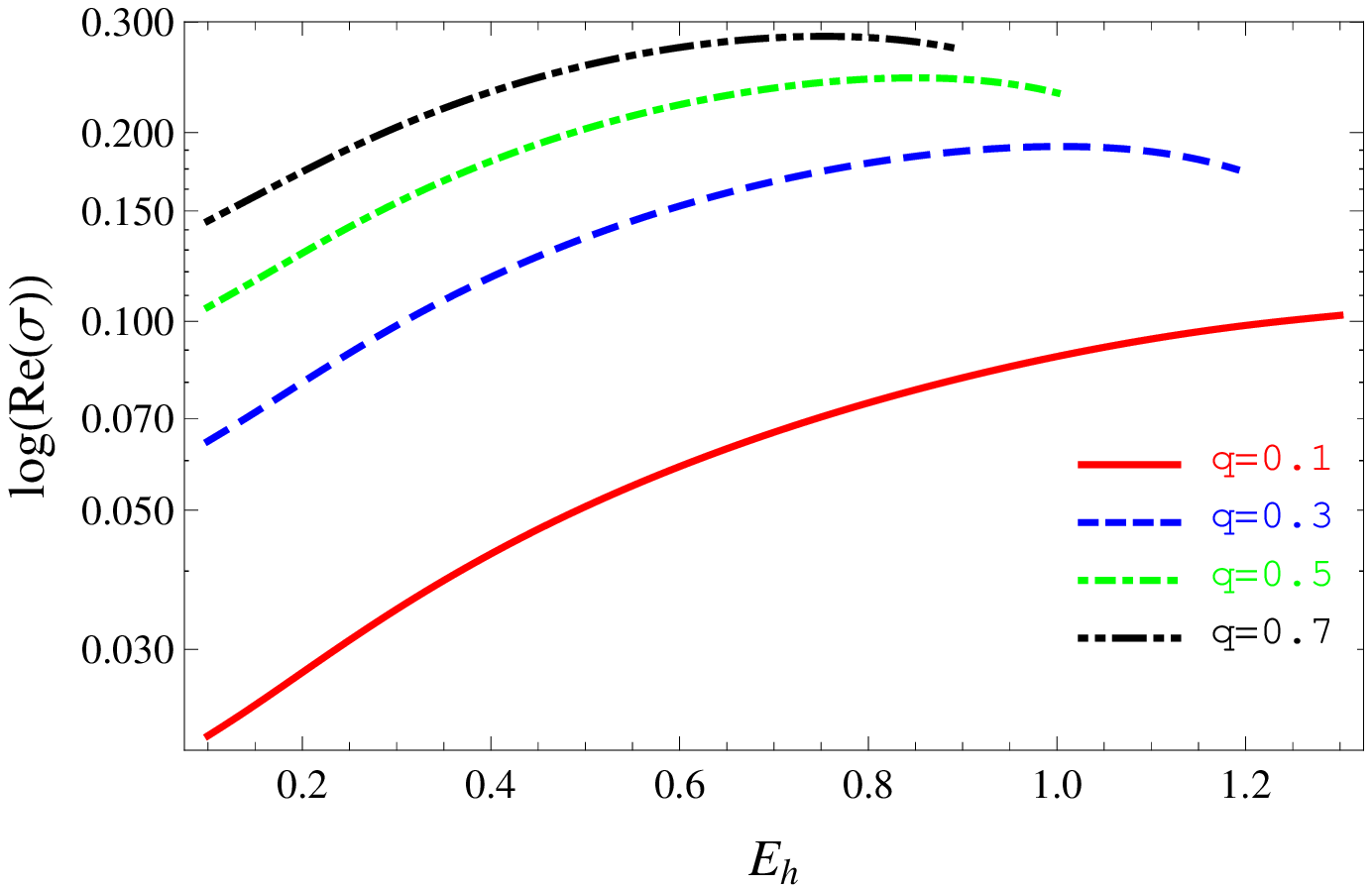}
\includegraphics[width=0.95\columnwidth]{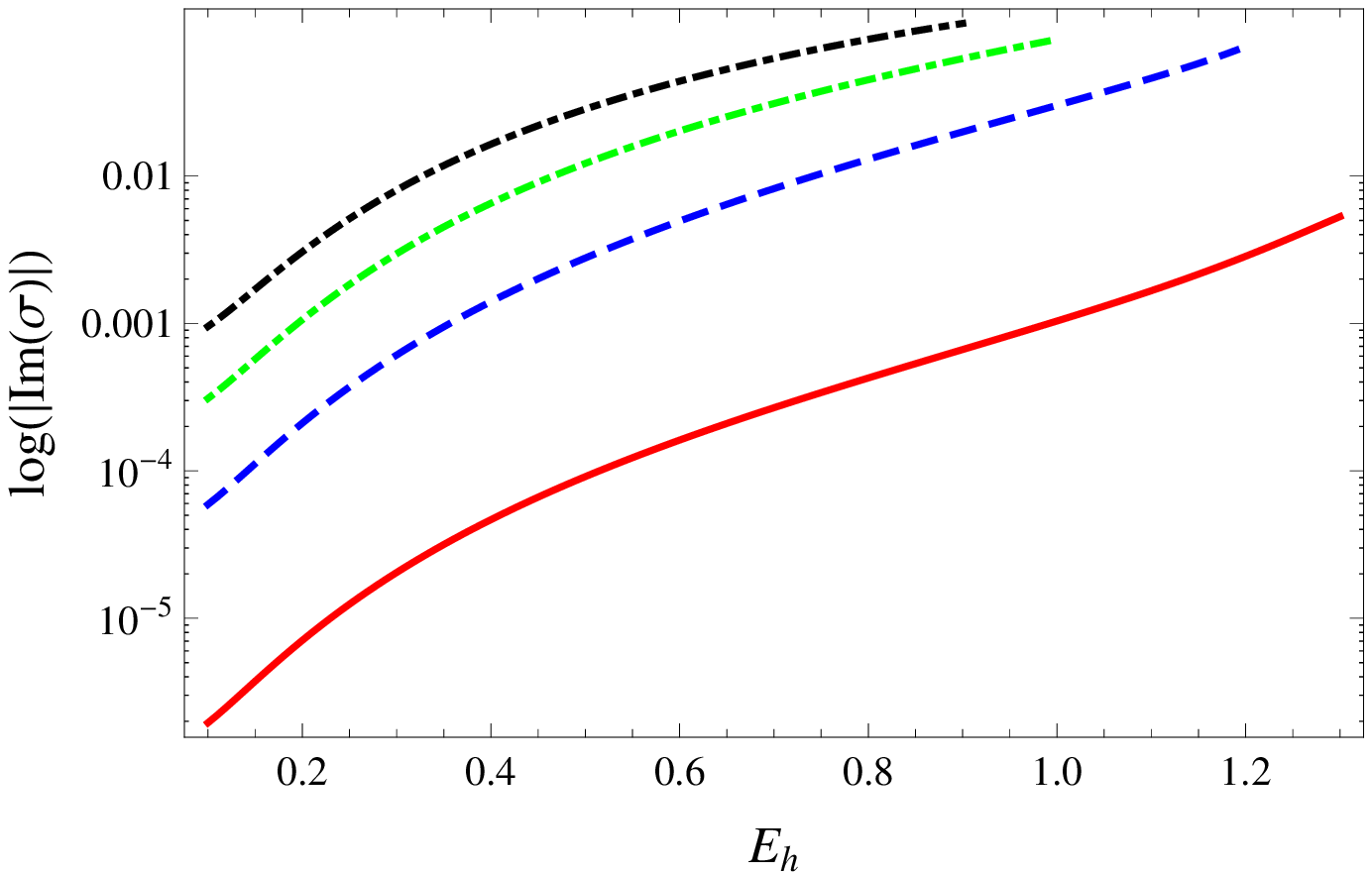}
\end{tabular}
\caption{The real (left) and imaginary (right) part of the mode frequency $\sigma$ is plotted as a function of the electric field at the horizon $E_h$, for fixed $\phi_h=0.2$ and various values of $q$. The mirror is located at the first zero of the equilibrium scalar field. All values of $\text{Im}(\sigma )$ shown in the right-hand plot are negative; we have plotted the logarithm of the modulus of $\text{Im}(\sigma )$.}
\label{fig:fixedPhihqvaryEh}
\end{figure*}

\begin{figure*}
\begin{tabular}{c}
\includegraphics[width=0.95\columnwidth]{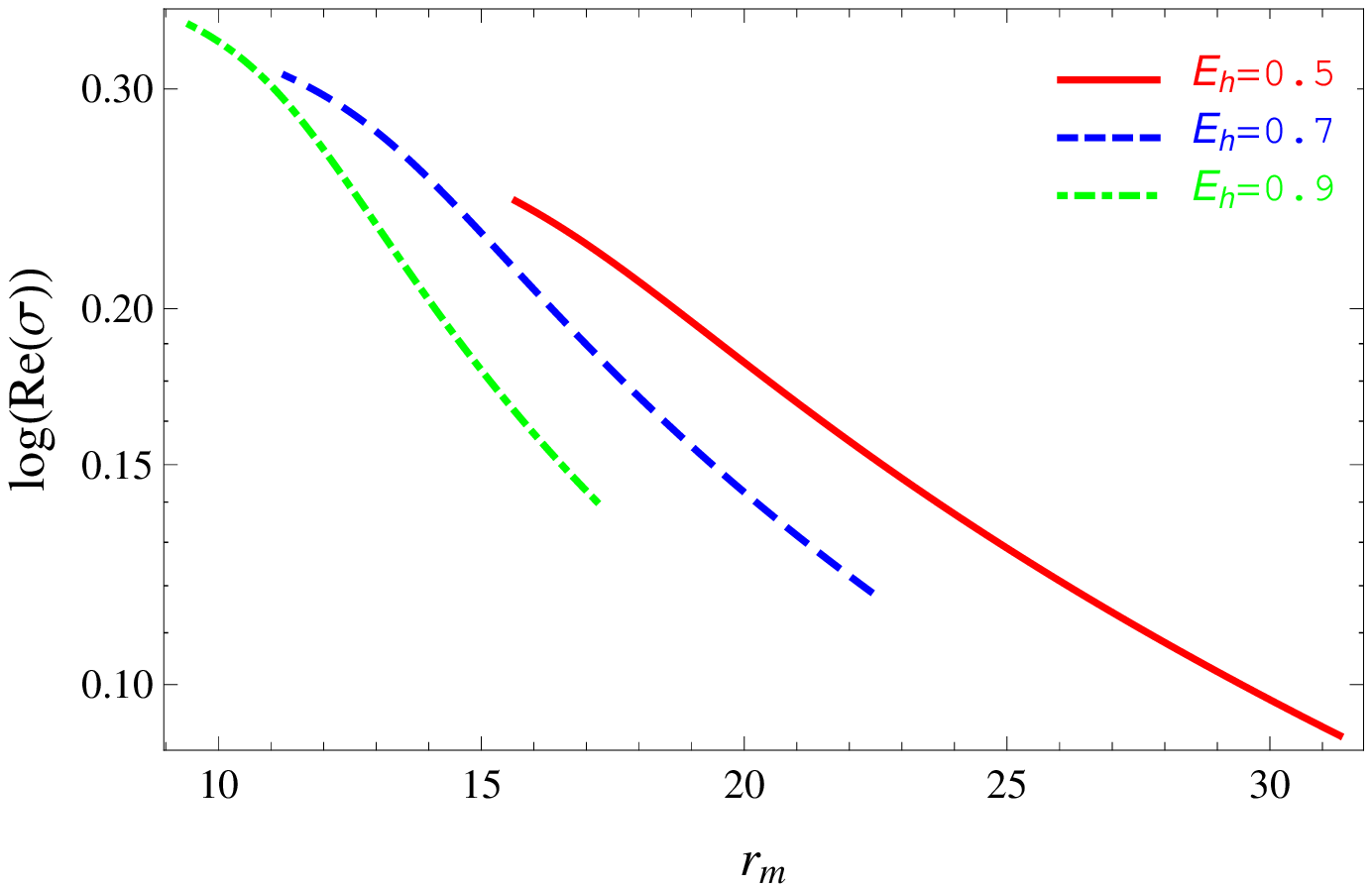}
\includegraphics[width=0.95\columnwidth]{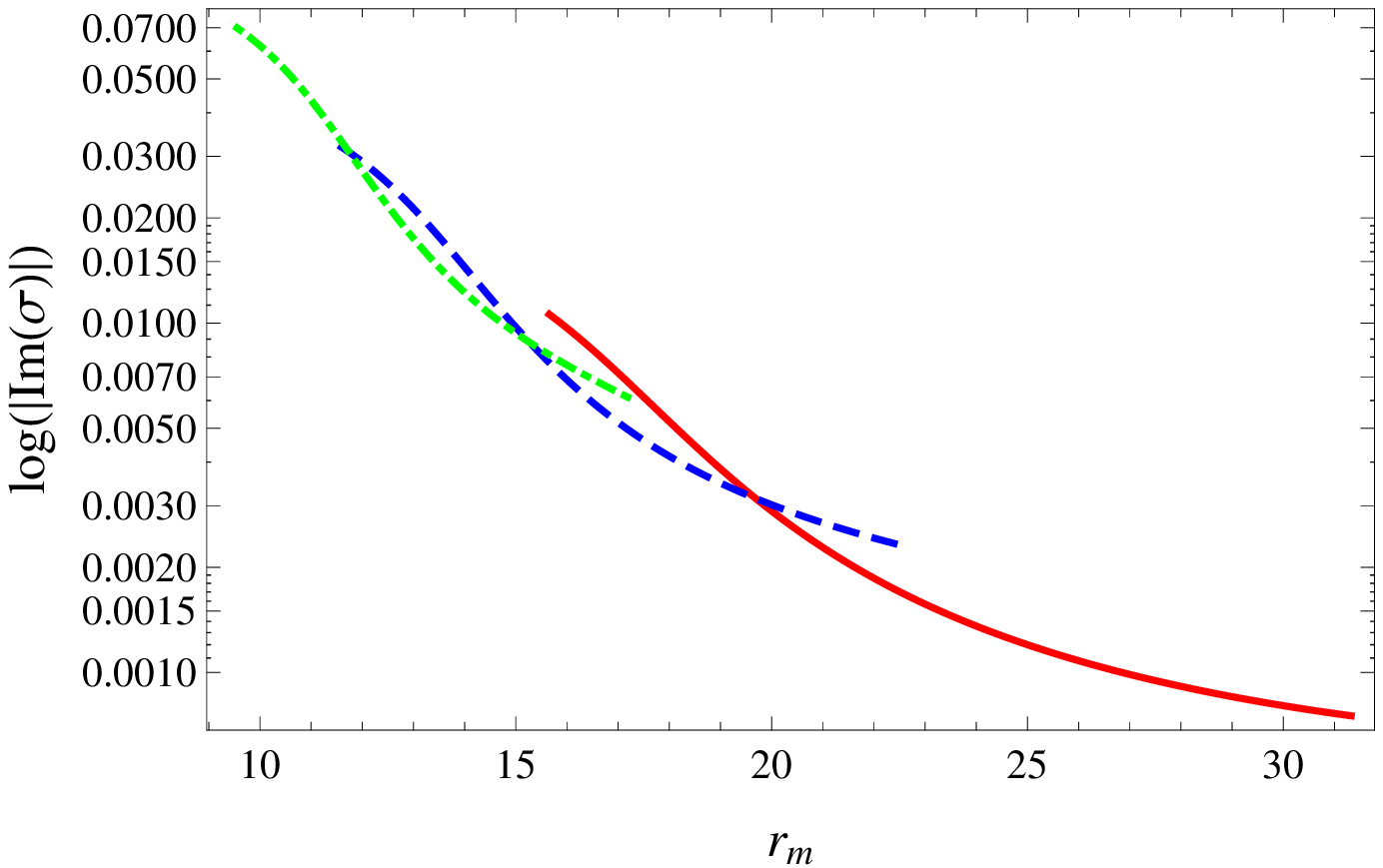} \\
\includegraphics[width=0.95\columnwidth]{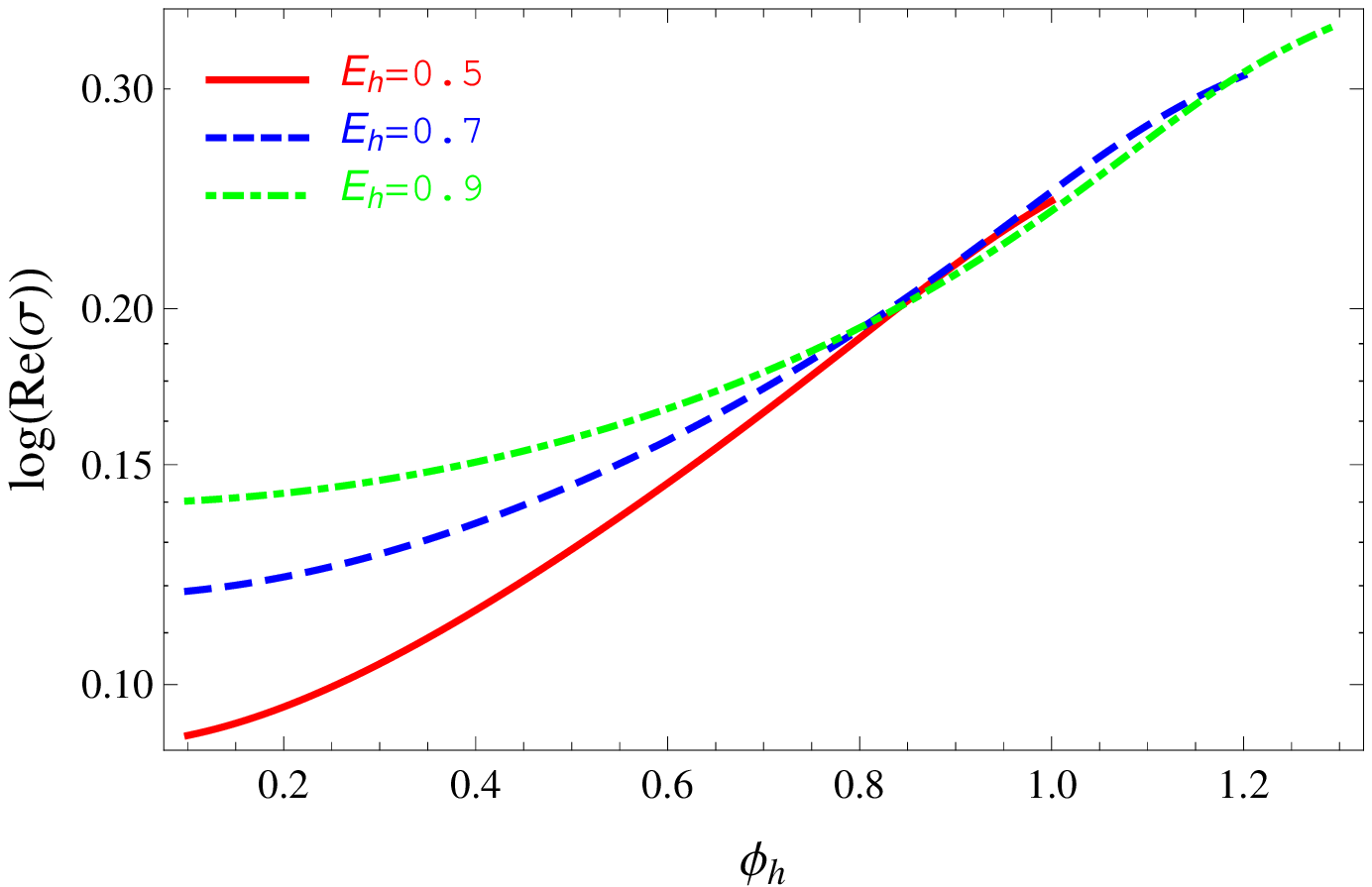}
\includegraphics[width=0.95\columnwidth]{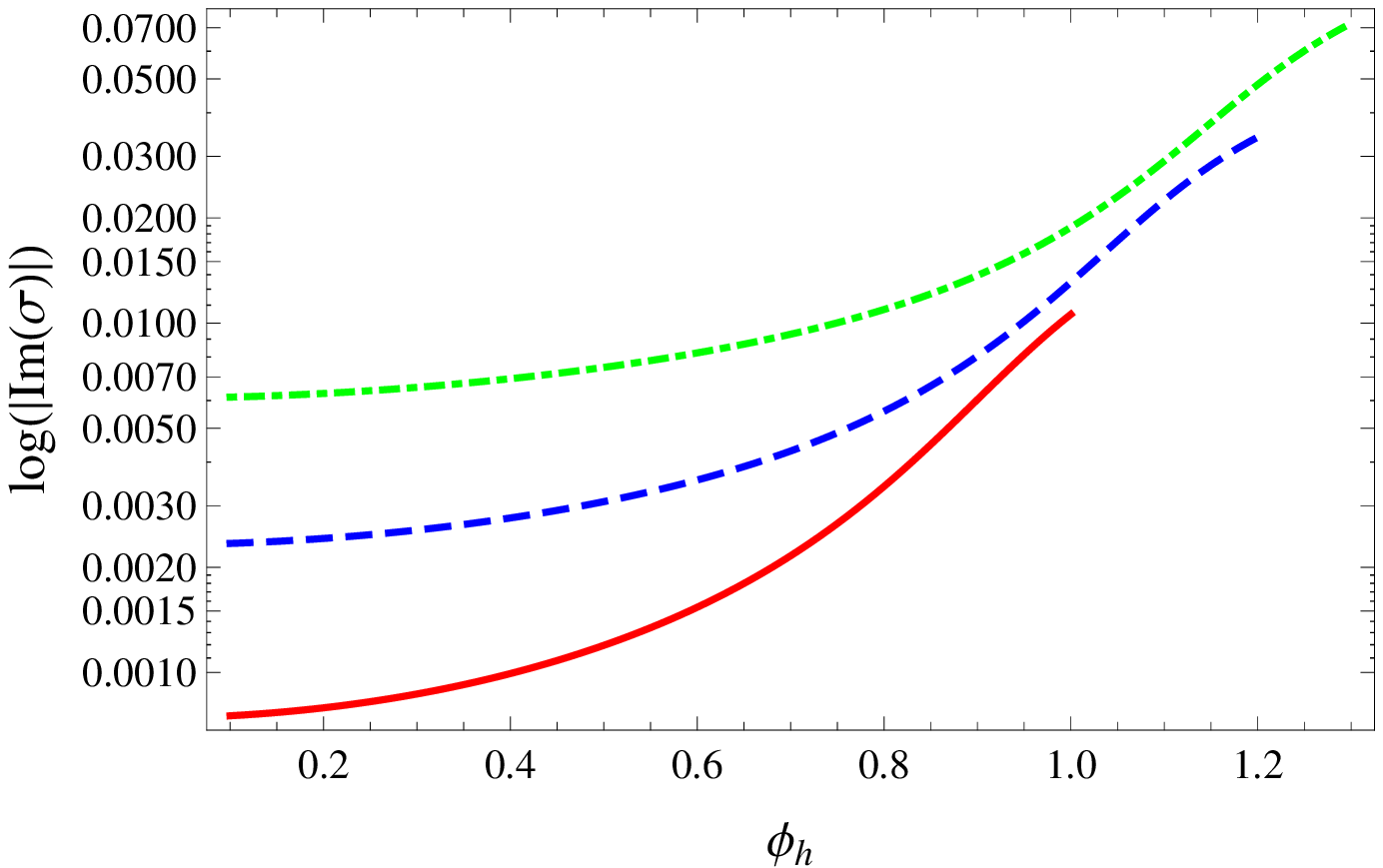} \\
\end{tabular}
\caption{The real (left) and imaginary (right) part of the mode frequency $\sigma$ is plotted as a function of (top row) the mirror radius $r_m$ and (bottom row) the equilibrium scalar field on the horizon $\phi_h$, for $q=0.2$, $\phi_h \in (0.1,1.3)$ and various values of $E_h$. The mirror is located at the first zero of the equilibrium scalar field.  In the right-hand plots, we plot the logarithm of the modulus of $\text{Im}(\sigma )$; note however that all values of $\text{Im}(\sigma )$ shown are negative.}
\label{fig:q2fixedEhvaryPhih}
\end{figure*}

\begin{figure*}
\begin{tabular}{c}
\includegraphics[width=0.95\columnwidth]{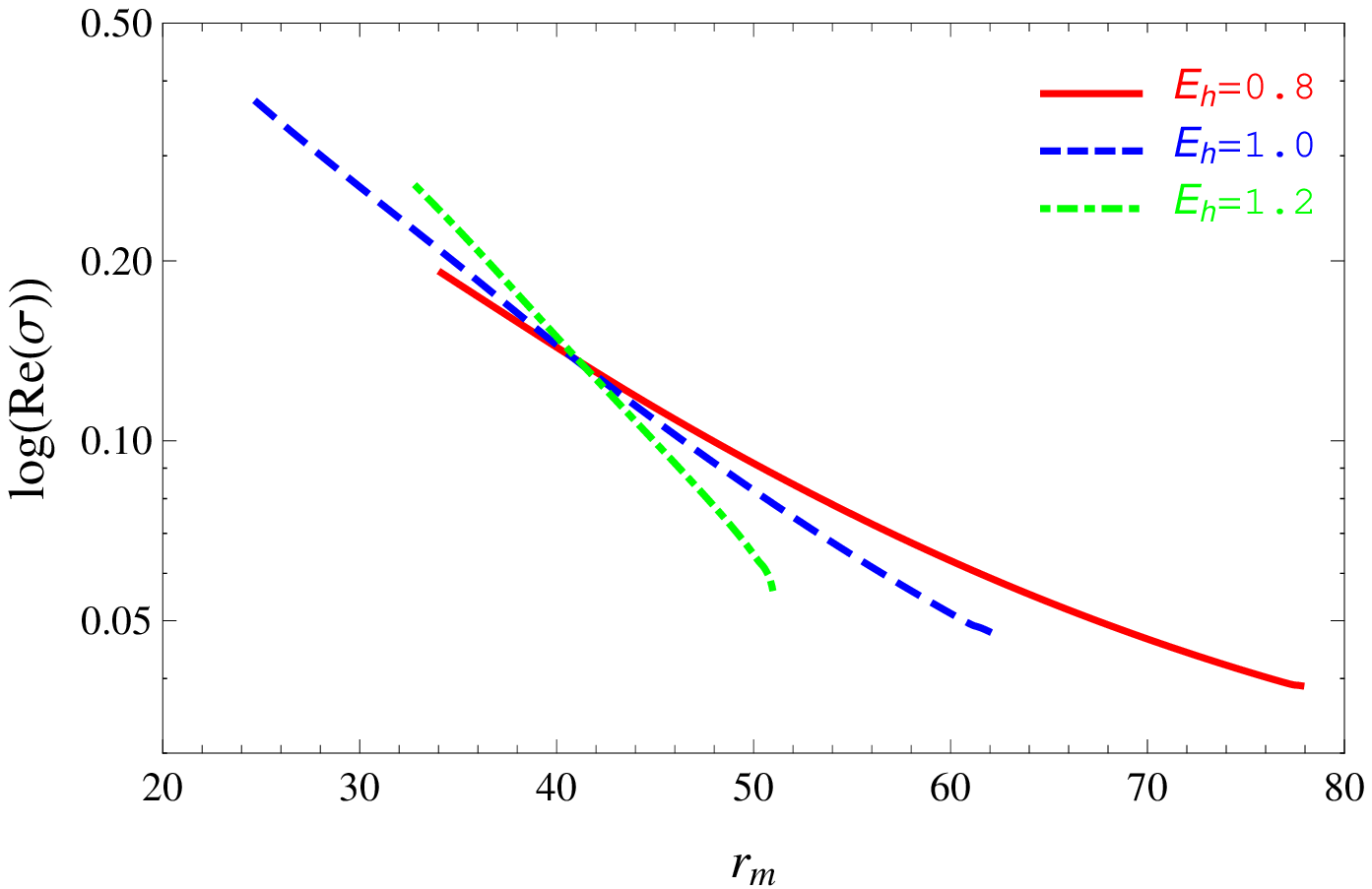}
\includegraphics[width=0.95\columnwidth]{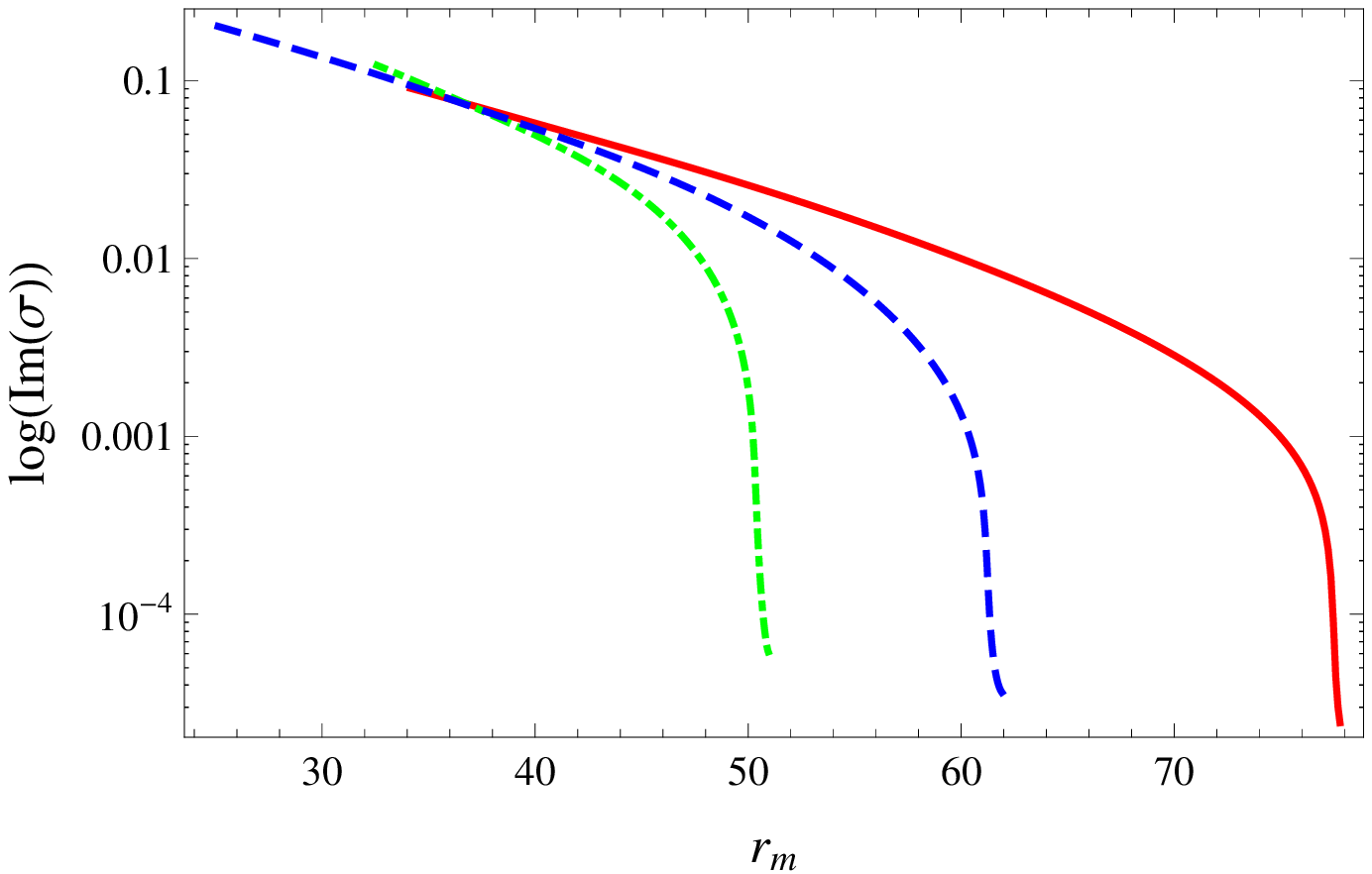} \\
\includegraphics[width=0.95\columnwidth]{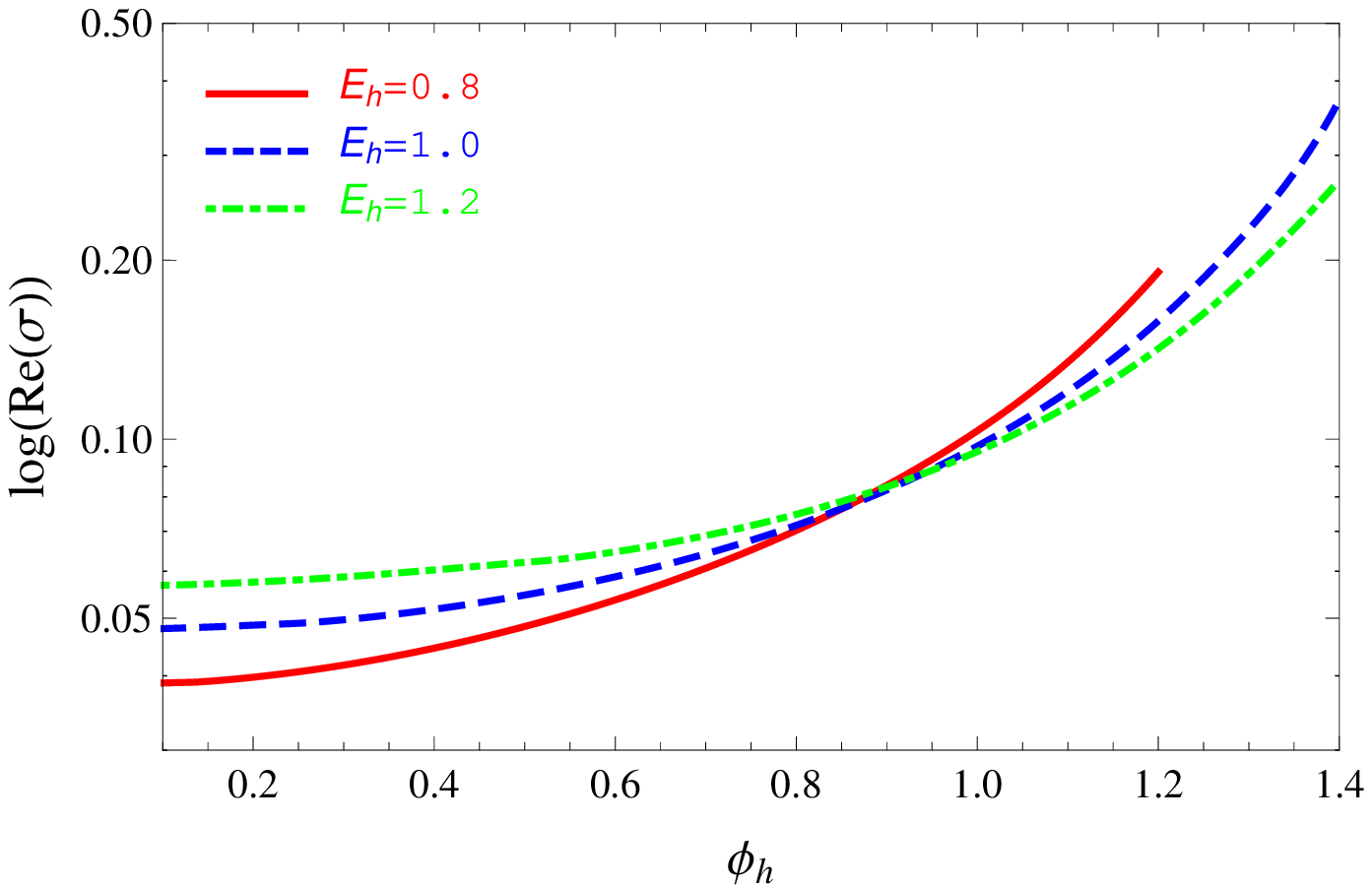}
\includegraphics[width=0.95\columnwidth]{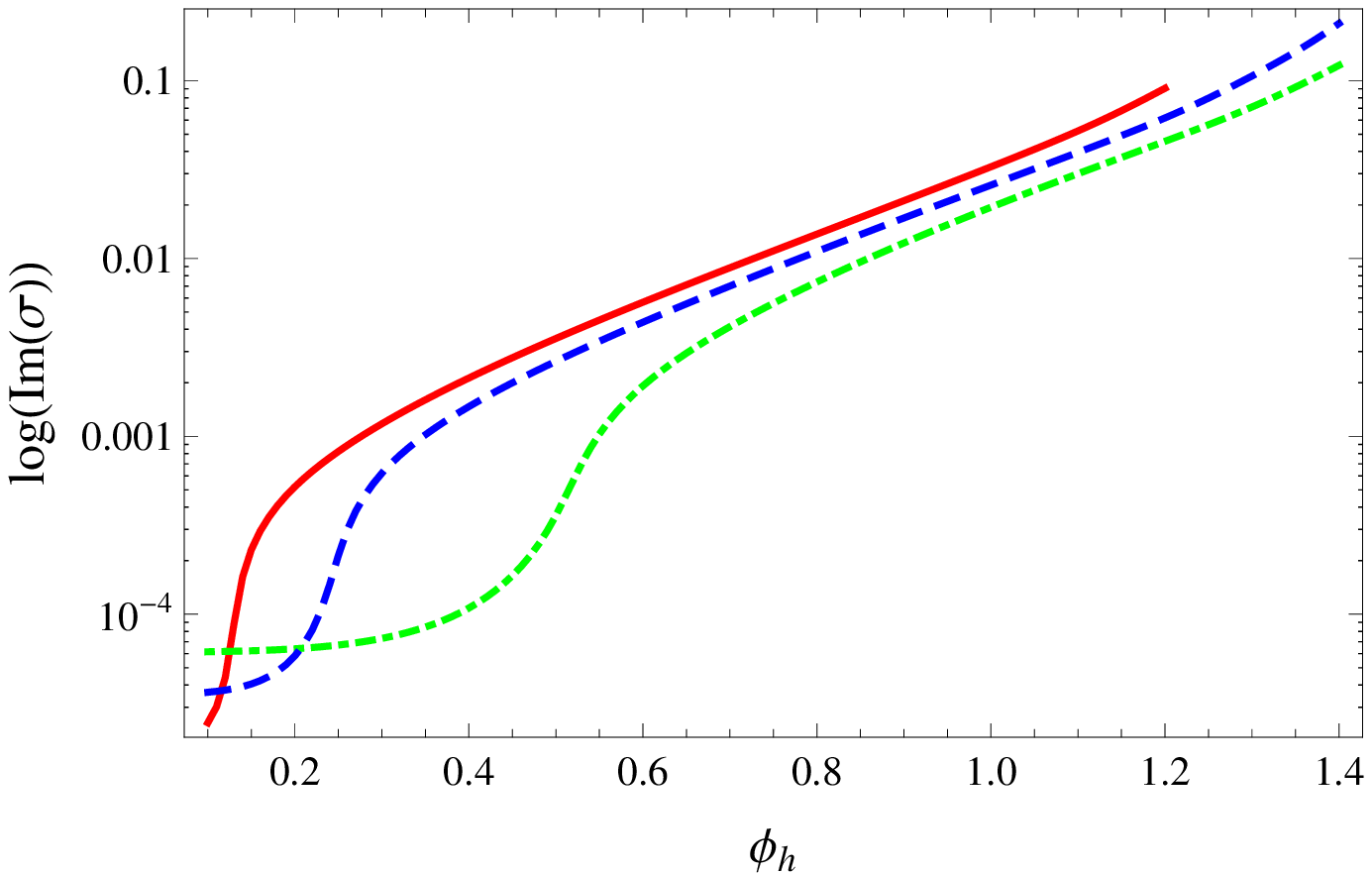}\\
\end{tabular}
\caption{Instability of perturbations when the mirror is placed at the \emph{second} zero of the static scalar field.   The real (left) and imaginary (right) part of the mode frequency $\sigma$ is plotted as a function of (top row) the mirror radius $r_m$ and (bottom row) the equilibrium scalar field on the horizon $\phi_h$, for $q=0.1$, $\phi_h \in (0.1, 1.4)$ and various values of $E_h$. Note that in these plots, unlike the plots for the first-zero case, the imaginary part of the frequency is positive, implying that the perturbations are exponentially growing in time.}
\label{fig:q1fixedEhvaryPhih2ndnode}
\end{figure*}

As an example, we plot in Fig.~\ref{fig:pert_functions} the behaviour of $\tilde{u}$, $\tilde{w}$ and $\tilde{A_0}$ for scalar charge $q=0.1$, with horizon values for the electric field $E_h=0.8$ and scalar field $\phi_h=1.2$. The values of the two shooting parameters are $\sigma=0.1731-0.0038i$ and ${\tilde {u}}_0 = 0.4397+0.0231i$. This figure clearly demonstrates that both the real and imaginary parts of the field variables $\tilde{u}$ and $\tilde{w}$ vanish at the location of the mirror. By contrast, the perturbation of the electric potential ${\tilde {A}}_{0}$ does not vanish on the mirror. This particular perturbation mode decays exponentially with time since the frequency $\sigma $ satisfies $\text{Im}(\sigma )<0$.

The key question we explore in this section is whether the perturbation mode shown in Fig.~\ref{fig:pert_functions} is typical in having $\text{Im}(\sigma )<0$.  As illustrated by Fig.~\ref{fig:solution-space}, we have a three-dimensional parameter space of equilibrium solutions, governed by the parameters $q$ (the scalar-field charge), $\phi _{h}$ (the value of the scalar field on the horizon) and $E_{h}$ (the electric field on the horizon).  It is clearly impractical to test every possible solution in this phase space, and therefore in this section we present a selection of results probing various parts of the phase space.

With the mirror at the first zero of the equilibrium scalar field, for fixed values of the parameters $q$, $\phi _{h}$ and $E_{h}$, we search for perturbations solving the equations (\ref{perteqnsfinal}), subject to the boundary conditions (\ref{Pt-BCHorizon}, \ref{Pt-BCmirror}).
For each fixed equilibrium solution we found a single value of $\sigma$ (together with one value of the other shooting parameter ${\tilde {u}}_{0}$)
such that the corresponding solution of the perturbation equations satisfies the boundary conditions.
In \emph{all} cases examined with the mirror at the first node, we found that $\text{Im}(\sigma )<0$, implying that the perturbations exponentially decay in time, suggesting the equilibrium solutions are stable.

We now present a selection of numerical results. In Figs.~\ref{fig:q1fixedEhvaryPhih}--\ref{fig:q2fixedEhvaryPhih} we fix two of the parameters $q$, $\phi _{h}$ and $E_{h}$ and vary the third.

Figure \ref{fig:q1fixedEhvaryPhih} shows the real (left) and imaginary (right) parts of the frequency $\sigma$ as a function of mirror radius $r_{m}$ (top row) and the value of the equilibrium scalar field $\phi _{h}$ (bottom row) for fixed $q=0.1$. In each plot, three different curves represent three distinct values of $E_h$, and on each curve $\phi_h$ varies from $0.1$ to $1.4$. Our numerical method breaks down when $\phi _{h}$ is very small and the mirror location is far from the black hole event horizon. When $\phi _{h}$ is greater than $\sim 1.4$, no static black hole solutions exist with nontrivial scalar field hair (see Fig.~\ref{fig:solution-space}); in the case where $E_h=0.8$, static hairy black holes cannot be found for $\phi_h$ larger than $\sim 1.2$. As can be clearly seen in the $\text{Im}(\sigma)$ plots, all black hole solutions examined appear to be linearly stable against spherically symmetric perturbations.

Figure \ref{fig:q1fixedEhvaryPhih} shows how $\text{Im}(\sigma )$ decreases (so that the perturbations decay more rapidly) and $\text{Re}(\sigma)$ increases as $\phi _{h}$ is increased, and the mirror moves closer to the black hole horizon. For the values of $E_{h}$ shown in this figure, $\text{Im}(\sigma )$ also decreases as $E_{h}$ increases for fixed $\phi _{h}$.

Figure \ref{fig:q1fixedPhihvaryEh} illustrates the real (left) and imaginary (right) part of the frequency $\sigma$ against the electric field at the horizon $E_h$ for three distinct values of $\phi_h$. It remains the case that $\text{Im}(\sigma )<0$; as it can take a small value, we plot the logarithm of the modulus of $\text{Im}(\sigma )$. As the electric field on the horizon, $E_{h}$, increases for fixed $\phi _{h}$, then $\text{Re}(\sigma )$ increases and $\text{Im}(\sigma )$ decreases. Furthermore, as $\phi_h$ increases, $\text{Re}(\sigma )$ increases and $\text{Im}(\sigma )$ decreases.

Let us now explore the effect of changing the scalar-field charge $q$.
We may fix the values of $\phi_h$ and $E_h$, and vary $q$. Figure \ref{fig:fixedPhihEhvaryq} shows $\text{Re}(\sigma)$ (left) and $\text{Im}(\sigma)$ (right) as functions of the charge of the scalar field $q$. Once again, we find only stable modes with $\text{Im}(\sigma )<0$. As the scalar-field charge $q$ increases, we see that $\text{Re}(\sigma )$ increases and $\text{Im}(\sigma )$ decreases.

In Fig.~\ref{fig:fixedPhihqvaryEh}, four different scalar-field charges are chosen, and the value of the equilibrium scalar field on the horizon is fixed to be $\phi_h=0.2$, with $E_{h}$ varying. Notice that as $q$ increases the real part of the frequency $\sigma $ increases and the imaginary part decreases (we plot the magnitude of $\text{Im}(\sigma )$ on a logarithmic scale as it is small). Moreover, from Fig.~\ref{fig:fixedPhihqvaryEh}, we observe that as $q$ decreases, the curves for $\text{Re}(\sigma)$ and $\text{Im}(\sigma)$ cover a greater range of $E_h$ values. This is due to the fact that as $q$ becomes smaller the two-dimensional phase space of static solutions expands (see Fig.~\ref{fig:solution-space}).

Finally, in Fig.~\ref{fig:q2fixedEhvaryPhih} we display the real (left) and imaginary (right) part of the frequency $\sigma$ as a function of mirror radius $r_{m}$ (top row) and the value of the equilibrium scalar field on the horizon $\phi _{h}$ (bottom row) for fixed scalar charge $q=0.2$. This should be compared with the corresponding plot in Fig.~\ref{fig:q1fixedEhvaryPhih} for $q=0.1$. We see in Fig.~\ref{fig:q2fixedEhvaryPhih} that as $E_h$ increases the range of values of $\phi_h$ also increases. This can be understood from the phase space plot of the static solutions (upper-right-hand plot in  Fig.~\ref{fig:solution-space}). Again, $\text{Im}(\sigma )$ remains negative. As in previous figures, with fixed $E_{h}$, increasing $\phi_{h}$ increases $\text{Re}(\sigma )$ and decreases $\text{Im}(\sigma )$, while increasing $E_{h}$ for fixed $\phi _{h}$ decreases $\text{Im}(\sigma )$.

In all the figures considered so far in this section, the mirror was located at the first zero of the equilibrium scalar field.  We have found a consistent picture: for each static hairy black hole, we can only find perturbations which decay exponentially in time. We conclude that the static hairy black holes with the mirror at the first zero of the equilibrium scalar field appear to be stable.

We close this section by considering some results when the mirror is located at the \emph{second} zero of the equilibrium scalar field, providing an example plot in Fig.~\ref{fig:q1fixedEhvaryPhih2ndnode}. The real (left) and imaginary (right) part of the frequency $\sigma$ are plotted against mirror radius $r_{m}$ (top row) and the value of the scalar field on the horizon $\phi _{h}$ (bottom row) for fixed $q=0.1$.
In contrast to the first-zero case, we now find perturbations with $\text{Im}(\sigma )>0$ for all the static black hole solutions considered in Fig.~\ref{fig:q1fixedEhvaryPhih2ndnode}, so that the perturbations grow exponentially in time. We conclude that static hairy  black holes with the mirror at the \emph{second} zero of the equilibrium scalar field are unstable. We conjecture that if the mirror was located at a node after the second zero of the equilibrium scalar field, then the hairy black holes would remain unstable.

\section{Conclusions}
\label{sec:conclusions}

When a charged scalar field interacts with an electrically charged Reissner-Nordstr\"om black hole surrounded by a reflecting mirror,
superradiantly-unstable modes exist \cite{Herdeiro:2013pia,Degollado:2013bha}.
A natural question arises: what is the end-point of this instability? This question has been the focus of our work in this paper.

Working in the frequency domain, we have confirmed the time-domain results of \cite{Degollado:2013bha}, namely that the charged scalar field, linearized around the $\Phi = 0$ background, has spherically symmetric unstable modes, if the mirror is sufficiently far from the black hole horizon.
This led us to consider non-linear spherically symmetric solutions of the fully coupled Einstein-charged scalar field theory as possible end-points of this superradiant instability.
In the `scalar electrodynamics' model, a charged complex scalar field is coupled to an electromagnetic field and the usual Einstein-Hilbert gravitational Lagrangian.
Solving the equilibrium field equations, we found black hole solutions with a nontrivial scalar field which oscillates about zero. We may place a reflecting mirror at any one of the nodes of the scalar field, to obtain a black hole in a cavity. It is important to note that the solutions we find do not contradict the no-hair theorem of Bekenstein \cite{Bekenstein:1971hc} which applies in the absence of a mirror-like boundary.

To investigate whether these hairy black holes could be possible end-points of the superradiant instability, we have considered spherically symmetric perturbations of the hairy black hole solutions. These perturbations satisfy ingoing boundary conditions on the horizon.
Furthermore, the perturbations of the charged scalar field vanish on the mirror.
The resulting perturbation equations, though linear, are highly coupled and can (we believe) only be solved numerically.

With the mirror placed at the first zero of the equilibrium scalar field, we find no evidence of any instabilities: all such equilibrium solutions appear to be stable under small perturbations.
On the other hand, if the mirror is placed at the second zero of the equilibrium scalar field, we find unstable perturbations which grow with time. We conjecture that the same result would hold if the mirror were situated at the third or subsequent zeros of the equilibrium scalar field.

Let us now consider the implications of these results in a wider context.
It is known that a superradiant instability can also arise when a charged black hole is embedded in a asymptotically Anti-de Sitter (AdS) spacetime \cite{Cardoso:2004nk,Cardoso:2004hs}. In such a scenario, the timelike boundary of spacetime itself acts as the mirror, providing a natural reflecting boundary condition (or `Dirichlet wall'). In Refs.~\cite{Bhattacharyya:2010yg, Dias:2011tj} asymptotically AdS `hairy' black holes were constructed within the context of supergravity/higher-dimensional theories. Two influential ideas underpinning this work can be traced through Refs.~\cite{Cardoso:2004nk,Cardoso:2004hs, Bhattacharyya:2010yg, Dias:2011tj, Dias:2011at}: (i) `hairy' stationary solutions are plausible endpoints of the superradiant instability; and (ii) in the limit of small field amplitude, the hairy solutions connect to (perturbed) vacuum solutions endowed with linear perturbations at the critical superradiant frequency $\sigma_c$. It is plausible that the stability properties of the four-dimensional hairy black holes, considered here, may be shared by their higher dimensional, non-asymptotically flat cousins. However, this remains to be investigated.

Our view is that the `scalar electrodynamics' model may also provide insight into the development of superradiant instabilities in astrophysical systems. However, we should proceed with an element of caution, for two reasons. First, in the Kerr black hole case, superradiance is promoted by angular momentum, rather than charge. Thus, the Kerr instability does \emph{not} appear in the spherically-symmetric sector, making any stability analysis considerably more involved. Second, in the Kerr case, instabilities can arise spontaneously in bound states of an (ultra-light) massive bosonic field. By contrast, in the charged case, the competition between gravitational attraction and electrostatic repulsion means that bound states cannot form in the superradiant frequency regime; thus the artifice of a mirror is necessary. These factors suggest that one should be cautious when attempting to infer from analogy.

This work was motivated, in part, by the recent discovery of a Kerr-scalar family of (asymptotically-flat and four-dimensional) black hole solutions \cite{Herdeiro:2014goa, Herdeiro:2015gia}. The discovery inspired Herdeiro and Radu \cite{Herdeiro:2014ima} to propose the following conjecture: `a (hairless) black hole which is afflicted by the superradiant instability of a given field must allow hairy generalizations with that field'. (As noted in Ref.~\cite{Herdeiro:2015gia}, the given field should also generate a stress-energy with appropriate Killing symmetries; this excludes, for example, a real scalar field). Our equilibrium charged-scalar black hole solutions (Sec.~\ref{sec:solutions}) are in accord with Herdeiro and Radu's conjecture.
One could also envisage a stronger conjecture: `a (hairless) black hole solution afflicted by a superradiant instability of a given field may naturally evolve towards a hairy black hole solution which is \emph{stable} under perturbations in that field'.
When the mirror is located at the first zero of the equilibrium scalar field, our charged-scalar black hole solutions are stable (at least to time-periodic, linear, spherically symmetric perturbations).
We have therefore conjectured that they are possible end-points of the superradiant instability discovered in \cite{Herdeiro:2013pia,Degollado:2013bha}, in accord with our stronger version of the Herdeiro/Radu conjecture. We should mention that the stronger conjecture is somewhat provocative, as it is apparently in tension with numerical simulations in the Kerr case which suggest that nonlinear effects lead to collapse of the field, and a subsequent explosive phenomenon, known as a `bosenova' \cite{Yoshino:2012kn} (for other perspectives see Refs.~\cite{Yoshino:2013ofa, Brito:2014wla, Okawa:2014nda, Sanchis-Gual:2014ewa, Zilhao:2015tya}).

In order to ascertain whether these hairy charged-scalar black holes are indeed (i) stable under more generic perturbations, and (ii) natural end-points of the vacuum superradiant instability, a full nonlinear time-domain numerical simulation would be required. One could start with a small perturbation of a Reissner-Nordstr\"om black hole in a cavity, and track the development of the instability into the nonlinear regime. Such a nonlinear simulation would be a technical achievement, as (i) the horizon would be dynamical, and (ii) the growth rate of the superradiantly unstable modes of the Reissner-Nordstr\"om black hole in a cavity is about two orders of magnitude smaller than their frequency. Here, the methods of numerical relativity may find a further application \cite{Witek:2010qc}. Such simulations could be greatly simplified by restricting to spherical symmetry, which is not possible in the Kerr context. Alternatively, as a starting point, it would be instructive to perform a time-domain analysis of \emph{linear}, spherically symmetric perturbations of the equilibrium charged-scalar hairy black holes presented in this paper, in order to confirm the frequency-domain stability results presented here.
Either approach would surely lead towards a fuller understanding of the generic features of superradiant instabilities, and their relevance (or otherwise) in astrophysics. \\

\begin{acknowledgments}
The work of SRD and EW is supported by the Lancaster-Manchester-Sheffield Consortium for Fundamental Physics under STFC grant ST/L000520/1. The work of SRD is also supported by EPSRC grant EP/M025802/1. The work of SP is supported by the 90$^{th}$ Anniversary of Chulalongkorn University Fund (Ratchadaphiseksomphot Endowment Fund). We thank the anonymous referee for helpful feedback.
\end{acknowledgments}

\bibliographystyle{apsrev4-1}

\bibliography{refs}

\begin{thebibliography}{59}%
\makeatletter
\providecommand \@ifxundefined [1]{%
 \@ifx{#1\undefined}
}%
\providecommand \@ifnum [1]{%
 \ifnum #1\expandafter \@firstoftwo
 \else \expandafter \@secondoftwo
 \fi
}%
\providecommand \@ifx [1]{%
 \ifx #1\expandafter \@firstoftwo
 \else \expandafter \@secondoftwo
 \fi
}%
\providecommand \natexlab [1]{#1}%
\providecommand \enquote  [1]{``#1''}%
\providecommand \bibnamefont  [1]{#1}%
\providecommand \bibfnamefont [1]{#1}%
\providecommand \citenamefont [1]{#1}%
\providecommand \href@noop [0]{\@secondoftwo}%
\providecommand \href [0]{\begingroup \@sanitize@url \@href}%
\providecommand \@href[1]{\@@startlink{#1}\@@href}%
\providecommand \@@href[1]{\endgroup#1\@@endlink}%
\providecommand \@sanitize@url [0]{\catcode `\\12\catcode `\$12\catcode
  `\&12\catcode `\#12\catcode `\^12\catcode `\_12\catcode `\%12\relax}%
\providecommand \@@startlink[1]{}%
\providecommand \@@endlink[0]{}%
\providecommand \url  [0]{\begingroup\@sanitize@url \@url }%
\providecommand \@url [1]{\endgroup\@href {#1}{\urlprefix }}%
\providecommand \urlprefix  [0]{URL }%
\providecommand \Eprint [0]{\href }%
\providecommand \doibase [0]{http://dx.doi.org/}%
\providecommand \selectlanguage [0]{\@gobble}%
\providecommand \bibinfo  [0]{\@secondoftwo}%
\providecommand \bibfield  [0]{\@secondoftwo}%
\providecommand \translation [1]{[#1]}%
\providecommand \BibitemOpen [0]{}%
\providecommand \bibitemStop [0]{}%
\providecommand \bibitemNoStop [0]{.\EOS\space}%
\providecommand \EOS [0]{\spacefactor3000\relax}%
\providecommand \BibitemShut  [1]{\csname bibitem#1\endcsname}%
\let\auto@bib@innerbib\@empty
%</preamble>
\bibitem [{\citenamefont {King}(2003)}]{King:2003ix}%
  \BibitemOpen
  \bibfield  {author} {\bibinfo {author} {\bibfnamefont {A.}~\bibnamefont
  {King}},\ }\href {\doibase 10.1086/379143} {\bibfield  {journal} {\bibinfo
  {journal} {Astrophys.~J.}\ }\textbf {\bibinfo {volume} {596}},\ \bibinfo
  {pages} {L27} (\bibinfo {year} {2003})},\ \Eprint
  {http://arxiv.org/abs/astro-ph/0308342} {astro-ph/0308342} \BibitemShut
  {NoStop}%
%%CITATION = ASTRO-PH/0308342;%%
\bibitem [{\citenamefont {Risaliti}\ \emph {et~al.}(2013)\citenamefont
  {Risaliti}, \citenamefont {Harrison}, \citenamefont {Madsen}, \citenamefont
  {Walton}, \citenamefont {Boggs} \emph {et~al.}}]{Risaliti:2013cga}%
  \BibitemOpen
  \bibfield  {author} {\bibinfo {author} {\bibfnamefont {G.}~\bibnamefont
  {Risaliti}}, \bibinfo {author} {\bibfnamefont {F.~A.}\ \bibnamefont
  {Harrison}}, \bibinfo {author} {\bibfnamefont {K.~K.}\ \bibnamefont
  {Madsen}}, \bibinfo {author} {\bibfnamefont {D.~J.}\ \bibnamefont {Walton}},
  \bibinfo {author} {\bibfnamefont {S.~E.}\ \bibnamefont {Boggs}},  \emph
  {et~al.},\ }\href {\doibase 10.1038/nature11938} {\bibfield  {journal}
  {\bibinfo  {journal} {Nature}\ }\textbf {\bibinfo {volume} {494}},\ \bibinfo
  {pages} {449} (\bibinfo {year} {2013})},\ \Eprint
  {http://arxiv.org/abs/1302.7002} {arXiv:1302.7002 [astro-ph.HE]} \BibitemShut
  {NoStop}%
%%CITATION = ARXIV:1302.7002;%%
\bibitem [{\citenamefont {Reynolds}(2013)}]{Reynolds:2013rva}%
  \BibitemOpen
  \bibfield  {author} {\bibinfo {author} {\bibfnamefont {C.~S.}\ \bibnamefont
  {Reynolds}},\ }\href {\doibase 10.1088/0264-9381/30/24/244004} {\bibfield
  {journal} {\bibinfo  {journal} {Class.~Quant.~Grav.}\ }\textbf {\bibinfo
  {volume} {30}},\ \bibinfo {pages} {244004} (\bibinfo {year} {2013})},\
  \Eprint {http://arxiv.org/abs/1307.3246} {arXiv:1307.3246 [astro-ph.HE]}
  \BibitemShut {NoStop}%
%%CITATION = ARXIV:1307.3246;%%
\bibitem [{\citenamefont {Blandford}\ and\ \citenamefont
  {Znajek}(1977)}]{Blandford:1977ds}%
  \BibitemOpen
  \bibfield  {author} {\bibinfo {author} {\bibfnamefont {R.~D.}\ \bibnamefont
  {Blandford}}\ and\ \bibinfo {author} {\bibfnamefont {R.~L.}\ \bibnamefont
  {Znajek}},\ }\href@noop {} {\bibfield  {journal} {\bibinfo  {journal}
  {Mon.~Not.~Roy.~Astron.~Soc.}\ }\textbf {\bibinfo {volume} {179}},\ \bibinfo
  {pages} {433} (\bibinfo {year} {1977})}\BibitemShut {NoStop}%
%%CITATION = MNRAA,179,433;%%
\bibitem [{\citenamefont {Gralla}\ and\ \citenamefont
  {Jacobson}(2014)}]{Gralla:2014yja}%
  \BibitemOpen
  \bibfield  {author} {\bibinfo {author} {\bibfnamefont {S.~E.}\ \bibnamefont
  {Gralla}}\ and\ \bibinfo {author} {\bibfnamefont {T.}~\bibnamefont
  {Jacobson}},\ }\href {\doibase 10.1093/mnras/stu1690} {\bibfield  {journal}
  {\bibinfo  {journal} {Mon.~Not.~Roy.~Astron.~Soc.}\ }\textbf {\bibinfo
  {volume} {445}},\ \bibinfo {pages} {2500} (\bibinfo {year} {2014})},\ \Eprint
  {http://arxiv.org/abs/1401.6159} {arXiv:1401.6159 [astro-ph.HE]} \BibitemShut
  {NoStop}%
%%CITATION = ARXIV:1401.6159;%%
\bibitem [{\citenamefont {Penrose}\ and\ \citenamefont
  {Floyd}(1971)}]{Penrose:1971uk}%
  \BibitemOpen
  \bibfield  {author} {\bibinfo {author} {\bibfnamefont {R.}~\bibnamefont
  {Penrose}}\ and\ \bibinfo {author} {\bibfnamefont {R.~M.}\ \bibnamefont
  {Floyd}},\ }\href@noop {} {\bibfield  {journal} {\bibinfo  {journal}
  {Nature}\ }\textbf {\bibinfo {volume} {229}},\ \bibinfo {pages} {177}
  (\bibinfo {year} {1971})}\BibitemShut {NoStop}%
%%CITATION = NATUA,229,177;%%
\bibitem [{\citenamefont {Bardeen}\ \emph {et~al.}(1973)\citenamefont
  {Bardeen}, \citenamefont {Carter},\ and\ \citenamefont
  {Hawking}}]{Bardeen:1973gs}%
  \BibitemOpen
  \bibfield  {author} {\bibinfo {author} {\bibfnamefont {J.~M.}\ \bibnamefont
  {Bardeen}}, \bibinfo {author} {\bibfnamefont {B.}~\bibnamefont {Carter}}, \
  and\ \bibinfo {author} {\bibfnamefont {S.~W.}\ \bibnamefont {Hawking}},\
  }\href {\doibase 10.1007/BF01645742} {\bibfield  {journal} {\bibinfo
  {journal} {Commun.~Math.~Phys.}\ }\textbf {\bibinfo {volume} {31}},\ \bibinfo
  {pages} {161} (\bibinfo {year} {1973})}\BibitemShut {NoStop}%
%%CITATION = CMPHA,31,161;%%
\bibitem [{\citenamefont {Hawking}(1975)}]{Hawking:1974sw}%
  \BibitemOpen
  \bibfield  {author} {\bibinfo {author} {\bibfnamefont {S.~W.}\ \bibnamefont
  {Hawking}},\ }\href {\doibase 10.1007/BF02345020} {\bibfield  {journal}
  {\bibinfo  {journal} {Commun.~Math.~Phys.}\ }\textbf {\bibinfo {volume}
  {43}},\ \bibinfo {pages} {199} (\bibinfo {year} {1975})}\BibitemShut
  {NoStop}%
%%CITATION = CMPHA,43,199;%%
\bibitem [{\citenamefont {Bekenstein}(1974)}]{Bekenstein:1974ax}%
  \BibitemOpen
  \bibfield  {author} {\bibinfo {author} {\bibfnamefont {J.~D.}\ \bibnamefont
  {Bekenstein}},\ }\href {\doibase 10.1103/PhysRevD.9.3292} {\bibfield
  {journal} {\bibinfo  {journal} {Phys.~Rev.}\ }\textbf {\bibinfo {volume} {D
  9}},\ \bibinfo {pages} {3292} (\bibinfo {year} {1974})}\BibitemShut {NoStop}%
%%CITATION = PHRVA,D9,3292;%%
\bibitem [{\citenamefont {Starobinskii}\ and\ \citenamefont
  {Churilov}(1973)}]{Starobinskii:1973}%
  \BibitemOpen
  \bibfield  {author} {\bibinfo {author} {\bibfnamefont {A.~A.}\ \bibnamefont
  {Starobinskii}}\ and\ \bibinfo {author} {\bibfnamefont {S.~M.}\ \bibnamefont
  {Churilov}},\ }\href@noop {} {\bibfield  {journal} {\bibinfo  {journal}
  {Zh.~Eksp.~Teor.~Fiz}\ }\textbf {\bibinfo {volume} {65}},\ \bibinfo {pages}
  {3} (\bibinfo {year} {1973})}\BibitemShut {NoStop}%
\bibitem [{\citenamefont {Brito}\ \emph
  {et~al.}(2015{\natexlab{a}})\citenamefont {Brito}, \citenamefont {Cardoso},\
  and\ \citenamefont {Pani}}]{Brito:2015oca}%
  \BibitemOpen
  \bibfield  {author} {\bibinfo {author} {\bibfnamefont {R.}~\bibnamefont
  {Brito}}, \bibinfo {author} {\bibfnamefont {V.}~\bibnamefont {Cardoso}}, \
  and\ \bibinfo {author} {\bibfnamefont {P.}~\bibnamefont {Pani}},\ }\href
  {\doibase 10.1007/978-3-319-19000-6} {\emph {\bibinfo {title}
  {{Superradiance}}}},\ Vol.\ \bibinfo {volume} {906}\ (\bibinfo  {publisher}
  {Springer},\ \bibinfo {year} {2015})\ \Eprint
  {http://arxiv.org/abs/1501.06570} {arXiv:1501.06570 [gr-qc]} \BibitemShut
  {NoStop}%
%%CITATION = ARXIV:1501.06570;%%
\bibitem [{\citenamefont {Press}\ and\ \citenamefont
  {Teukolsky}(1972)}]{Press:1972zz}%
  \BibitemOpen
  \bibfield  {author} {\bibinfo {author} {\bibfnamefont {W.~H.}\ \bibnamefont
  {Press}}\ and\ \bibinfo {author} {\bibfnamefont {S.~A.}\ \bibnamefont
  {Teukolsky}},\ }\href {\doibase 10.1038/238211a0} {\bibfield  {journal}
  {\bibinfo  {journal} {Nature}\ }\textbf {\bibinfo {volume} {238}},\ \bibinfo
  {pages} {211} (\bibinfo {year} {1972})}\BibitemShut {NoStop}%
%%CITATION = NATUA,238,211;%%
\bibitem [{\citenamefont {Arvanitaki}\ and\ \citenamefont
  {Dubovsky}(2011)}]{Arvanitaki:2010sy}%
  \BibitemOpen
  \bibfield  {author} {\bibinfo {author} {\bibfnamefont {A.}~\bibnamefont
  {Arvanitaki}}\ and\ \bibinfo {author} {\bibfnamefont {S.}~\bibnamefont
  {Dubovsky}},\ }\href {\doibase 10.1103/PhysRevD.83.044026} {\bibfield
  {journal} {\bibinfo  {journal} {Phys.~Rev.}\ }\textbf {\bibinfo {volume} {D
  83}},\ \bibinfo {pages} {044026} (\bibinfo {year} {2011})},\ \Eprint
  {http://arxiv.org/abs/1004.3558} {arXiv:1004.3558 [hep-th]} \BibitemShut
  {NoStop}%
%%CITATION = ARXIV:1004.3558;%%
\bibitem [{\citenamefont {Pani}\ \emph {et~al.}(2012)\citenamefont {Pani},
  \citenamefont {Cardoso}, \citenamefont {Gualtieri}, \citenamefont {Berti},\
  and\ \citenamefont {Ishibashi}}]{Pani:2012vp}%
  \BibitemOpen
  \bibfield  {author} {\bibinfo {author} {\bibfnamefont {P.}~\bibnamefont
  {Pani}}, \bibinfo {author} {\bibfnamefont {V.}~\bibnamefont {Cardoso}},
  \bibinfo {author} {\bibfnamefont {L.}~\bibnamefont {Gualtieri}}, \bibinfo
  {author} {\bibfnamefont {E.}~\bibnamefont {Berti}}, \ and\ \bibinfo {author}
  {\bibfnamefont {A.}~\bibnamefont {Ishibashi}},\ }\href {\doibase
  10.1103/PhysRevLett.109.131102} {\bibfield  {journal} {\bibinfo  {journal}
  {Phys.~Rev.~Lett.}\ }\textbf {\bibinfo {volume} {109}},\ \bibinfo {pages}
  {131102} (\bibinfo {year} {2012})},\ \Eprint {http://arxiv.org/abs/1209.0465}
  {arXiv:1209.0465 [gr-qc]} \BibitemShut {NoStop}%
%%CITATION = ARXIV:1209.0465;%%
\bibitem [{\citenamefont {Damour}\ \emph {et~al.}(1976)\citenamefont {Damour},
  \citenamefont {Deruelle},\ and\ \citenamefont {Ruffini}}]{Damour:1976kh}%
  \BibitemOpen
  \bibfield  {author} {\bibinfo {author} {\bibfnamefont {T.}~\bibnamefont
  {Damour}}, \bibinfo {author} {\bibfnamefont {N.}~\bibnamefont {Deruelle}}, \
  and\ \bibinfo {author} {\bibfnamefont {R.}~\bibnamefont {Ruffini}},\ }\href
  {\doibase 10.1007/BF02725534} {\bibfield  {journal} {\bibinfo  {journal}
  {Lett.~Nuovo Cim.}\ }\textbf {\bibinfo {volume} {15}},\ \bibinfo {pages}
  {257} (\bibinfo {year} {1976})}\BibitemShut {NoStop}%
%%CITATION = NCLTA,15,257;%%
\bibitem [{\citenamefont {Zouros}\ and\ \citenamefont
  {Eardley}(1979)}]{Zouros:1979iw}%
  \BibitemOpen
  \bibfield  {author} {\bibinfo {author} {\bibfnamefont {T.~J.~M.}\
  \bibnamefont {Zouros}}\ and\ \bibinfo {author} {\bibfnamefont {D.~M.}\
  \bibnamefont {Eardley}},\ }\href {\doibase 10.1016/0003-4916(79)90237-9}
  {\bibfield  {journal} {\bibinfo  {journal} {Annals Phys.}\ }\textbf {\bibinfo
  {volume} {118}},\ \bibinfo {pages} {139} (\bibinfo {year}
  {1979})}\BibitemShut {NoStop}%
%%CITATION = APNYA,118,139;%%
\bibitem [{\citenamefont {Detweiler}(1980)}]{Detweiler:1980uk}%
  \BibitemOpen
  \bibfield  {author} {\bibinfo {author} {\bibfnamefont {S.~L.}\ \bibnamefont
  {Detweiler}},\ }\href {\doibase 10.1103/PhysRevD.22.2323} {\bibfield
  {journal} {\bibinfo  {journal} {Phys.~Rev.}\ }\textbf {\bibinfo {volume} {D
  22}},\ \bibinfo {pages} {2323} (\bibinfo {year} {1980})}\BibitemShut
  {NoStop}%
%%CITATION = PHRVA,D22,2323;%%
\bibitem [{\citenamefont {Cardoso}\ and\ \citenamefont
  {Yoshida}(2005)}]{Cardoso:2005vk}%
  \BibitemOpen
  \bibfield  {author} {\bibinfo {author} {\bibfnamefont {V.}~\bibnamefont
  {Cardoso}}\ and\ \bibinfo {author} {\bibfnamefont {S.}~\bibnamefont
  {Yoshida}},\ }\href {\doibase 10.1088/1126-6708/2005/07/009} {\bibfield
  {journal} {\bibinfo  {journal} {JHEP}\ }\textbf {\bibinfo {volume} {0507}},\
  \bibinfo {pages} {009} (\bibinfo {year} {2005})},\ \Eprint
  {http://arxiv.org/abs/hep-th/0502206} {hep-th/0502206} \BibitemShut {NoStop}%
%%CITATION = HEP-TH/0502206;%%
\bibitem [{\citenamefont {Furuhashi}\ and\ \citenamefont
  {Nambu}(2004)}]{Furuhashi:2004jk}%
  \BibitemOpen
  \bibfield  {author} {\bibinfo {author} {\bibfnamefont {H.}~\bibnamefont
  {Furuhashi}}\ and\ \bibinfo {author} {\bibfnamefont {Y.}~\bibnamefont
  {Nambu}},\ }\href {\doibase 10.1143/PTP.112.983} {\bibfield  {journal}
  {\bibinfo  {journal} {Prog.~Theor.~Phys.}\ }\textbf {\bibinfo {volume}
  {112}},\ \bibinfo {pages} {983} (\bibinfo {year} {2004})},\ \Eprint
  {http://arxiv.org/abs/gr-qc/0402037} {gr-qc/0402037} \BibitemShut {NoStop}%
%%CITATION = GR-QC/0402037;%%
\bibitem [{\citenamefont {Dolan}(2007)}]{Dolan:2007mj}%
  \BibitemOpen
  \bibfield  {author} {\bibinfo {author} {\bibfnamefont {S.~R.}\ \bibnamefont
  {Dolan}},\ }\href {\doibase 10.1103/PhysRevD.76.084001} {\bibfield  {journal}
  {\bibinfo  {journal} {Phys.~Rev.}\ }\textbf {\bibinfo {volume} {D 76}},\
  \bibinfo {pages} {084001} (\bibinfo {year} {2007})},\ \Eprint
  {http://arxiv.org/abs/0705.2880} {arXiv:0705.2880 [gr-qc]} \BibitemShut
  {NoStop}%
%%CITATION = ARXIV:0705.2880;%%
\bibitem [{\citenamefont {Rosa}(2010)}]{Rosa:2009ei}%
  \BibitemOpen
  \bibfield  {author} {\bibinfo {author} {\bibfnamefont {J.~G.}\ \bibnamefont
  {Rosa}},\ }\href {\doibase 10.1007/JHEP06(2010)015} {\bibfield  {journal}
  {\bibinfo  {journal} {JHEP}\ }\textbf {\bibinfo {volume} {1006}},\ \bibinfo
  {pages} {015} (\bibinfo {year} {2010})},\ \Eprint
  {http://arxiv.org/abs/0912.1780} {arXiv:0912.1780 [hep-th]} \BibitemShut
  {NoStop}%
%%CITATION = ARXIV:0912.1780;%%
\bibitem [{\citenamefont {Kodama}\ and\ \citenamefont
  {Yoshino}(2012)}]{Kodama:2011zc}%
  \BibitemOpen
  \bibfield  {author} {\bibinfo {author} {\bibfnamefont {H.}~\bibnamefont
  {Kodama}}\ and\ \bibinfo {author} {\bibfnamefont {H.}~\bibnamefont
  {Yoshino}},\ }\href {\doibase 10.1142/S2010194512004199} {\bibfield
  {journal} {\bibinfo  {journal} {Int.~J.~Mod.~Phys.~Conf.~Ser.}\ }\textbf
  {\bibinfo {volume} {7}},\ \bibinfo {pages} {84} (\bibinfo {year} {2012})},\
  \Eprint {http://arxiv.org/abs/1108.1365} {arXiv:1108.1365 [hep-th]}
  \BibitemShut {NoStop}%
%%CITATION = ARXIV:1108.1365;%%
\bibitem [{\citenamefont {Witek}\ \emph {et~al.}(2013)\citenamefont {Witek},
  \citenamefont {Cardoso}, \citenamefont {Ishibashi},\ and\ \citenamefont
  {Sperhake}}]{Witek:2012tr}%
  \BibitemOpen
  \bibfield  {author} {\bibinfo {author} {\bibfnamefont {H.}~\bibnamefont
  {Witek}}, \bibinfo {author} {\bibfnamefont {V.}~\bibnamefont {Cardoso}},
  \bibinfo {author} {\bibfnamefont {A.}~\bibnamefont {Ishibashi}}, \ and\
  \bibinfo {author} {\bibfnamefont {U.}~\bibnamefont {Sperhake}},\ }\href
  {\doibase 10.1103/PhysRevD.87.043513} {\bibfield  {journal} {\bibinfo
  {journal} {Phys.~Rev.}\ }\textbf {\bibinfo {volume} {D 87}},\ \bibinfo
  {pages} {043513} (\bibinfo {year} {2013})},\ \Eprint
  {http://arxiv.org/abs/1212.0551} {arXiv:1212.0551 [gr-qc]} \BibitemShut
  {NoStop}%
%%CITATION = ARXIV:1212.0551;%%
\bibitem [{\citenamefont {Dolan}(2013)}]{Dolan:2012yt}%
  \BibitemOpen
  \bibfield  {author} {\bibinfo {author} {\bibfnamefont {S.~R.}\ \bibnamefont
  {Dolan}},\ }\href {\doibase 10.1103/PhysRevD.87.124026} {\bibfield  {journal}
  {\bibinfo  {journal} {Phys.~Rev.}\ }\textbf {\bibinfo {volume} {D 87}},\
  \bibinfo {pages} {124026} (\bibinfo {year} {2013})},\ \Eprint
  {http://arxiv.org/abs/1212.1477} {arXiv:1212.1477 [gr-qc]} \BibitemShut
  {NoStop}%
%%CITATION = ARXIV:1212.1477;%%
\bibitem [{\citenamefont
  {Shlapentokh-Rothman}(2014)}]{Shlapentokh-Rothman:2013ysa}%
  \BibitemOpen
  \bibfield  {author} {\bibinfo {author} {\bibfnamefont {Y.}~\bibnamefont
  {Shlapentokh-Rothman}},\ }\href {\doibase 10.1007/s00220-014-2033-x}
  {\bibfield  {journal} {\bibinfo  {journal} {Commun.~Math.~Phys.}\ }\textbf
  {\bibinfo {volume} {329}},\ \bibinfo {pages} {859} (\bibinfo {year}
  {2014})},\ \Eprint {http://arxiv.org/abs/1302.3448} {arXiv:1302.3448 [gr-qc]}
  \BibitemShut {NoStop}%
%%CITATION = ARXIV:1302.3448;%%
\bibitem [{\citenamefont {Berti}\ \emph {et~al.}(2015)\citenamefont {Berti}
  \emph {et~al.}}]{Berti:2015itd}%
  \BibitemOpen
  \bibfield  {author} {\bibinfo {author} {\bibfnamefont {E.}~\bibnamefont
  {Berti}} \emph {et~al.},\ }\href {\doibase 10.1088/0264-9381/32/24/243001}
  {\bibfield  {journal} {\bibinfo  {journal} {Class. Quant. Grav.}\ }\textbf
  {\bibinfo {volume} {32}},\ \bibinfo {pages} {243001} (\bibinfo {year}
  {2015})},\ \Eprint {http://arxiv.org/abs/1501.07274} {arXiv:1501.07274
  [gr-qc]} \BibitemShut {NoStop}%
%%CITATION = ARXIV:1501.07274;%%
\bibitem [{\citenamefont {van Putten}(1999)}]{VanPutten:1999vda}%
  \BibitemOpen
  \bibfield  {author} {\bibinfo {author} {\bibfnamefont {M.~H. P.~M.}\
  \bibnamefont {van Putten}},\ }\href {\doibase 10.1126/science.284.5411.115}
  {\bibfield  {journal} {\bibinfo  {journal} {Science}\ }\textbf {\bibinfo
  {volume} {284}},\ \bibinfo {pages} {115} (\bibinfo {year}
  {1999})}\BibitemShut {NoStop}%
%%CITATION = SCIEA,284,115;%%
\bibitem [{\citenamefont {Israel}(1968)}]{Israel:1967za}%
  \BibitemOpen
  \bibfield  {author} {\bibinfo {author} {\bibfnamefont {W.}~\bibnamefont
  {Israel}},\ }\href {\doibase 10.1007/BF01645859} {\bibfield  {journal}
  {\bibinfo  {journal} {Commun.~Math.~Phys.}\ }\textbf {\bibinfo {volume}
  {8}},\ \bibinfo {pages} {245} (\bibinfo {year} {1968})}\BibitemShut {NoStop}%
%%CITATION = CMPHA,8,245;%%
\bibitem [{\citenamefont {Carter}(1971)}]{Carter:1971zc}%
  \BibitemOpen
  \bibfield  {author} {\bibinfo {author} {\bibfnamefont {B.}~\bibnamefont
  {Carter}},\ }\href {\doibase 10.1103/PhysRevLett.26.331} {\bibfield
  {journal} {\bibinfo  {journal} {Phys.~Rev.~Lett.}\ }\textbf {\bibinfo
  {volume} {26}},\ \bibinfo {pages} {331} (\bibinfo {year} {1971})}\BibitemShut
  {NoStop}%
%%CITATION = PRLTA,26,331;%%
\bibitem [{\citenamefont {Ruffini}\ and\ \citenamefont
  {Wheeler}(1971)}]{Ruffini:1971bza}%
  \BibitemOpen
  \bibfield  {author} {\bibinfo {author} {\bibfnamefont {R.}~\bibnamefont
  {Ruffini}}\ and\ \bibinfo {author} {\bibfnamefont {J.~A.}\ \bibnamefont
  {Wheeler}},\ }\href {\doibase 10.1063/1.3022513} {\bibfield  {journal}
  {\bibinfo  {journal} {Phys.~Today}\ }\textbf {\bibinfo {volume} {24}},\
  \bibinfo {pages} {30} (\bibinfo {year} {1971})}\BibitemShut {NoStop}%
%%CITATION = PHTOA,24,30;%%
\bibitem [{\citenamefont {Bekenstein}(1972{\natexlab{a}})}]{Bekenstein:1971hc}%
  \BibitemOpen
  \bibfield  {author} {\bibinfo {author} {\bibfnamefont {J.~D.}\ \bibnamefont
  {Bekenstein}},\ }\href {\doibase 10.1103/PhysRevD.5.1239} {\bibfield
  {journal} {\bibinfo  {journal} {Phys.~Rev.}\ }\textbf {\bibinfo {volume} {D
  5}},\ \bibinfo {pages} {1239} (\bibinfo {year}
  {1972}{\natexlab{a}})}\BibitemShut {NoStop}%
%%CITATION = PHRVA,D5,1239;%%
\bibitem [{\citenamefont {Bekenstein}(1972{\natexlab{b}})}]{Bekenstein:1972ky}%
  \BibitemOpen
  \bibfield  {author} {\bibinfo {author} {\bibfnamefont {J.~D.}\ \bibnamefont
  {Bekenstein}},\ }\href {\doibase 10.1103/PhysRevD.5.2403} {\bibfield
  {journal} {\bibinfo  {journal} {Phys.~Rev.}\ }\textbf {\bibinfo {volume} {D
  5}},\ \bibinfo {pages} {2403} (\bibinfo {year}
  {1972}{\natexlab{b}})}\BibitemShut {NoStop}%
%%CITATION = PHRVA,D5,2403;%%
\bibitem [{\citenamefont {Bekenstein}(1995)}]{Bekenstein:1995un}%
  \BibitemOpen
  \bibfield  {author} {\bibinfo {author} {\bibfnamefont {J.~D.}\ \bibnamefont
  {Bekenstein}},\ }\href {\doibase 10.1103/PhysRevD.51.R6608} {\bibfield
  {journal} {\bibinfo  {journal} {Phys.~Rev.}\ }\textbf {\bibinfo {volume} {D
  51}},\ \bibinfo {pages} {6608} (\bibinfo {year} {1995})}\BibitemShut
  {NoStop}%
%%CITATION = PHRVA,D51,6608;%%
\bibitem [{\citenamefont {Bekenstein}(1996)}]{Bekenstein:1996pn}%
  \BibitemOpen
  \bibfield  {author} {\bibinfo {author} {\bibfnamefont {J.~D.}\ \bibnamefont
  {Bekenstein}},\ }\href@noop {} {\  (\bibinfo {year} {1996})},\ \Eprint
  {http://arxiv.org/abs/gr-qc/9605059} {gr-qc/9605059} \BibitemShut {NoStop}%
%%CITATION = GR-QC/9605059;%%
\bibitem [{\citenamefont {Chrusciel}\ \emph {et~al.}(2012)\citenamefont
  {Chrusciel}, \citenamefont {Costa}, \citenamefont {Heusler} \emph
  {et~al.}}]{Chrusciel:2012}%
  \BibitemOpen
  \bibfield  {author} {\bibinfo {author} {\bibfnamefont {P.~T.}\ \bibnamefont
  {Chrusciel}}, \bibinfo {author} {\bibfnamefont {J.~L.}\ \bibnamefont
  {Costa}}, \bibinfo {author} {\bibfnamefont {M.}~\bibnamefont {Heusler}},
  \emph {et~al.},\ }\href@noop {} {\bibfield  {journal} {\bibinfo  {journal}
  {Living Rev.~Relativity}\ }\textbf {\bibinfo {volume} {15}},\ \bibinfo
  {pages} {7} (\bibinfo {year} {2012})}\BibitemShut {NoStop}%
\bibitem [{\citenamefont {Whiting}(1989)}]{Whiting:1988vc}%
  \BibitemOpen
  \bibfield  {author} {\bibinfo {author} {\bibfnamefont {B.~F.}\ \bibnamefont
  {Whiting}},\ }\href {\doibase 10.1063/1.528308} {\bibfield  {journal}
  {\bibinfo  {journal} {J.~Math.~Phys.}\ }\textbf {\bibinfo {volume} {30}},\
  \bibinfo {pages} {1301} (\bibinfo {year} {1989})}\BibitemShut {NoStop}%
%%CITATION = JMAPA,30,1301;%%
\bibitem [{\citenamefont
  {Shlapentokh-Rothman}(2015)}]{Shlapentokh-Rothman:2013hza}%
  \BibitemOpen
  \bibfield  {author} {\bibinfo {author} {\bibfnamefont {Y.}~\bibnamefont
  {Shlapentokh-Rothman}},\ }\href {\doibase 10.1007/s00023-014-0315-7}
  {\bibfield  {journal} {\bibinfo  {journal} {Annales Henri Poincare}\ }\textbf
  {\bibinfo {volume} {16}},\ \bibinfo {pages} {289} (\bibinfo {year} {2015})},\
  \Eprint {http://arxiv.org/abs/1302.6902} {arXiv:1302.6902 [gr-qc]}
  \BibitemShut {NoStop}%
%%CITATION = ARXIV:1302.6902;%%
\bibitem [{\citenamefont {Dafermos}\ \emph {et~al.}(2014)\citenamefont
  {Dafermos}, \citenamefont {Rodnianski},\ and\ \citenamefont
  {Shlapentokh-Rothman}}]{Dafermos:2014cua}%
  \BibitemOpen
  \bibfield  {author} {\bibinfo {author} {\bibfnamefont {M.}~\bibnamefont
  {Dafermos}}, \bibinfo {author} {\bibfnamefont {I.}~\bibnamefont
  {Rodnianski}}, \ and\ \bibinfo {author} {\bibfnamefont {Y.}~\bibnamefont
  {Shlapentokh-Rothman}},\ }\href@noop {} {\  (\bibinfo {year} {2014})},\
  \Eprint {http://arxiv.org/abs/1402.7034} {arXiv:1402.7034 [gr-qc]}
  \BibitemShut {NoStop}%
%%CITATION = ARXIV:1402.7034;%%
\bibitem [{\citenamefont {Herdeiro}\ and\ \citenamefont
  {Radu}(2014{\natexlab{a}})}]{Herdeiro:2014goa}%
  \BibitemOpen
  \bibfield  {author} {\bibinfo {author} {\bibfnamefont {C.~A.~R.}\
  \bibnamefont {Herdeiro}}\ and\ \bibinfo {author} {\bibfnamefont
  {E.}~\bibnamefont {Radu}},\ }\href {\doibase 10.1103/PhysRevLett.112.221101}
  {\bibfield  {journal} {\bibinfo  {journal} {Phys.~Rev.~Lett.}\ }\textbf
  {\bibinfo {volume} {112}},\ \bibinfo {pages} {221101} (\bibinfo {year}
  {2014}{\natexlab{a}})},\ \Eprint {http://arxiv.org/abs/1403.2757}
  {arXiv:1403.2757 [gr-qc]} \BibitemShut {NoStop}%
%%CITATION = ARXIV:1403.2757;%%
\bibitem [{\citenamefont {Herdeiro}\ and\ \citenamefont
  {Radu}(2015{\natexlab{a}})}]{Herdeiro:2015gia}%
  \BibitemOpen
  \bibfield  {author} {\bibinfo {author} {\bibfnamefont {C.~A.~R.}\
  \bibnamefont {Herdeiro}}\ and\ \bibinfo {author} {\bibfnamefont
  {E.}~\bibnamefont {Radu}},\ }\href {\doibase 10.1088/0264-9381/32/14/144001}
  {\bibfield  {journal} {\bibinfo  {journal} {Class.~Quant.~Grav.}\ }\textbf
  {\bibinfo {volume} {32}},\ \bibinfo {pages} {144001} (\bibinfo {year}
  {2015}{\natexlab{a}})},\ \Eprint {http://arxiv.org/abs/1501.04319}
  {arXiv:1501.04319 [gr-qc]} \BibitemShut {NoStop}%
%%CITATION = ARXIV:1501.04319;%%
\bibitem [{\citenamefont {Dias}\ \emph {et~al.}(2011)\citenamefont {Dias},
  \citenamefont {Horowitz},\ and\ \citenamefont {Santos}}]{Dias:2011at}%
  \BibitemOpen
  \bibfield  {author} {\bibinfo {author} {\bibfnamefont {O.~J.~C.}\
  \bibnamefont {Dias}}, \bibinfo {author} {\bibfnamefont {G.~T.}\ \bibnamefont
  {Horowitz}}, \ and\ \bibinfo {author} {\bibfnamefont {J.~E.}\ \bibnamefont
  {Santos}},\ }\href {\doibase 10.1007/JHEP07(2011)115} {\bibfield  {journal}
  {\bibinfo  {journal} {JHEP}\ }\textbf {\bibinfo {volume} {07}},\ \bibinfo
  {pages} {115} (\bibinfo {year} {2011})},\ \Eprint
  {http://arxiv.org/abs/1105.4167} {arXiv:1105.4167 [hep-th]} \BibitemShut
  {NoStop}%
%%CITATION = ARXIV:1105.4167;%%
\bibitem [{\citenamefont {Herdeiro}\ and\ \citenamefont
  {Radu}(2015{\natexlab{b}})}]{Herdeiro:2015waa}%
  \BibitemOpen
  \bibfield  {author} {\bibinfo {author} {\bibfnamefont {C.~A.~R.}\
  \bibnamefont {Herdeiro}}\ and\ \bibinfo {author} {\bibfnamefont
  {E.}~\bibnamefont {Radu}},\ }\href {\doibase 10.1142/S0218271815420146}
  {\bibfield  {journal} {\bibinfo  {journal} {Int.~J.~Mod.~Phys.}\ }\textbf
  {\bibinfo {volume} {D 24}},\ \bibinfo {pages} {1542014} (\bibinfo {year}
  {2015}{\natexlab{b}})},\ \Eprint {http://arxiv.org/abs/1504.08209}
  {arXiv:1504.08209 [gr-qc]} \BibitemShut {NoStop}%
%%CITATION = ARXIV:1504.08209;%%
\bibitem [{\citenamefont {Hod}(2012)}]{Hod:2012px}%
  \BibitemOpen
  \bibfield  {author} {\bibinfo {author} {\bibfnamefont {S.}~\bibnamefont
  {Hod}},\ }\href {\doibase 10.1103/PhysRevD.86.129902,
  10.1103/PhysRevD.86.104026} {\bibfield  {journal} {\bibinfo  {journal} {Phys.
  Rev.}\ }\textbf {\bibinfo {volume} {D 86}},\ \bibinfo {pages} {104026}
  (\bibinfo {year} {2012})},\ \bibinfo {note} {[Erratum: Phys.
  Rev.D86,129902(2012)]},\ \Eprint {http://arxiv.org/abs/1211.3202}
  {arXiv:1211.3202 [gr-qc]} \BibitemShut {NoStop}%
%%CITATION = ARXIV:1211.3202;%%
\bibitem [{\citenamefont {Gundlach}\ and\ \citenamefont
  {Martin-Garcia}(1996)}]{Gundlach:1996vv}%
  \BibitemOpen
  \bibfield  {author} {\bibinfo {author} {\bibfnamefont {C.}~\bibnamefont
  {Gundlach}}\ and\ \bibinfo {author} {\bibfnamefont {J.~M.}\ \bibnamefont
  {Martin-Garcia}},\ }\href {\doibase 10.1103/PhysRevD.54.7353} {\bibfield
  {journal} {\bibinfo  {journal} {Phys.~Rev.}\ }\textbf {\bibinfo {volume} {D
  54}},\ \bibinfo {pages} {7353} (\bibinfo {year} {1996})},\ \Eprint
  {http://arxiv.org/abs/gr-qc/9606072} {gr-qc/9606072} \BibitemShut {NoStop}%
%%CITATION = GR-QC/9606072;%%
\bibitem [{\citenamefont {Degollado}\ \emph {et~al.}(2013)\citenamefont
  {Degollado}, \citenamefont {Herdeiro},\ and\ \citenamefont
  {Runarsson}}]{Herdeiro:2013pia}%
  \BibitemOpen
  \bibfield  {author} {\bibinfo {author} {\bibfnamefont {J.~C.}\ \bibnamefont
  {Degollado}}, \bibinfo {author} {\bibfnamefont {C.~A.~R.}\ \bibnamefont
  {Herdeiro}}, \ and\ \bibinfo {author} {\bibfnamefont {H.~F.}\ \bibnamefont
  {Runarsson}},\ }\href {\doibase 10.1103/PhysRevD.88.063003} {\bibfield
  {journal} {\bibinfo  {journal} {Phys.~Rev.}\ }\textbf {\bibinfo {volume} {D
  88}},\ \bibinfo {pages} {063003} (\bibinfo {year} {2013})},\ \Eprint
  {http://arxiv.org/abs/1305.5513} {arXiv:1305.5513 [gr-qc]} \BibitemShut
  {NoStop}%
%%CITATION = ARXIV:1305.5513;%%
\bibitem [{\citenamefont {Degollado}\ and\ \citenamefont
  {Herdeiro}(2014)}]{Degollado:2013bha}%
  \BibitemOpen
  \bibfield  {author} {\bibinfo {author} {\bibfnamefont {J.~C.}\ \bibnamefont
  {Degollado}}\ and\ \bibinfo {author} {\bibfnamefont {C.~A.~R.}\ \bibnamefont
  {Herdeiro}},\ }\href {\doibase 10.1103/PhysRevD.89.063005} {\bibfield
  {journal} {\bibinfo  {journal} {Phys.~Rev.}\ }\textbf {\bibinfo {volume} {D
  89}},\ \bibinfo {pages} {063005} (\bibinfo {year} {2014})},\ \Eprint
  {http://arxiv.org/abs/1312.4579} {arXiv:1312.4579 [gr-qc]} \BibitemShut
  {NoStop}%
%%CITATION = ARXIV:1312.4579;%%
\bibitem [{\citenamefont {Hod}(2013)}]{Hod:2013fvl}%
  \BibitemOpen
  \bibfield  {author} {\bibinfo {author} {\bibfnamefont {S.}~\bibnamefont
  {Hod}},\ }\href {\doibase 10.1103/PhysRevD.88.064055} {\bibfield  {journal}
  {\bibinfo  {journal} {Phys. Rev.}\ }\textbf {\bibinfo {volume} {D 88}},\
  \bibinfo {pages} {064055} (\bibinfo {year} {2013})},\ \Eprint
  {http://arxiv.org/abs/1310.6101} {arXiv:1310.6101 [gr-qc]} \BibitemShut
  {NoStop}%
%%CITATION = ARXIV:1310.6101;%%
\bibitem [{\citenamefont {Cardoso}\ \emph {et~al.}(2004)\citenamefont
  {Cardoso}, \citenamefont {Dias}, \citenamefont {Lemos},\ and\ \citenamefont
  {Yoshida}}]{Cardoso:2004nk}%
  \BibitemOpen
  \bibfield  {author} {\bibinfo {author} {\bibfnamefont {V.}~\bibnamefont
  {Cardoso}}, \bibinfo {author} {\bibfnamefont {O.~J.~C.}\ \bibnamefont
  {Dias}}, \bibinfo {author} {\bibfnamefont {J.~P.~S.}\ \bibnamefont {Lemos}},
  \ and\ \bibinfo {author} {\bibfnamefont {S.}~\bibnamefont {Yoshida}},\ }\href
  {\doibase 10.1103/PhysRevD.70.049903, 10.1103/PhysRevD.70.044039} {\bibfield
  {journal} {\bibinfo  {journal} {Phys.~Rev.}\ }\textbf {\bibinfo {volume} {D
  70}},\ \bibinfo {pages} {044039} (\bibinfo {year} {2004})},\ \Eprint
  {http://arxiv.org/abs/hep-th/0404096} {hep-th/0404096} \BibitemShut {NoStop}%
%%CITATION = HEP-TH/0404096;%%
\bibitem [{\citenamefont {Cardoso}\ and\ \citenamefont
  {Dias}(2004)}]{Cardoso:2004hs}%
  \BibitemOpen
  \bibfield  {author} {\bibinfo {author} {\bibfnamefont {V.}~\bibnamefont
  {Cardoso}}\ and\ \bibinfo {author} {\bibfnamefont {O.~J.~C.}\ \bibnamefont
  {Dias}},\ }\href {\doibase 10.1103/PhysRevD.70.084011} {\bibfield  {journal}
  {\bibinfo  {journal} {Phys. Rev.}\ }\textbf {\bibinfo {volume} {D 70}},\
  \bibinfo {pages} {084011} (\bibinfo {year} {2004})},\ \Eprint
  {http://arxiv.org/abs/hep-th/0405006} {arXiv:hep-th/0405006 [hep-th]}
  \BibitemShut {NoStop}%
%%CITATION = HEP-TH/0405006;%%
\bibitem [{\citenamefont {Bhattacharyya}\ \emph {et~al.}(2011)\citenamefont
  {Bhattacharyya}, \citenamefont {Minwalla},\ and\ \citenamefont
  {Papadodimas}}]{Bhattacharyya:2010yg}%
  \BibitemOpen
  \bibfield  {author} {\bibinfo {author} {\bibfnamefont {S.}~\bibnamefont
  {Bhattacharyya}}, \bibinfo {author} {\bibfnamefont {S.}~\bibnamefont
  {Minwalla}}, \ and\ \bibinfo {author} {\bibfnamefont {K.}~\bibnamefont
  {Papadodimas}},\ }\href {\doibase 10.1007/JHEP11(2011)035} {\bibfield
  {journal} {\bibinfo  {journal} {JHEP}\ }\textbf {\bibinfo {volume} {11}},\
  \bibinfo {pages} {035} (\bibinfo {year} {2011})},\ \Eprint
  {http://arxiv.org/abs/1005.1287} {arXiv:1005.1287 [hep-th]} \BibitemShut
  {NoStop}%
%%CITATION = ARXIV:1005.1287;%%
\bibitem [{\citenamefont {Dias}\ \emph {et~al.}(2012)\citenamefont {Dias},
  \citenamefont {Figueras}, \citenamefont {Minwalla}, \citenamefont {Mitra},
  \citenamefont {Monteiro},\ and\ \citenamefont {Santos}}]{Dias:2011tj}%
  \BibitemOpen
  \bibfield  {author} {\bibinfo {author} {\bibfnamefont {O.~J.~C.}\
  \bibnamefont {Dias}}, \bibinfo {author} {\bibfnamefont {P.}~\bibnamefont
  {Figueras}}, \bibinfo {author} {\bibfnamefont {S.}~\bibnamefont {Minwalla}},
  \bibinfo {author} {\bibfnamefont {P.}~\bibnamefont {Mitra}}, \bibinfo
  {author} {\bibfnamefont {R.}~\bibnamefont {Monteiro}}, \ and\ \bibinfo
  {author} {\bibfnamefont {J.~E.}\ \bibnamefont {Santos}},\ }\href {\doibase
  10.1007/JHEP08(2012)117} {\bibfield  {journal} {\bibinfo  {journal} {JHEP}\
  }\textbf {\bibinfo {volume} {08}},\ \bibinfo {pages} {117} (\bibinfo {year}
  {2012})},\ \Eprint {http://arxiv.org/abs/1112.4447} {arXiv:1112.4447
  [hep-th]} \BibitemShut {NoStop}%
%%CITATION = ARXIV:1112.4447;%%
\bibitem [{\citenamefont {Herdeiro}\ and\ \citenamefont
  {Radu}(2014{\natexlab{b}})}]{Herdeiro:2014ima}%
  \BibitemOpen
  \bibfield  {author} {\bibinfo {author} {\bibfnamefont {C.~A.~R.}\
  \bibnamefont {Herdeiro}}\ and\ \bibinfo {author} {\bibfnamefont
  {E.}~\bibnamefont {Radu}},\ }\href {\doibase 10.1142/S0218271814420140}
  {\bibfield  {journal} {\bibinfo  {journal} {Int.~J.~Mod.~Phys.}\ }\textbf
  {\bibinfo {volume} {D 23}},\ \bibinfo {pages} {1442014} (\bibinfo {year}
  {2014}{\natexlab{b}})},\ \Eprint {http://arxiv.org/abs/1405.3696}
  {arXiv:1405.3696 [gr-qc]} \BibitemShut {NoStop}%
%%CITATION = ARXIV:1405.3696;%%
\bibitem [{\citenamefont {Yoshino}\ and\ \citenamefont
  {Kodama}(2012)}]{Yoshino:2012kn}%
  \BibitemOpen
  \bibfield  {author} {\bibinfo {author} {\bibfnamefont {H.}~\bibnamefont
  {Yoshino}}\ and\ \bibinfo {author} {\bibfnamefont {H.}~\bibnamefont
  {Kodama}},\ }\href {\doibase 10.1143/PTP.128.153} {\bibfield  {journal}
  {\bibinfo  {journal} {Prog.~Theor.~Phys.}\ }\textbf {\bibinfo {volume}
  {128}},\ \bibinfo {pages} {153} (\bibinfo {year} {2012})},\ \Eprint
  {http://arxiv.org/abs/1203.5070} {arXiv:1203.5070 [gr-qc]} \BibitemShut
  {NoStop}%
%%CITATION = ARXIV:1203.5070;%%
\bibitem [{\citenamefont {Yoshino}\ and\ \citenamefont
  {Kodama}(2014)}]{Yoshino:2013ofa}%
  \BibitemOpen
  \bibfield  {author} {\bibinfo {author} {\bibfnamefont {H.}~\bibnamefont
  {Yoshino}}\ and\ \bibinfo {author} {\bibfnamefont {H.}~\bibnamefont
  {Kodama}},\ }\href {\doibase 10.1093/ptep/ptu029} {\bibfield  {journal}
  {\bibinfo  {journal} {Prog.~Theor.~Exp.~Phys.}\ }\textbf {\bibinfo {volume}
  {2014}},\ \bibinfo {pages} {043E02} (\bibinfo {year} {2014})},\ \Eprint
  {http://arxiv.org/abs/1312.2326} {arXiv:1312.2326 [gr-qc]} \BibitemShut
  {NoStop}%
%%CITATION = ARXIV:1312.2326;%%
\bibitem [{\citenamefont {Brito}\ \emph
  {et~al.}(2015{\natexlab{b}})\citenamefont {Brito}, \citenamefont {Cardoso},\
  and\ \citenamefont {Pani}}]{Brito:2014wla}%
  \BibitemOpen
  \bibfield  {author} {\bibinfo {author} {\bibfnamefont {R.}~\bibnamefont
  {Brito}}, \bibinfo {author} {\bibfnamefont {V.}~\bibnamefont {Cardoso}}, \
  and\ \bibinfo {author} {\bibfnamefont {P.}~\bibnamefont {Pani}},\ }\href
  {\doibase 10.1088/0264-9381/32/13/134001} {\bibfield  {journal} {\bibinfo
  {journal} {Class.~Quant.~Grav.}\ }\textbf {\bibinfo {volume} {32}},\ \bibinfo
  {pages} {134001} (\bibinfo {year} {2015}{\natexlab{b}})},\ \Eprint
  {http://arxiv.org/abs/1411.0686} {arXiv:1411.0686 [gr-qc]} \BibitemShut
  {NoStop}%
%%CITATION = ARXIV:1411.0686;%%
\bibitem [{\citenamefont {Okawa}\ \emph {et~al.}(2014)\citenamefont {Okawa},
  \citenamefont {Witek},\ and\ \citenamefont {Cardoso}}]{Okawa:2014nda}%
  \BibitemOpen
  \bibfield  {author} {\bibinfo {author} {\bibfnamefont {H.}~\bibnamefont
  {Okawa}}, \bibinfo {author} {\bibfnamefont {H.}~\bibnamefont {Witek}}, \ and\
  \bibinfo {author} {\bibfnamefont {V.}~\bibnamefont {Cardoso}},\ }\href
  {\doibase 10.1103/PhysRevD.89.104032} {\bibfield  {journal} {\bibinfo
  {journal} {Phys.~Rev.}\ }\textbf {\bibinfo {volume} {D 89}},\ \bibinfo
  {pages} {104032} (\bibinfo {year} {2014})},\ \Eprint
  {http://arxiv.org/abs/1401.1548} {arXiv:1401.1548 [gr-qc]} \BibitemShut
  {NoStop}%
%%CITATION = ARXIV:1401.1548;%%
\bibitem [{\citenamefont {Sanchis-Gual}\ \emph {et~al.}(2015)\citenamefont
  {Sanchis-Gual}, \citenamefont {Degollado}, \citenamefont {Montero},\ and\
  \citenamefont {Font}}]{Sanchis-Gual:2014ewa}%
  \BibitemOpen
  \bibfield  {author} {\bibinfo {author} {\bibfnamefont {N.}~\bibnamefont
  {Sanchis-Gual}}, \bibinfo {author} {\bibfnamefont {J.~C.}\ \bibnamefont
  {Degollado}}, \bibinfo {author} {\bibfnamefont {P.~J.}\ \bibnamefont
  {Montero}}, \ and\ \bibinfo {author} {\bibfnamefont {J.~A.}\ \bibnamefont
  {Font}},\ }\href {\doibase 10.1103/PhysRevD.91.043005} {\bibfield  {journal}
  {\bibinfo  {journal} {Phys.~Rev.}\ }\textbf {\bibinfo {volume} {D 91}},\
  \bibinfo {pages} {043005} (\bibinfo {year} {2015})},\ \Eprint
  {http://arxiv.org/abs/1412.8304} {arXiv:1412.8304 [gr-qc]} \BibitemShut
  {NoStop}%
%%CITATION = ARXIV:1412.8304;%%
\bibitem [{\citenamefont {Zilh\~ao}\ \emph {et~al.}(2015)\citenamefont
  {Zilh\~ao}, \citenamefont {Witek},\ and\ \citenamefont
  {Cardoso}}]{Zilhao:2015tya}%
  \BibitemOpen
  \bibfield  {author} {\bibinfo {author} {\bibfnamefont {M.}~\bibnamefont
  {Zilh\~ao}}, \bibinfo {author} {\bibfnamefont {H.}~\bibnamefont {Witek}}, \
  and\ \bibinfo {author} {\bibfnamefont {V.}~\bibnamefont {Cardoso}},\ }\href
  {\doibase 10.1088/0264-9381/32/23/234003} {\bibfield  {journal} {\bibinfo
  {journal} {Class. Quant. Grav.}\ }\textbf {\bibinfo {volume} {32}},\ \bibinfo
  {pages} {234003} (\bibinfo {year} {2015})},\ \Eprint
  {http://arxiv.org/abs/1505.00797} {arXiv:1505.00797 [gr-qc]} \BibitemShut
  {NoStop}%
%%CITATION = ARXIV:1505.00797;%%
\bibitem [{\citenamefont {Witek}\ \emph {et~al.}(2010)\citenamefont {Witek},
  \citenamefont {Cardoso}, \citenamefont {Herdeiro}, \citenamefont {Nerozzi},
  \citenamefont {Sperhake} \emph {et~al.}}]{Witek:2010qc}%
  \BibitemOpen
  \bibfield  {author} {\bibinfo {author} {\bibfnamefont {H.}~\bibnamefont
  {Witek}}, \bibinfo {author} {\bibfnamefont {V.}~\bibnamefont {Cardoso}},
  \bibinfo {author} {\bibfnamefont {C.~A.~R.}\ \bibnamefont {Herdeiro}},
  \bibinfo {author} {\bibfnamefont {A.}~\bibnamefont {Nerozzi}}, \bibinfo
  {author} {\bibfnamefont {U.}~\bibnamefont {Sperhake}},  \emph {et~al.},\
  }\href {\doibase 10.1103/PhysRevD.82.104037} {\bibfield  {journal} {\bibinfo
  {journal} {Phys.~Rev.}\ }\textbf {\bibinfo {volume} {D 82}},\ \bibinfo
  {pages} {104037} (\bibinfo {year} {2010})},\ \Eprint
  {http://arxiv.org/abs/1004.4633} {arXiv:1004.4633 [hep-th]} \BibitemShut
  {NoStop}%
%%CITATION = ARXIV:1004.4633;%%
\end{thebibliography}%

\end{document}